{}%disable hyper at arxiv
{}%use pdflatex at arxiv

\documentclass[11pt,a4paper]{article}

\newif\ifpublic\publictrue

\ifpublic\else\usepackage{showkeys}\fi
\usepackage[bookmarks=true,hyperfigures=true,colorlinks=true,linkcolor=black,citecolor=black,urlcolor=black,bookmarksnumbered]{hyperref}
\usepackage[nosort]{cite}
\usepackage[parsep]{collref}
\usepackage[english]{babel}
\usepackage[T1]{fontenc}
\usepackage[latin1]{inputenc}
\usepackage{amsfonts,amsbsy,dsfont,euscript,mathrsfs,fixmath}
\usepackage{amsmath,amssymb}
\usepackage{enumerate}
\usepackage{array}

\def\showkeysrefformat#1{{\normalfont\tiny\ttfamily#1}}
\makeatletter
\def\SK@@ref#1>#2\SK@{{\@inlabelfalse\leavevmode\vbox to\z@{\vss\SK@refcolor\rlap{\vrule\raise .75em \hbox{\showkeysrefformat{#2}}}}}}
\makeatother

\usepackage[a4paper,text={160mm,247mm},centering]{geometry}
\allowdisplaybreaks[3]
\numberwithin{equation}{section}
\def\[{\begin{equation}\begin{aligned}}
\def\]{\end{aligned}\end{equation}}

\usepackage[font=small,labelfont=bf,width=0.85\textwidth]{caption}

\expandafter\def\expandafter\bfseries\expandafter{\bfseries\ifmmode\else\boldmath\fi}
\expandafter\def\expandafter\mdseries\expandafter{\mdseries\ifmmode\else\unboldmath\fi}
\expandafter\def\expandafter\normalfont\expandafter{\normalfont\ifmmode\else\unboldmath\fi}

\RequirePackage{verbatim}

\makeatletter
\newwrite\bibinl@out
\newenvironment{bibtex}[1][\jobname]{%
\immediate\openout\bibinl@out #1.bib%
\immediate\write\bibinl@out{\@percentchar generated from `\jobname' starting line \the\inputlineno^^J}%
\def\verbatim@processline{\immediate\write\bibinl@out{\the\verbatim@line}}%
\@bsphack\let\do\@makeother\dospecials\catcode`\^^M\active\verbatim@start%
}
{\immediate\closeout\bibinl@out\@esphack}
\makeatother

\let\barefrac=\frac
\renewcommand{\frac}[2]{\mathinner{\barefrac{#1}{#2}}}

\let\baresqrt=\sqrt
\makeatletter
\renewcommand{\sqrt}{\@ifnextchar[\@sqrt@space@a\@sqrt@space@b}
\def\@sqrt@space@a[#1]#2{\mathinner{\mathchoice{\mkern-3mu}{\mkern-3mu}{}{}\baresqrt[#1]{#2}}}
\def\@sqrt@space@b#1{\mathinner{\mathchoice{\mkern-3mu}{\mkern-3mu}{}{}\baresqrt{#1}}}
\makeatother

\makeatletter
\let\per@dot@old=\.
\def\.{\ifmmode\def\per@dot@sel{\mkern3mu}\else\def\per@dot@sel{\per@dot@old}\fi\per@dot@sel}
\makeatother

\let\barefootnote=\footnote
\renewcommand{\footnote}[1]{\barefootnote{#1\vspace{3pt}}}

\newcommand{\sfrac}[2]{{\textstyle\frac{#1}{#2}}}
\newcommand{\half}{\sfrac{1}{2}}

\newcommand{\vfrac}[2]{\ifmmode\mathinner{\textstyle^{#1}\!/\!_{#2}}\else$^{#1}\!/\!_{#2}$\fi}

\newcommand{\identity}{\mathds{1}}
\DeclareMathOperator{\Mat}{Mat}
\DeclareMathOperator{\diag}{diag}

\DeclareMathOperator{\Tr}{Tr}

\DeclareMathOperator{\STr}{STr}

\newcommand{\set}[1]{\{#1\}}

\newcommand{\Real}{\mathds{R}}
\newcommand{\Complex}{\mathds{C}}
\newcommand{\Integer}{\mathds{Z}}

\let\Re\relax\DeclareMathOperator{\Re}{Re}
\let\Im\relax\DeclareMathOperator{\Im}{Im}

\newcommand{\ind}[1]{{\scriptscriptstyle{#1}}}

\newcommand{\alg}[1]{\mathfrak{#1}}
\newcommand{\grp}[1]{\mathrm{#1}}
\DeclareMathOperator{\rank}{rank}
\DeclareMathOperator{\Lie}{Lie}
\DeclareMathOperator{\ad}{ad}
\DeclareMathOperator{\Ad}{Ad}

\newcommand{\dsum}{\oplus}

\newcommand{\com}[2]{[#1,#2]}
\newcommand{\anticom}[2]{\{#1,#2\}}

\newcommand{\Proj}[1]{\text{P}_{#1}}
\newcommand{\proj}{\text{P}}

\def\<{\big\langle}
\def\>{\big\rangle}
\newcommand\gen{\mathbb}
\newcommand\Fermion{\ind{\text{F}}}
\newcommand\Boson{\ind{\text{B}}}

\newcommand{\geom}[1]{\mathrm{#1}}

\newcommand{\AdS}{\geom{AdS}}

\newcommand{\Sp}{\geom{S}}

\newcommand{\To}{\geom{T}}

\newcommand{\extder}{\mathrm{d}}

\newcommand{\dx}{\extder^2 \sigma}

\newcommand{\Lag}{\mathcal{L}}
\newcommand{\Act}{\mathcal{S}}

\def\wasyfamily{\fontencoding{U}\fontfamily{wasy}\selectfont}
\def\Circle{\mbox{\wasyfamily\char35}}
\def\En{F_0}
\def\Gn{G_0}
\def\Bn{B_0}
\def\EP{F_\ind{\text{P}}}
\def\ER{F_\ind{\text{R}}}
\def\Ent{{\En^\prime}}

\def\lambdastar{\lambda^\star}
\def\rot{\mathrm{rot}}
\def\WZ{\mathrm{WZ}}
\def\kst{\tilde{k}}
\def\lst{\tilde{\lambda}}

\providecommand{\href}[2]{#2}

\makeatletter
\def\mr@ignsp#1 {\ifx\:#1\@empty\else #1\expandafter\mr@ignsp\fi}
\newcommand{\multiref}[1]{\begingroup%
\xdef\mr@no@sparg{\expandafter\mr@ignsp#1 \: }%
\def\mr@comma{}\def\mr@dash{-}%
\@for\mr@refs:=\mr@no@sparg\do{%
\ifx\mr@refs\mr@dash\def\mr@comma{}--\else%
\mr@comma\def\mr@comma{,}\ref{\mr@refs}\fi}%
\endgroup}
\renewcommand{\eqref}[1]{(\multiref{#1})}
\makeatother

\makeatletter
\newcommand{\namedref}[2]{\hyperref[#2]{#1~\ref*{#2}}}
\newcommand{\secref}{\@ifstar{\namedref{Section}}{\namedref{sec.}}}
\newcommand{\appref}{\@ifstar{\namedref{Appendix}}{\namedref{app.}}}
\newcommand{\tabref}{\@ifstar{\namedref{Table}}{\namedref{tab.}}}
\newcommand{\figref}{\@ifstar{\namedref{Figure}}{\namedref{fig.}}}
\makeatother

\let\oldbib=\thebibliography
\def\thebibliography{\phantomsection\addcontentsline{toc}{section}{\refname}\oldbib}

\let\oldtoc=\tableofcontents
\def\tableofcontents{\phantomsection\addcontentsline{toc}{section}{\contentsname}\oldtoc}

\providecommand{\hypersetup}[1]{}
\providecommand{\texorpdfstring}[2]{#1}
\hypersetup{plainpages=false}
\hypersetup{pdfpagemode=UseNone}
\hypersetup{bookmarksnumbered=true}
\hypersetup{pdfstartview=FitH}
\hypersetup{colorlinks=false}
\hypersetup{citebordercolor={1 1 1}}
\hypersetup{urlbordercolor={1 1 1}}
\hypersetup{linkbordercolor={1 1 1}}
\makeatletter
\let\@keywords\@empty
\let\@subject\@empty
\providecommand{\keywords}[1]{\gdef\@keywords{#1}}
\providecommand{\subject}[1]{\gdef\@subject{#1}}
\def\thetitle{\@title}
\def\theauthor{\@author}
\def\thesubject{\@subject}
\def\thedate{\@date}
\def\thekeywords{\@keywords}
\makeatother
\AtBeginDocument{
\hypersetup{pdftitle={\thetitle}}
\hypersetup{pdfauthor={\theauthor}}
\hypersetup{pdfsubject={\thesubject}}
\hypersetup{pdfkeywords={\thekeywords}}}

\newif\ifshownote
\shownotetrue

\ifpublic\shownotefalse\fi

\ifshownote

\ifpdf\else\RequirePackage[active]{srcltx}\fi
\RequirePackage{xcolor}
\RequirePackage{ulem}

\newcommand{\remark}[2][]{{\normalfont\sffamily\hspace{1ex}
\def\emph{\textsl}\def\textbullet{$\bullet$}
\def\tmparga{#1}
\def\tmpargb{BH}\ifx\tmparga\tmpargb\color[rgb]{0.5,0,0}\fi
\def\tmpargb{FS}\ifx\tmparga\tmpargb\color[rgb]{0,0.5,0}\fi
\def\tmpargb{}\ifx\tmparga\tmpargb\color{red}\fi
\def\tmpargb{}\ifx\tmparga\tmpargb\else \textbf{#1:}\fi
#2\hspace{1ex}}}

\else
\newcommand{\remark}[2][]{\ignorespaces}

\fi

\title{Poisson-Lie duals of the \texorpdfstring{\\}{} \texorpdfstring{$\eta$}{eta}-deformed \texorpdfstring{$\AdS_2 \times \Sp^2 \times \To^6$}{AdS2 x S2 x T6} superstring}
\author{Ben~Hoare and Fiona~K.~Seibold}

\begin{document}

\pdfbookmark[1]{Title Page}{title}
\thispagestyle{empty}

%\begingroup\raggedleft\footnotesize\ttfamily
%\arxivlink{yymm.nnnnnn}
%\par\endgroup

\vspace*{2cm}
\begin{center}
\begingroup\Large\bfseries\thetitle\par\endgroup
\vspace{1cm}

\begingroup\theauthor\par\endgroup
\vspace{1cm}

\textit{
Institut f\"ur Theoretische Physik,\\
Eidgen\"ossische Technische Hochschule Z\"urich,\\
Wolfgang-Pauli-Strasse 27, 8093 Z\"urich, Switzerland}
\vspace{5mm}

\begingroup\ttfamily\small
\verb+{+bhoare,fseibold\verb+}+@itp.phys.ethz.ch\par
\endgroup
\vspace{5mm}

\vfill

\textbf{Abstract}\vspace{5mm}

\begin{minipage}{12.5cm}\small
We investigate Poisson-Lie duals of the $\eta$-deformed $\AdS_2 \times \Sp^2 \times \To^6$ superstring.
The $\eta$-deformed background satisfies a generalisation of the type II supergravity equations.
We discuss three Poisson-Lie duals, with respect to \textit{(i)} the full $\mathfrak{psu}(1,1|2)$ superalgebra, \textit{(ii)} the full bosonic subalgebra and \textit{(iii)} the Cartan subalgebra, for which the corresponding backgrounds are expected to satisfy the standard type II supergravity equations.
The metrics and B-fields for the first two cases are the same and given by an analytic continuation of the $\lambda$-deformed model on $\AdS_2 \times \Sp^2 \times \To^6$ with the torus undeformed.
However, the RR fluxes and dilaton will differ.
Focusing on the second case we explicitly derive the background and show agreement with an analytic continuation of a known embedding of the $\lambda$-deformed model on $\AdS_2 \times \Sp^2$ in type II supergravity.
\end{minipage}

\vspace*{2cm}

\end{center}

\newpage

%%%%%%%%%%%%%%%%%%%%%%%%%%%%%%%%%%%%%%%%%%%%%%%%%%%%%%%%%%%%%%%%%%%%%%%%%%%%%%%%
\tableofcontents

%%%%%%%%%%%%%%%%%%%%%%%%%%%%%%%%%%%%%%%%%%%%%%%%%%%%%%%%%%%%%%%%%%%%%%%%%%%%%%%%
\section{Introduction}

In this paper we continue the exploration of Poisson-Lie duals of $\eta$-deformed sigma models initiated in \cite{Hoare:2017ukq}.
In \cite{Hoare:2017ukq} we investigated the Poisson-Lie duals \cite{Klimcik:1995ux,Klimcik:1995jn} of the $\eta$-deformation \cite{Klimcik:2002zj,Klimcik:2008eq,Delduc:2013fga} of the bosonic symmetric space sigma model on $\grp{G}/\grp{H}$ \cite{Eichenherr:1979ci,Eichenherr:1979hz} for compact groups $\grp{G}$.
Here we focus on the $\eta$-deformation of the $\AdS_2 \times \Sp^2 \times \To^6$ superstring.
To study this model we consider the semi-symmetric space sigma model \cite{Metsaev:1998it,Zhou:1999sm,Berkovits:1999zq,Zarembo:2010sg} on the supercoset
\begin{equation}\label{eq:supercoset}
\frac{\grp{PSU}(1,1|2)}{\grp{SO}(1,1)\times\grp{SO}(2)} ~,
\end{equation}
and its $\eta$-deformation \cite{Delduc:2013qra,Delduc:2014kha}.
The bosonic part of this model is the symmetric space sigma model on the coset
\begin{equation}
\frac{\grp{SU}(1,1)}{\grp{SO}(1,1)} \times \frac{\grp{SU}(2)}{\grp{SO}(2)} ~,
\end{equation}
that is with target space $\AdS_2 \times \Sp^2$.
The semi-symmetric space sigma model then describes a truncation of the type II Green-Schwarz superstring \cite{Green:1983wt,Green:1983sg,Witten:1985nt,Grisaru:1985fv} on certain $\AdS_2 \times \Sp^2 \times \To^6$ supergravity backgrounds \cite{Sorokin:2011rr}.
This truncation is well-understood for both the two-dimensional worldsheet sigma model and the supergravity background.

To define a particular $\eta$-deformation of the semi-symmetric space sigma model, we first need to specify an antisymmetric operator $R$ satisfying the non-split modified classical Yang-Baxter equation on the superalgebra $\alg{psu}(1,1|2)$.
We will take this R-matrix to be given by the canonical Drinfel'd-Jimbo solution associated to a particular Dynkin diagram and Cartan-Weyl basis of the superalgebra.
\unskip\footnote{The relation between $\eta$-deformations corresponding to inequivalent Cartan-Weyl bases and the associated Drinfel'd-Jimbo R-matrices is not fully understood.
These can exist for non-compact real forms of bosonic Lie algebras and have been partially investigated for the $\eta$-deformations of the sigma model on $\AdS_5$, for which the relevant Lie algebra is $\alg{so}(2,4)$, in \cite{Delduc:2014kha,Hoare:2016ibq,Araujo:2017enj}.
They can also exist for Lie superalgebras, for which there exist different Dynkin diagrams.}
For such a choice of R-matrix the manifest symmetry algebra of the deformed model is broken to the Cartan subalgebra.
Together with the remaining charges, which are hidden, the isometry algebra is $q$-deformed \cite{Delduc:2013fga,Delduc:2014kha,Delduc:2016ihq,Delduc:2017brb,Arutyunov:2013ega} with $q \in \Real$ depending on the string tension and the deformation parameter $\eta$.

The Poisson-Lie duals of the $\eta$-deformed $\AdS_2 \times \Sp^2 \times \To^6$ superstring can be studied starting from a model on the complexified double
\begin{equation}
\frac{\grp{PSL}(2|2;\Complex)}{\grp{SO}(1,1)\times\grp{SO}(2)} ~,
\end{equation}
following the general construction of \cite{Klimcik:1995dy,Klimcik:1996nq}, which is extended to coset spaces in \cite{Klimcik:1996np,Squellari:2011dg,Sfetsos:1999zm}.
The model is constructed such that on integrating out the degrees of freedom associated to an appropriate Borel subalgebra (that correlates with the R-matrix) we recover the $\eta$-deformation of interest.
Following the results of \cite{Hoare:2017ukq}, for subalgebras $\alg{g}_0$ of $\alg{psu}(1,1|2)$ corresponding to sub-Dynkin diagrams we can construct subalgebras of the complexified double $\alg{psl}(2|2;\Complex)$ whose associated degrees of freedom can be integrated out to give the Poisson-Lie dual of the $\eta$-deformed $\AdS_2 \times \Sp^2 \times \To^6$ superstring with respect to $\alg{g}_0$.
Any additional Cartan generators not covered by the sub-Dynkin diagram can also be included in $\alg{g}_0$.
It is likely that this is not a complete list of possible Poisson-Lie duals of the $\eta$-deformed $\AdS_2 \times \Sp^2 \times \To^6$ superstring (see, for example, \cite{Lust:2018jsx}).

In this paper we mostly work with the Dynkin diagram $\Circle - \otimes - \Circle$.
A discussion of the other possible Dynkin diagrams is given in \appref{app:Different_Dynkin}.
Let us briefly outline three possible Poisson-Lie duals that one can consider based on this choice:
\begin{enumerate}
\item First, one can consider the Poisson-Lie dual with respect to the full $\alg{psu}(1,1|2)$ superalgebra.
This model is conjectured \cite{Vicedo:2015pna,Hoare:2015gda,Sfetsos:2015nya,Klimcik:2015gba,Hoare:2017ukq,Driezen:2018glg} to be an analytic continuation of the $\lambda$-deformation of the $\AdS_2 \times \Sp^2 \times \To^6$ superstring \cite{Hollowood:2014qma} (generalising the bosonic $\lambda$-deformed models of \cite{Sfetsos:2013wia,Hollowood:2014rla}), which, following the terminology of \cite{Hoare:2017ukq} we refer to as the $\lambdastar$-deformed model.
\item Second, one can take the sub-Dynkin diagram formed of the two bosonic nodes.
This corresponds to dualising with respect to the full bosonic subalgebra $\alg{su}(1,1) \dsum \alg{su}(2)$.
The bosonic part of this model coincides with the $\lambdastar$-deformed model, however they differ in the fermionic part.
\item Finally, one can consider just the $\alg{u}(1) \oplus \alg{u}(1)$ subalgebra associated to the two Cartan generators.
This model is conjectured to be equivalent to taking the two-fold T-dual of the $\eta$-deformed $\AdS_2 \times \Sp^2 \times \To^6$ superstring.
\end{enumerate}

There is substantial evidence \cite{Alvarez:1994np,Elitzur:1994ri,Tyurin:1995bu,Bossard:2001au,VonUnge:2002xjf,Hlavaty:2004jp,Hlavaty:2004mr,Jurco:2017gii} that a Weyl anomaly is associated to integrating out the degrees of freedom of a non-unimodular algebra, that is when the trace of the structure constants is non-vanishing, $f_{ab}{}^b \neq 0$.
In this case, rather than solving the standard supergravity equations, the background solves a generalisation thereof \cite{Arutyunov:2015mqj,Wulff:2016tju} (as discussed in the context of non-abelian duality in \cite{Hoare:2016wsk,Hong:2018tlp,Wulff:2018aku}).
These generalised supergravity equations are equivalent to the $\kappa$-symmetry of the Green-Schwarz superstring \cite{Wulff:2016tju}.
They are also related to the standard supergravity equations by T-dualising a supergravity background in a $\grp{U}(1)$ isometry, $y \to y + c$, which is a symmetry of all the fields except the dilaton, $\Phi ~ \sim y + \ldots$ \cite{Arutyunov:2015mqj}.
The relation with dualities has been explored further in the context of generalised geometry, double field theory and exceptional field theory \cite{Sakatani:2016fvh,Baguet:2016prz,Sakamoto:2017wor,Fernandez-Melgarejo:2017oyu,Lust:2018jsx,Sakamoto:2018krs}.

The $\eta$-deformation of $\Sp^2$ \cite{Delduc:2013fga,Arutyunov:2013ega} is equivalent \cite{Hoare:2014pna} to the sausage model of \cite{Fateev:1992tk}.
A proposal for the $\eta$-deformation of the $\AdS_2 \times \Sp^2 \times \To^6$ supergravity background solving the generalised supergravity equations is given in app.~F of \cite{Arutyunov:2015mqj} (see also \cite{Borsato:2016ose,Bakhmatov:2017joy,Araujo:2018rbc}).
This is consistent as the Borel subalgebra, whose degrees of freedom we integrate out to give the $\eta$-deformed $\AdS_2 \times \Sp^2 \times \To^6$ superstring, is not unimodular.
For the three duals listed above the subalgebras whose degrees of freedom we integrate out are all unimodular, and hence the corresponding backgrounds are expected to solve the standard supergravity equations.
Note that, since all three models involve dualising in a timelike direction, these solutions may actually be of type II or II$^\star$ supergravity \cite{Hull:1998vg}. 
For the Poisson-Lie dual with respect to the full $\alg{psu}(1,1|2)$ superalgebra, assuming the conjectured relation to the $\lambda$-deformation \cite{Vicedo:2015pna,Hoare:2015gda,Sfetsos:2015nya,Klimcik:2015gba}, this is indeed the case \cite{Borsato:2016zcf}.
It is also true for the two-fold T-dual \cite{Hoare:2015gda,Hoare:2015wia,Arutyunov:2015mqj}.
For the remaining case, we recall that an alternative, arguably simpler, embedding of the metric of the $\lambda$-deformed model in supergravity to that of \cite{Borsato:2016zcf} is given in \cite{Sfetsos:2014cea}.
The analytic continuation of this background therefore provides a natural conjecture for the Poisson-Lie dual of the $\eta$-deformed $\AdS_2 \times \Sp^2 \times \To^6$ superstring with respect to the full bosonic subalgebra.
The main result of this paper is to confirm this proposal.

That the same metric and B-field can be supported by different RR fluxes is known in the literature.
Indeed, it is the case for the different embeddings of the metric of the $\lambda$-deformed model on $\AdS_2 \times \Sp^2$ \cite{Sfetsos:2014cea,Borsato:2016zcf} and is also discussed in \cite{Lunin:2014tsa} in the context of the $\eta$-deformed models.
Here, one mechanism by which this may happen, that is duality transformations with respect to different subalgebras, is studied.

The layout of this paper is as follows.
In \secref{sec:PLduality} we review the model on the Drinfel'd double and the formalism for constructing Poisson-Lie dual sigma models.
We then focus on the $\eta$-deformation of the $\AdS_2 \times \Sp^2$ supercoset in \secref{sec:etadeformation}, writing its action in a manifestly Poisson-Lie symmetric form.
Using these results, in \secref{sec:PLdualityBosonic} the background of the Poisson-Lie dual with respect to the full bosonic subalgebra is derived.
In \appref{app:Different_Dynkin} we discuss the different Dynkin diagrams of $\alg{psu}(1,1|2)$ in the context of $\eta$-deformations and Poisson-Lie duality.
In \appref{app:sl_generators} and \appref{app:gamma_matrices} we give our conventions for the superalgebras $\alg{psu}(1,1|2)$ and $\alg{pb}(1,1|2)$ and 4-dimensional and 32-dimensional gamma matrices respectively, while \appref{app:field_redef} contains technical details of the derivation of the background of the Poisson-Lie dual with respect to the full bosonic subalgebra.

%%%%%%%%%%%%%%%%%%%%%%%%%%%%%%%%%%%%%%%%%%%%%%%%%%%%%%%%%%%%%%%%%%%%%%%%%%%%%%%%
\section{Poisson-Lie duality and the Drinfel'd double}
\label{sec:PLduality}

\paragraph{The Drinfel'd double.}
Poisson-Lie duality \cite{Klimcik:1995ux,Klimcik:1995jn} is a generalisation of non-abelian duality to certain backgrounds that do not necessarily possess manifest isometries.
The underlying algebraic structure is a Drinfel'd double, defined as a $2n$-dimensional real connected Lie group $\grp{D}$ whose Lie algebra $\alg{d} = \Lie(\grp{D})$ can be decomposed as
\[
\label{eq:decompo_dgtg}
\alg{d} = \alg{g} \dsum \tilde{\alg{g}} ~,
\]
where $\alg{g}$ and $\tilde{\alg{g}}$ are two $n$-dimensional real Lie subalgebras, maximally isotropic with respect to a non-degenerate ad-invariant inner product $\beta(\cdot , \cdot )$ on $\alg{d}$,
\[
\beta(\alg{g}, \alg{g})=0~, \qquad \beta(\tilde{\alg{g}}, \tilde{\alg{g}}) =0~.
\]
When $\alg{d}$, $\alg{g}$ and $\tilde{\alg{g}}$ are Lie superalgebras, the $\Integer_2$ grading allows one to decompose them as
\[
\alg{d} = \alg{d}_\Boson \dsum \alg{d}_\Fermion~, \qquad \alg{g} = \alg{g}_\Boson \dsum \alg{g}_\Fermion~, \qquad \tilde{\alg{g}} = \tilde{\alg{g}}_\Boson \dsum \tilde{\alg{g}}_\Fermion~,
\]
where $\alg{d}_\Boson$ and $\alg{d}_\Fermion$ contain the elements of grade zero and one respectively.
The inner product should also be consistent with the $\Integer_2$ grading, supersymmetric
\[
\beta(X,Y) & = \beta(Y,X) = 0 ~, & \qquad X \in \alg{d}_\Boson, Y \in \alg{d}_\Fermion ~,
\\
\beta(X,Y) & = \beta(Y,X) ~, & \qquad X,Y \in \alg{d}_\Boson ~,
\\
\beta(X,Y) & = - \beta(Y,X) ~, & \qquad X,Y \in \alg{d}_\Fermion ~,
\]
and ad-invariant
\[
\beta( [X,Y\}, Z ) = \beta( X, [Y,Z\}) ~, \qquad X,Y,Z \in \alg{d} ~,
\]
where $[\cdot, \cdot \}$ is the $\Integer_2$-graded commutator.

\paragraph{First-order action on the Drinfel'd double.}
As for abelian and non-abelian duality, dual models can be obtained starting from an action for a dynamical field in the Drinfel'd double and integrating out half the degrees of freedom.
To define this first-order action on the Drinfel'd double we extend the definition of the inner-product to the Grassmann envelope of the superalgebra as
\[
\< X, Y\> &\equiv \beta(X, Y) ~, & \qquad X,Y &\in \alg{d}_\Boson ~, \\
\<\theta_1 X, \theta_2 Y \> &\equiv \textsf{c} \, \theta_1 \theta_2 \, \beta( X, Y) ~, & \qquad X, Y &\in \alg{d}_\Fermion ~, \\
\< X, \textsf{c} \,\theta_1 \theta_2 Y \> &\equiv \textsf{c} \,\theta_1 \theta_2 \, \beta(X, Y)~, & \qquad X, Y &\in \alg{d}_\Boson ~, \\
\]
and so on, where $\theta_1, \theta_2$ are real Grassmann variables and $ \textsf{c} \in \Complex$ is such that $|\textsf{c}|=1$ and $ \textsf{c}\, \theta_1 \theta_2 \in \Real$.
If $c$ is a complex number with complex conjugate $c^\star$, then we take the conjugation operation on Grassmann variables to be given by
\[\label{eq:conjugation}
(c\theta)^\star = c^\star \theta^\star~, \qquad (\theta^\star)^\star=\theta~, \qquad (\theta_1 \theta_2)^\star = \theta_2^\star \theta_1^\star~,
\]
and correspondingly
\[ \label{eq:cei} \textsf{c} = i ~.\]
The inner product between two elements of different grading vanishes.
The first-order action for the dynamical field $l \in \grp{D}$ that gives the Poisson-Lie dual models is \cite{Klimcik:1995dy,Klimcik:1996nq}
\[\label{eq:action_double}
\Act_\grp{D}(l) = \int \extder \tau \extder \sigma \, \Big[ \half \< l^{-1} \partial_\sigma l , l^{-1} \partial_\tau l \> - \half K(l^{-1} \partial_\sigma l ) \Big] +\WZ(l)~,
\]
where
\[
\WZ(l) = - \sfrac{1}{12} \int \, \extder^{-1} \< l^{-1} \extder l , \com{l^{-1} \extder l}{l^{-1} \extder l} \>~,
\]
is the standard Wess-Zumino term.
The quadratic form $K$, whose explicit form is discussed below, is model dependent and acts on the current, which takes values in the Grassmann envelope of the algebra $\alg{d}$.
Henceforth, we will use $\alg{d}$, $\alg{g}$, $\tilde{\alg{g}}$ and so on to refer to both the algebra and its Grassmann envelope.

\paragraph{Canonical Poisson-Lie duality.}
Starting from the action \eqref{eq:action_double} and using the decomposition of the Drinfel'd double \eqref{eq:decompo_dgtg}, where we recall both $\alg{g}$ and $\tilde{\alg{g}}$ are subalgebras, we recover the Poisson-Lie dual models on $\grp{G}$ and $\tilde{\grp{G}}$ by integrating out the degrees of freedom associated to $\tilde{\alg{g}}$ and $\alg{g}$ respectively \cite{Klimcik:1995dy,Klimcik:1996nq}.
To obtain the explicit form of the dual models we need to specify the action of the bilinear form $K$ on an arbitrary element $x \in \alg{d}$.
Such an element admits a unique decomposition $x = y+z$ where $y \in \alg{g}$ and $z \in \tilde{\alg{g}}$.
Without loss of generality one may then define the action of the bilinear form as
\[
K(x) = \< z, \Gn z \> + \< (y+\Bn z), \Gn^{-1} (y+\Bn z) \>~,
\]
where $\Gn$ and $\Bn$ are the symmetric and antisymmetric parts of some operator $\En: \tilde{\alg{g}} \rightarrow \alg{g}$ with respect to the inner product $\< \cdot , \cdot \>$.
To integrate out the degrees of freedom associated to $\tilde{\alg{g}}$, we parametrise the field $l \in \grp{D}$ as $l=\tilde{g} g$, where $\tilde{g} \in \tilde{\grp{G}}$ and $g \in \grp{G}$.
Taking $x = l^{-1} \partial_\sigma l = g^{-1} \partial_\sigma g + \Ad_g^{-1} \tilde{g}^{-1} \partial_\sigma \tilde{g}$ we then have
\[
y & = \Proj{\alg{g}} \, g^{-1} \partial_\sigma g + \Proj{\alg{g}} \Ad_g^{-1} \Proj{\tilde{\alg{g}}}\, \tilde{g}^{-1} \partial_\sigma \tilde{g} ~, \\
z & = \Proj{\tilde{\alg{g}}} \Ad_g^{-1} \Proj{\tilde{\alg{g}}}\, \tilde{g}^{-1} \partial_\sigma \tilde{g} ~,
\]
where $\proj_\alg{g}$ (respectively $\proj_{\tilde{\alg{g}}}$) takes an element of the Drinfel'd double and projects it onto $\alg{g}$ (respectively $\tilde{\alg{g}}$).
The operator $\Proj{\tilde{\alg{g}}} \Ad_g^{-1} \Proj{\tilde{\alg{g}}}$ is invertible on $\tilde{\alg{g}}$ and hence it is possible to eliminate $y$ in favour of $z$,
\[
y = \Proj{\alg{g}} \, g^{-1} \partial_\sigma g + \Proj{\alg{g}} \Ad_g^{-1} \Proj{\tilde{\alg{g}}} ( \Proj{\tilde{\alg{g}}} \Ad_g^{-1} \Proj{\tilde{\alg{g}}} )^{-1} z ~.
\]
The action then becomes quadratic in $z$, which can be integrated out to give
\[\label{eq:dual_actions}
\Act_{\grp{G}}(g) = \frac{1}{2} \int \dx \, \< g^{-1} \partial_+ g , (\En + \Pi(g))^{-1} \, g^{-1} \partial_- g \> ~, \qquad \Pi(g) = \proj_{\alg{g}} \Ad_g^{-1} \proj_{\tilde{\alg{g}}} \Ad_g \proj_{\tilde{\alg{g}}}~,
\]
where the worldsheet light-cone coordinates are defined as
\[\label{eq:wslcg}
\sigma^\pm = \frac{1}{2}(\tau \pm \sigma) ~, \qquad \partial_\pm = \partial_\tau \pm \partial_\sigma~, \qquad \dx = \extder \tau \extder \sigma = 2 \extder \sigma^+ \extder \sigma^- ~.
\]
To obtain the dual model, we parametrise the field $l \in \grp{D}$ as $l=g \tilde{g} $ and integrate out the degrees of freedom associated to $\alg{g}$.
After exchanging the role of $\alg{g}$ and $\tilde{\alg{g}}$ in the above derivation one obtains the dual sigma model
\[
\label{eq:dual_actions_2}
\Act_{\tilde{\grp{G}}}(\tilde{g}) = \frac{1}{2} \int \dx \, \< \tilde{g}^{-1} \partial_+ \tilde{g} , (\En^{-1} + \tilde{\Pi}(\tilde{g}))^{-1} \, \tilde{g}^{-1} \partial_- \tilde{g} \> ~, \qquad \tilde{\Pi}(\tilde{g}) &= \proj_{\tilde{\alg{g}}} \Ad_{\tilde{g}}^{-1} \proj_{\alg{g}} \Ad_{\tilde{g}} \proj_{\alg{g}}~.
\]
The two sigma models \eqref{eq:dual_actions} and \eqref{eq:dual_actions_2} are described by the same set of equations after appropriate non-local field and parameter redefinitions.
The model corresponding to the action \eqref{eq:dual_actions_2} is said to be the dual of \eqref{eq:dual_actions} with respect to $\alg{g}$.
Non-abelian duality is a special case of Poisson-Lie duality, and corresponds to the case in which the dual algebra $\tilde{\alg{g}}$ is abelian and hence $\Pi(g)=0$.

\paragraph{Poisson-Lie duality with respect to a subalgebra.}
Besides the canonical decomposition $\alg{d}=\alg{g} \dsum \tilde{\alg{g}}$ of \eqref{eq:decompo_dgtg}, it is also possible to consider more general maximally isotropic decompositions of the Drinfel'd double of the type $\alg{d} = \alg{k} \dsum \tilde{\alg{k}}$ where $\alg{k}$ is not necessarily an algebra \cite{Klimcik:1995dy,Klimcik:1996nq,Klimcik:2002zj,Klimcik:2015gba}.
On the other hand, $\tilde{\alg{k}}$ is still a subalgebra of $\alg{d}$ and its associated degrees of freedom can be integrated out to yield a model on the coset space $\tilde{\grp{K}}\backslash\grp{D}$.
Starting with the field $l \in \grp{D}$ parametrised as $l=\tilde{k} k$, where $\tilde{k} \in \tilde{\grp{K}}= \exp[\tilde{\alg{k}}]$ and $k \in \tilde{\grp{K}}\backslash \grp{D}$, and integrating out the degrees of freedom associated to $\tilde{\alg{k}}$ following the steps outlined above, we find the following Lorentz-invariant action for the field $k$
\[\label{eq:actionS1}
\Act_{\tilde{\grp{K}}\backslash\grp{D}}(k) = \frac{1}{2}  \int \dx \, \< k^{-1} \partial_+ k, (\half \Proj{\tilde{\alg{k}}}^\rot - \half \Proj{\alg{k}}^\rot + \Proj{\tilde{\alg{k}}}^\rot (\Ent+\Pi(k))^{-1} \, \Proj{\alg{k}}^\rot) k^{-1} \partial_- k \> + \WZ(k) ~,
\]
where
\[
\Pi(k)= \proj_{\alg{k}} \Ad_k^{-1} \proj_{\tilde{\alg{k}}} (\proj_{\tilde{\alg{k}}} \Ad_k^{-1} \proj_{\tilde{\alg{k}}})^{-1}~, \qquad \Proj{\alg{k}}^\rot = \Proj{\alg{k}} - \Pi(k) \Proj{\tilde{\alg{k}}}~, \qquad \Proj{\tilde{\alg{k}}}^\rot = \Proj{\tilde{\alg{k}}} + \Pi(k) \Proj{\tilde{\alg{k}}}~.
\]
The operator $\Ent: \tilde{\alg{k}} \rightarrow \alg{k}$ is related to $\En:\tilde{\alg{g}} \rightarrow \alg{g}$ via the non-linear transformation
\[\label{eq:E0t}
\Ent = (\proj_{\tilde{\alg{g}}} (\identity + \En^{-1} \proj_{\alg{g}}) \proj_{\alg{k}})^{-1} (\proj_{\tilde{\alg{g}}}(\identity + \En^{-1} \proj_{\alg{g}}) \proj_{\tilde{\alg{k}}}) ~.
\]
In the particular case where the intersection of $\tilde{\alg{k}}$ with $\alg{g}$ defines a common subalgebra $\alg{g}_0$ and one has the decomposition \cite{Hoare:2017ukq,Lust:2018jsx}
\[
\label{eq:decomposition_gk}
\alg{g} & = \alg{g}_0 \dsum \alg{m} ~, & \qquad \tilde{\alg{g}} & = \tilde{\alg{g}}_0 \dsum \tilde{\alg{m}} ~,
\\
\alg{k} & = \tilde{\alg{g}}_0 \dsum \alg{m} ~, & \qquad \tilde{\alg{k}} & = \alg{g}_0 \dsum \tilde{\alg{m}} ~,
\]
we can interpret \eqref{eq:actionS1} as the Poisson-Lie dual of the model on $\grp{G}$ with respect to $\alg{g}_0$.
The requirement that $\tilde{\alg{k}}$ forms an algebra imposes restrictions with respect to which subalgebras it is possible to dualise.

%%%%%%%%%%%%%%%%%%%%%%%%%%%%%%%%%%%%%%%%%%%%%%%%%%%%%%%%%%%%%%%%%%%%%%%%%%%%%%%%
\section{The \texorpdfstring{$\eta$}{eta}-deformed \texorpdfstring{$\AdS_2 \times \Sp^2$}{AdS2 x S2} supercoset}
\label{sec:etadeformation}

In this section we turn our attention to the $\eta$-deformed $\AdS_2 \times \Sp^2 \times \To^6$ superstring and its Poisson-Lie duals.
To this end we consider the $\eta$-deformed semi-symmetric space sigma model \cite{Metsaev:1998it,Zhou:1999sm,Berkovits:1999zq,Zarembo:2010sg,Delduc:2013qra,Delduc:2014kha} on the supercoset
\[\label{eq:sc}
\frac{\grp{PSU}(1,1|2)}{\grp{SO}(1,2) \times \grp{SO}(2)} ~,
\]
which we will refer to as the $\eta$-deformed $\AdS_2 \times \Sp^2$ supercoset model, and investigate Poisson-Lie duals with respect to various subalgebras of $\alg{psu}(1,1|2)$.
We start by explaining how to write the $\eta$-deformed semi-symmetric space sigma model in the manifestly Poisson-Lie symmetric form \eqref{eq:dual_actions}.
For this we need to specify the Drinfel'd double together with its invariant inner product, as well as the specific form of the operator $\En$.
We then specialise to the supercoset \eqref{eq:sc}, discussing the possible Poisson-Lie duals of the $\eta$-deformed $\AdS_2 \times \Sp^2$ supercoset model.

%%%%%%%%%%%%%%%%%%%%%%%%%%%%%%%%%%%%%%%%%%%%%%%%%%%%%%%%%%%%%%%%%%%%%%%%%%%%%%%%
\subsection{The \texorpdfstring{$\eta$}{eta}-deformed semi-symmetric space sigma model}

\paragraph{The action of the deformed model.}
Let $g$ be an element of a supergroup $\grp{G}$ whose corresponding Lie superalgebra $\alg{g}$ is basic and admits a $\Integer_4$ grading consistent with the commutation relations
\[
\alg{g} = \alg{g}^{(0)} \dsum \alg{g}^{(1)} \dsum \alg{g}^{(2)} \dsum \alg{g}^{(3)}~, \qquad
\com{\alg{g}^{(k)}}{\alg{g}^{(l)}} \subset \alg{g}^{(k+l \mod 4)}~.
\]
The $\Integer_4$ grading follows from the existence of a linear automorphism of the complexified superalgebra $\Omega: \alg{g}^\Complex \rightarrow \alg{g}^\Complex$ satisfying
\[
\Omega^4 = \identity ~, \qquad \Omega(\alg{g}^{(k)})= i^k \alg{g}^{(k)}~.
\]
The bosonic subalgebra of $\alg{g}$ is given by $\alg{g}^{(0)} \dsum \alg{g}^{(2)}$, while the fermionic generators belong to either $\alg{g}^{(1)}$ or $\alg{g}^{(3)}$.
We introduce the projectors $P_k \, \alg{g} = \alg{g}^{(k)}$, $k=0,1,2,3$, and denote the group corresponding to the grade 0 subalgebra by $\grp{H}=\grp{G}^{(0)}=\exp[\alg{g}^{(0)}]$.
The $\eta$-deformed model describes a deformation of the semi-symmetric space sigma model on the supercoset $\grp{G}/\grp{H}$.
Its action is \cite{Delduc:2013qra}
\[
\label{eq:eta_deformation}
\Act_\eta(g) =  T \int \dx \, \STr \Big( g^{-1} \partial_+ g , P \frac{1}{\identity - \frac{2\eta}{1-\eta^2} R_g P} g^{-1} \partial_- g\Big) ~,
\]
where the supertrace $\STr$ denotes an ad-invariant and $\Integer_4$ invariant bilinear form on $\alg{g}$, which is symmetric (respectively antisymmetric) on the bosonic (respectively fermionic) subspace of $\alg{g}$ and hence it is symmetric on the Grassmann envelope.
The operators $P$ and $R_g$ are given by
\[
P= P_2 + \frac{1-\eta^2}{2} ( P_1 -P_3) ~, \qquad R_g = \Ad_g^{-1} R \Ad_g ~,
\]
where the operator $R$ satisfies the non-split modified classical Yang Baxter equation
\[
\com{R X}{R Y} - R(\com{X}{R Y} + \com{R X}{Y} ) = \com{X}{Y} ~, \qquad X,Y \in \alg{g} ~,
\]
and is antisymmetric with respect to the supertrace $\STr( X, R Y ) = - \STr ( R X, Y )$.
We will take this R-matrix to be given by the canonical Drinfel'd-Jimbo solution associated to a particular Dynkin diagram and Cartan-Weyl basis of $\alg{g}$.
The overall coupling constant $T$ plays the role of the effective string tension and $\eta$ is the deformation parameter, flipping the sign of which is equivalent a parity transformation.
The global left-acting $\grp{G}$ symmetry is broken to the Cartan subgroup, while the right-acting gauge symmetry, $g \rightarrow g h$, where $h$ belongs to $\grp{H}$, is preserved.

\medskip

\paragraph{Poisson-Lie symmetric action.}
To write the action of the $\eta$-deformed semi-symmetric space sigma model in a manifestly Poisson-Lie symmetric form, we recall that for the $\eta$-deformed models the relevant Drinfel'd double is the complexified Lie algebra $\alg{d}=\alg{g}^\Complex$ \cite{Klimcik:2002zj,Vicedo:2015pna}, which as a real vector space admits the decomposition
\[
\alg{g}^\Complex = \alg{g} \dsum \tilde{\alg{g}}~,
\]
where $\tilde{\alg{g}}$ is the Borel subalgebra, formed by the Cartan generators and the positive roots of $\alg{g}^\Complex$.
More precisely, if $\set{h_i}$ ($i=1,\ldots,\rank \alg{d}$), $\set{e_\ind{M}}$ and $\set{f_\ind{M}}$, ($M=1,\ldots, \half(\dim_\Complex \alg{d}-\rank \alg{d})$) are the Cartan generators, positive and negative roots respectively, then the Borel subalgebra is spanned by $\tilde{\alg{g}}=\set{h_i, e_\ind{M}, ie_\ind{M}}$.
Furthermore, the action of the Drinfel'd Jimbo R-matrix on the Cartan-Weyl basis is given by
\[
\label{eq:DrinfeldRmatrix}
R(h_i) = 0 ~, \qquad R(e_\ind{M}) = - i e_\ind{M} ~, \qquad R(f_\ind{M}) = i f_\ind{M} ~.
\]

To specify the operator $\En$ we introduce bases of $\alg{g}$ and $\tilde{\alg{g}}$, denoting the generators of $\alg{g}$ (respectively $\tilde{\alg{g}}$) by $T_\ind{A}$, $A = 1, 2, \ldots , \dim \grp{G}$ (respectively $\tilde{T}^\ind{A}$).
We further assume that we have an inner product $\< \cdot, \cdot \>$ on $\alg{g}^\Complex$ with respect to which the two subalgebras are isotropic, $\<T_\ind{A}, T_\ind{B}\> = \<\tilde{T}^\ind{A}, \tilde{T}^\ind{B} \>=0$ and that provides a canonical pairing, that is $\<T_\ind{A}, \tilde{T}^\ind{B}\>=\delta_\ind{A}^\ind{B}$ for bosonic generators and $\<\theta_1 T_\ind{A}, \theta_2 \tilde{T}^\ind{B}\> = i \, \theta_1 \theta_2 \delta_\ind{A}^\ind{B}$ for fermionic ones.
Introducing $\kappa_\ind{AB} = \STr(T_\ind{A} T_\ind{B})$ and the auxiliary operator
\[
P_\epsilon : \left\{ \begin{aligned}
& T_\ind{A} \rightarrow \epsilon \kappa_\ind{BA} \tilde{T}^\ind{B}~, \quad &T_\ind{A} &\in \alg{g}^{(0)}~, \\
& T_\ind{A} \rightarrow -\sfrac{i}{2}(1-\eta^2)\kappa_\ind{BA} \tilde{T}^\ind{B}~, \quad &T_\ind{A} &\in \alg{g}^{(1)}~, \\
& T_\ind{A} \rightarrow \kappa_\ind{BA} \tilde{T}^\ind{B}~, \quad &T_\ind{A} &\in \alg{g}^{(2)} ~, \\
& T_\ind{A} \rightarrow \sfrac{i}{2}(1-\eta^2) \kappa_\ind{BA} \tilde{T}^\ind{B}~, \quad & T_\ind{A} &\in \alg{g}^{(3)} ~,
\end{aligned} \right.
\]
we define the operators $\EP : \tilde{\alg{g}} \rightarrow \alg{g}$ and $\ER : \tilde{\alg{g}} \rightarrow \alg{g}$ by
\[\label{eq:g0def}
\EP^{-1} &= -\frac{2 \eta}{1-\eta^2} \lim_{\epsilon \rightarrow 0} P_\epsilon~, \]
and
\[\label{eq:b0def}
\ER(\tilde{T}^\ind{B})&= \ER^\ind{AB} T_\ind{A}~, \qquad &\ER^\ind{AB} &= R^\ind{A}{}_\ind{C} \kappa^\ind{CB} ~, &\qquad T_\ind{A} &\in \alg{g}^{(0)}+\alg{g}^{(2)}~,\\
\ER(\theta \tilde{T}^\ind{B})&= \ER^\ind{AB}\theta T_\ind{A}~, \qquad &\ER^\ind{AB} &= i R^\ind{A}{}_\ind{C} \kappa^\ind{CB} ~, &\qquad T_\ind{A} &\in \alg{g}^{(1)}+\alg{g}^{(3)} ~,
\]
where $R(T_\ind{A}) = R^\ind{B}{}_\ind{A} T_\ind{B}$ and $\kappa^\ind{AB} \kappa_\ind{BC} = \delta^\ind{A}_\ind{C}$.
These operators preserve the $\Integer_2$ grading of the superalgebra.
The action \eqref{eq:eta_deformation} of the $\eta$-deformation is then equivalent to
\[
\Act_\eta(g) &= - T \frac{1-\eta^2}{2 \eta} \int \dx\,\< g^{-1} \partial_+ g , \EP^{-1} \frac{1}{ \proj_{\alg{g}} + (\ER + \Pi(g)) \EP^{-1}} g^{-1} \partial_- g \>~,
\]
which is of the form \eqref{eq:dual_actions} with $\En = \EP + \ER$.

%%%%%%%%%%%%%%%%%%%%%%%%%%%%%%%%%%%%%%%%%%%%%%%%%%%%%%%%%%%%%%%%%%%%%%%%%%%%%%%%
\subsection{Poisson-Lie duals of the \texorpdfstring{$\eta$}{eta}-deformed \texorpdfstring{$\AdS_2 \times \Sp^2$}{AdS2 x S2} supercoset}

\paragraph{The complex Drinfel'd double.}
The isometry algebra of the $\AdS_2 \times \Sp^2$ supercoset \eqref{eq:sc} is $\alg{g}=\alg{psu}(1,1|2)$, the bosonic subalgebra of which is $\alg{su}(1,1) \dsum \alg{su}(2)$.
As discussed above, the relevant Drinfel'd double for the $\eta$-deformed model is the complexified Lie superalgebra $\alg{d}=\alg{g}^\Complex= \alg{psl}(2|2 ; \Complex)$.
The Drinfel'd double can be decomposed into two real subalgebras
\[
\alg{d} = \alg{psl}(2|2 ; \Complex) = \alg{psu}(1,1|2) \dsum \alg{pb}(1,1|2) ~,
\]
where $\alg{pb}(1,1|2)$ is the projected Borel subalgebra spanned by the Cartan generators and positive roots of $\alg{psl}(2|2;\Complex)$.
It will often be convenient for us to work with the superalgebra $\alg{sl}(2|2;\Complex)$.
The superalgebra $\alg{psl}(2|2;\Complex)$ is then obtained by quotienting out the $\alg{u}(1)$ ideal, that is we identify elements of $\alg{sl}(2|2; \Complex)$ that differ by the central element.

At this point let us make a brief comment on the different Dynkin diagrams of $\alg{sl}(2|2;\Complex)$.
In general, for Lie superalgebras there are inequivalent choices for the set of Cartan generators and simple roots, where the latter can either be bosonic or fermionic.
Since the Drinfel'd Jimbo R-matrix \eqref{eq:DrinfeldRmatrix} is defined by its action on the Cartan generators and roots, it is possible that different choices of Dynkin diagrams and Cartan-Weyl bases define inequivalent deformations.
The three Dynkin diagrams of $\alg{sl}(2|2;\Complex)$ are $\Circle- \otimes- \Circle$, $\otimes-\Circle-\otimes$ and $\otimes-\otimes-\otimes$, where $\Circle$ represents a bosonic root and $\otimes$ a fermionic root.
In this paper we will focus on the distinguished Dynkin diagram $\Circle-\otimes-\Circle$.
A discussion of the other choices is given in \appref{app:Different_Dynkin}.

\paragraph{Matrix realisation.}
For calculations we will use a given matrix realisation of the superalgebras $\alg{psu}(1,1|2; \Complex)$ and $\alg{pb}(1,1|2; \Complex)$, which is presented explicitly in \appref{app:sl_generators}.
The generators of the grade 0 subalgebra $\alg{h}=\alg{so}(1,1)\dsum \alg{so}(2)$ are denoted $J_{01}$ and $J_{23}$ respectively, while the remaining bosonic generators are denoted $P_a$, with $a=0,1$ for $\alg{su}(1,1)$ and $a=2,3$ for $\alg{su}(2)$.
The supercharges are denoted by $Q^{\ind{I} \check{\alpha} \hat{\alpha}}$ where $I=1,2$ is the grading, $\check{\alpha}=1,2$ is the $\alg{su}(1,1)$ index and $\hat{\alpha}=1,2$ the $\alg{su}(2)$ index.
The dual generators $\tilde{P}^a, \tilde{J}^{ab}, \tilde{Q}^{\ind{I} \check{\alpha} \hat{\alpha}}$ are then identified using the inner product
\[\label{eq:innerproduct}
\< \cdot , \cdot \> = -\Im \STr (\cdot,\cdot) ~.
\]
The positive roots span the upper triangular matrices such that on an arbitrary $4 \times 4$ matrix $M$, the Drinfel'd Jimbo R-matrix \eqref{eq:DrinfeldRmatrix} acts as
\[
R(M)_{ij} = -i \epsilon_{ij} M_{ij}~, \qquad \epsilon_{ij} = \left\{ \begin{aligned} 1 &\text{ if } i < j \\ 0 &\text{ if } i = j \\ -1 &\text{ if } i > j \end{aligned}\right.~.
\]

\paragraph{Poisson-Lie duals and unimodularity.}
The background of the $\eta$-deformed $\AdS_2 \times \Sp^2 \times \To^6$ superstring has a non-vanishing, albeit closed, B-field, together with a three-form and five-form RR flux \cite{Arutyunov:2015mqj,Araujo:2018rbc}.
\unskip\footnote{Note that in the explicit formulae given in app.~F of \cite{Arutyunov:2015mqj} a gauge transformation has been used to set the B-field to vanish.}
This background does not solve the standard supergravity equations, rather a generalisation thereof \cite{Arutyunov:2015mqj,Wulff:2016tju}.
In the context of the first-order action on the Drinfel'd double and duality, the corresponding Weyl anomaly is expected to be associated to integrating out the degrees of freedom of a non-unimodular algebra, that is when the trace of the structure constants is non-vanishing, $f_{ab}{}^b \neq 0$.
Indeed, when starting from the first-order action on the Drinfel'd double, the $\eta$-deformed model is obtained by integrating out the degrees of freedom associated to the projected Borel algebra $\alg{pb}(1,1|2)$, which is indeed non-unimodular.

As the background of the $\eta$-deformed $\AdS_2 \times \Sp^2 \times \To^6$ superstring solves the generalised supergravity equations, the Weyl anomaly is of a particularly special type.
As a result, the background of the $\eta$-deformed model is expected to be related to solutions of the standard supergravity equations by Poisson-Lie duality.
These dual models are given by integrating out the degrees of freedom of unimodular algebras in the model on the Drinfel'd double.
We finish this section by listing three examples of such Poisson-Lie duals of the $\eta$-deformed $\AdS_2 \times \Sp^2$ supercoset and describe the associated unimodular subalgebras of $\alg{psl}(2|2;\Complex)$.
\begin{enumerate}
\item First, it is possible to dualise with respect to the full superalgebra $\alg{psu}(1,1|2)$.
In this case one integrates out the degrees of freedom associated to $\alg{psu}(1,1|2)$, a unimodular algebra.
In the terminology of \cite{Hoare:2017ukq} this gives the $\lambdastar$-deformed model and is conjectured to be an analytic continuation of the $\lambda$-deformation \cite{Vicedo:2015pna,Hoare:2015gda,Sfetsos:2015nya,Klimcik:2015gba,Hoare:2017ukq,Driezen:2018glg}.
The background of the $\lambda$-deformation of the $\AdS_2 \times \Sp^2 \times \To^6$ superstring has been derived in \cite{Borsato:2016zcf}.
It has a vanishing B-field and the metric is supported by a RR five-form flux and dilaton giving a solution of the standard supergravity equations.
\item Second, as we are considering the distinguished Dynkin diagram $\Circle - \otimes - \Circle$, the results of \cite{Hoare:2017ukq} tell us that we can dualise with respect to the full bosonic subalgebra of $\alg{psu}(1,1|2)$, that is $\alg{su}(1,1) \dsum \alg{su}(2)$, by considering the sub-Dynkin diagram formed of the two bosonic nodes.
In this case the degrees of freedom that are integrated out are associated to the algebra $\tilde{\alg{k}}=\alg{su}(1,1) \dsum \alg{su}(2) \dsum \set{\tilde{Q}^{\ind{I}\check{\alpha} \hat{\alpha}}}$, where $\set{\tilde{Q}^{\ind{I} \check{\alpha} \hat{\alpha}}}$ are the dual fermionic generators, that is the positive fermionic roots.
Since this is also a unimodular algebra we expect the resulting background to again solve the standard supergravity equations.

The bosonic part of this model, and hence the metric and B-field, coincides with the $\lambdastar$-deformation.
However, they differ in the fermionic part.
In \cite{Sfetsos:2014cea} an alternative embedding of the metric of the $\lambda$-deformed $\AdS_2 \times \Sp^2 \times \To^6$ superstring in supergravity was given.
The metric is again supported by a RR five-form flux and dilaton, however these are different to those found in \cite{Borsato:2016zcf}.
In \secref{sec:PLdualityBosonic} we show that this background corresponds to an analytic continuation of the Poisson-Lie dual of the $\eta$-deformed $\AdS_2 \times \Sp^2 \times \To^6$ superstring with respect to the full bosonic subalgebra.
\item Finally, we consider the two-fold T-dual of the $\eta$-deformed $\AdS_2 \times \Sp^2$ supercoset, equivalent to dualising with respect to the $\alg{u}(1) \dsum \alg{u}(1)$ Cartan subalgebra of $\alg{psu}(1,1|2)$.
As discussed in detail for the bosonic case in \cite{Hoare:2017ukq}, one can show that the algebra whose degrees of freedom are integrated out is unimodular.
Accordingly, the background of the two-fold T-dual of the $\eta$-deformed $\AdS_2 \times \Sp^2 \times \To^6$ superstring solves the standard supergravity equations, again supported by a RR five-form flux and dilaton \cite{Hoare:2015gda,Hoare:2015wia,Arutyunov:2015mqj}.

As shown in \cite{Hoare:2015gda} and \cite{Borsato:2016zcf} respectively, the two-fold T-dual can be found by analytically continuing and taking a scaling limit of the backgrounds of \cite{Sfetsos:2014cea} and \cite{Borsato:2016zcf}.
The analytic continuation amounts to considering the reality conditions relevant for the $\eta$-deformed models, and hence the two-fold T-dual should be given by a real scaling limit of the two Poisson-Lie duals of the $\eta$-deformed $\AdS_2 \times \Sp^2 \times \To^6$ superstring discussed above.
\end{enumerate}
These three examples of Poisson-Lie duals of the $\eta$-deformed $\AdS_2 \times \Sp^2 \times \To^6$ superstring all involve dualising in a timelike direction.
Abelian T-duality in a timelike direction maps solutions of type II supergravity to solutions of type II$^\star$ \cite{Hull:1998vg}.
As Poisson-Lie duality is a generalisation of abelian T-duality, the corresponding backgrounds are expected to solve the standard type II$^\star$ supergravity equations.

%%%%%%%%%%%%%%%%%%%%%%%%%%%%%%%%%%%%%%%%%%%%%%%%%%%%%%%%%%%%%%%%%%%%%%%%%%%%%%%%
\section{Poisson-Lie duality with respect to the full bosonic subalgebra}
\label{sec:PLdualityBosonic}

In this section we derive the background of the Poisson-Lie dual of the $\eta$-deformed $\AdS_2 \times \Sp^2 \times \To^6$ superstring with respect to the full bosonic subalgebra $\alg{su}(1,1) \dsum \alg{su}(2)$.
We start from the first-order action on the Drinfel'd double \eqref{eq:action_double} with $\En = \EP + \ER$ defined in eqs.~\eqref{eq:g0def} and \eqref{eq:b0def}.
We then consider the decomposition \eqref{eq:decomposition_gk} with
\[
\alg{g}_0 = \alg{g}_\Boson=\alg{su}(1,1) \dsum \alg{su}(2)~, \qquad \alg{m}=\alg{g}_\Fermion = \set{Q_{\ind{I} \check{\alpha} \hat{\alpha}} } ~,
\]
and integrate out the degrees of freedom associated to the algebra $\tilde{\alg{k}} = \alg{g}_0 \dsum \tilde{\alg{m}}$, where $\tilde{\alg{m}}$ is spanned by the positive fermionic roots.
These degrees of freedom are associated to a unimodular algebra and hence the corresponding background is expected to solve the standard supergravity equations.
Expanding the action \eqref{eq:actionS1} to quadratic order in fermions, we rewrite it in Green-Schwarz form and extract the background fields.
The resulting background indeed solves the standard supergravity equations and, as conjectured, is given by an analytic continuation of that constructed in \cite{Sfetsos:2014cea}.

%%%%%%%%%%%%%%%%%%%%%%%%%%%%%%%%%%%%%%%%%%%%%%%%%%%%%%%%%%%%%%%%%%%%%%%%%%%%%%%%
\subsection{Parametrisation}

In order to Poisson-Lie dualise with respect to the full bosonic subalgebra $\alg{g}_\Boson = \alg{su}(1,1) \dsum \alg{su}(2)$ we need to find a suitable parametrisation of the field $k \in \tilde{\grp{K}} \backslash \grp{D} / \grp{H}$ appearing in the action.
Starting with a group-valued field $k \in \grp{D}=\grp{PSL}(2|2; \Complex)$ and using the fermionic part of the left-acting $\tilde{\grp{K}}$ gauge symmetry we partially gauge fix
\[
k= \tilde{g}_0 \exp[ \theta^{\ind{I} \check{\alpha} \hat{\alpha}} Q_{\ind{I} \check{\alpha} \hat{\alpha}}]~,
\]
where
\[
\tilde{g}_0 \in \grp{G}_\Boson \backslash \grp{D}_\Boson / \grp{H} = (\grp{SU}(1,1) \times \grp{SU}(2))\backslash (\grp{SL}(2;\Complex) \times \grp{SL}(2;\Complex)) / (\grp{SO}(1,1) \times \grp{SO}(2)) ~.
\]
To gauge fix $\tilde{g}_0$ we write $\tilde{g}_0 = \check{g}_0 \dsum \hat{g}_0$, where $\check{g}_0$ corresponds to the $\AdS_2$ factor and $\hat{g}_0$ to the $\Sp^2$ factor and gauge fix in each sector separately.

\paragraph{Gauge fixing in the $\Sp^2$ sector.}
Let us introduce the generators $S_\ind{A}$ and $\tilde{S}^\ind{A}$, $A=4,5,6$, defined in terms of the Cartan generator $h=\sigma_3$ and the simple roots $e=\sigma_+$, $f=\sigma_-$ of $\alg{sl}(2;\Complex)$ as
\[
S_4 &= i(e+f)~, & \qquad S_5 &= -(e-f)~, & \qquad S_6 &= ih~, \\
\tilde{S}^4 &= (e+f)/2~, &\qquad \tilde{S}^5 &= -i(h-e+f)/2~, &\qquad \tilde{S}^6 &= (h-e+f)/2 ~.
\]
The right-acting gauge symmetry is generated by $S_4$, the adjoint action of which rotates $\tilde{S}^5$ and $\tilde{S}^6$ amongst themselves.
Therefore, using this right-acting gauge symmetry together with the left-acting gauge symmetry generated by $\set{S_4,S_5,S_6}$, we can partially gauge fix
\[
\hat{g}_0 \in \exp[\set{S_5,\tilde{S}^4,\tilde{S}^6}] \in \grp{SO}(2) \backslash \grp{SL}(2;\Real) ~,
\]
where the residual left-acting gauge symmetry is generated by $S_5$.
Using this residual gauge freedom we choose the familiar parametrisation of this coset
\[
\hat{g}_0= \exp[\phi_5 e] \exp[\phi_4 h/2] ~,
\]
which in terms of the generators $\{\tilde P^2, \tilde P^3, \tilde J^{23}\}$ defined in \appref{app:sl_generators} is given by
\[
\hat{g}_0= \exp[- \phi_5 \tilde{J}^{23}] \exp[-\phi_4 \tilde{P}^2] ~.
\]

\paragraph{Gauge fixing in the $\AdS_2$ sector.}
For the $\AdS_2$ sector we introduce the following generators $\set{S_\ind{A}, \tilde{S}^\ind{A}}$, $A=1,2,3$, of $\alg{sl}(2;\Complex)$
\[
S_1 &= -(e+f)~, & \qquad S_2 &= i(e-f)~, & \qquad S_3 &= ih~, \\
\tilde{S}^1 &= i(e+f)/2~, &\qquad \tilde{S}^2 &=i(h-i(e-f))/2~, &\qquad \tilde{S}^3 &= -(h+i(e-f))/2~.\\
\]
Here the right-acting gauge symmetry is generated by $S_1$, the adjoint action of which hyperbolically rotates $\tilde{S}^2$ and $\tilde{S}^3$ amongst themselves, while the left-acting gauge symmetry is generated by $\set{S_1,S_2,S_3}$.
Following the same logic as for the $\Sp^2$ sector we find it is possible to gauge fix
\[
\check{g}_0= \exp[\phi_2 ie] \exp[\phi_1 h/2] ~,
\]
which in terms of the generators $\set{\tilde P^0, \tilde P^1, \tilde J^{01}}$ defined in \appref{app:sl_generators} is given by
\[
\check{g}_0 = \exp[-\phi_2 \tilde{J}^{01}] \exp[\phi_1 \tilde{P}^0] ~.
\]

\paragraph{Full parametrisation.}
Finally, our parametrisation of the field $k \in \tilde{\grp{K}} \backslash \grp{D} / \grp{H}$ is given by
\[
\label{eq:param_gen}
k= \tilde{g}_0 \exp[ \theta^{\ind{I} \check{\alpha} \hat{\alpha}} Q_{\ind{I} \check{\alpha} \hat{\alpha}}]~,
\]
with
\[
\label{eq:paramg0}
\tilde{g}_0 = \exp[-\phi_2 \tilde{J}^{01}] \exp[\phi_1 \tilde{P}^0] \dsum \exp[- \phi_5 \tilde{J}^{23}] \exp[-\phi_4 \tilde{P}^2] ~.
\]

%%%%%%%%%%%%%%%%%%%%%%%%%%%%%%%%%%%%%%%%%%%%%%%%%%%%%%%%%%%%%%%%%%%%%%%%%%%%%%%%
\subsection{NSNS background}\label{ssec:nsns}

The NSNS fields (the metric and the B-field) are obtained by setting the fermions in the action \eqref{eq:actionS1} to zero and considering the bosonic Lagrangian
\unskip\footnote{In our conventions the bosonic part of the action is related to the metric by
\begin{equation*}
\Act = \int \dx \, \Lag^0 = \int \dx \, G_\ind{MN} \partial_+ X^\ind{M} \partial_- X^\ind{N} ~,
\end{equation*}
and the line element is $\extder s^2 = G_\ind{MN} \extder X^\ind{M} \extder X^\ind{N}$.
That is the tension $T$ is included in the metric.}
\[
\Lag^0 = - \frac{T}{\kappa} \< \tilde{g}_0^{-1} \partial_+ \tilde{g}_0, (\En^{-1} + \Pi(\tilde{g}))^{-1} \tilde{g}_0^{-1} \partial_- \tilde{g}_0 \> ~,
\]
which corresponds to the Lagrangian of the $\lambdastar$-deformation of the bosonic sigma model on $\AdS_2 \times \Sp^2$ and we have introduced the deformation parameter \cite{Arutyunov:2013ega}
\[ \kappa = \frac{2 \eta}{1-\eta^2} ~. \]
Using the parametrisation \eqref{eq:paramg0} we find the following metric
\[
2 T^{-1} \extder s^2 = & \frac{1}{\kappa^2} \Big(- \extder \phi_1^2 + \frac{4(e^{2 \phi_1} \kappa^2 - \phi_2^2)}{(1-e^{2 \phi_1} - \phi_2^2)^2} \extder \phi_2^2 - \frac{4 \phi_2 \extder \phi_1 \extder \phi_2}{1-e^{2 \phi_1} - \phi_2^2} \Big) \\
&+ \frac{1}{\kappa^2} \Big(\extder \phi_4^2 + \frac{4(e^{2 \phi_4} \kappa^2 + \phi_5^2)}{(1-e^{2 \phi_4} + \phi_5^2)^2} \extder \phi_5^2 - \frac{4 \phi_5 \extder \phi_4 \extder \phi_5}{1-e^{2 \phi_4}+\phi_5^2} \Big)~,
\]
The B-field vanishes.
After the coordinate redefinition
\unskip\footnote{The redefinitions $\phi_1 = \log |\check{p}+\sqrt{\check{p}^2+\check{q}^2-1}|$ and $\phi_4 = \log|\hat{p}+\sqrt{\hat{p}^2-\hat{q}^2-1}|$ give the same result.}
\[
\label{eq:coord_redef}
\phi_1 &= \log |\check{p}-\sqrt{\check{p}^2+\check{q}^2-1}|~, & \qquad \phi_2 &= \check{q}~, & \qquad \check{p}^2 + \check{q}^2 > 1~, \\
\phi_4 &= \log|\hat{p}-\sqrt{\hat{p}^2-\hat{q}^2-1}|~, & \qquad \phi_5 &= \hat{q}~, & \qquad \hat{p}^2 - \hat{q}^2 > 1~,
\]
the metric becomes conformally flat,
\[\label{eq:cfmet}
2 T^{-1} \extder s^2 = \frac{1}{\kappa^2} \frac{1}{\check{p}^2+\check{q}^2-1} \Big( - \extder \check{p}^2 + \kappa^2 \extder \check{q}^2 \Big) + \frac{1}{\kappa^2} \frac{1}{\hat{p}^2-\hat{q}^2-1} \Big( \extder \hat{p}^2 + \kappa^2 \extder\hat{q}^2 \Big) ~.
\]
We also introduce the vielbein
\[\label{eq:vielbein}
&E_{\check{p} \, 0} = \frac{1}{\kappa} \sqrt{\frac{T}{2} } \frac{1}{\sqrt{\check{p}^2 + \check{q}^2 -1}} ~, \qquad E_{\check{q}\, 1} = \sqrt{\frac{T}{2} } \frac{1}{\sqrt{\check{p}^2 + \check{q}^2 -1}} ~, \\
&E_{\hat{p}\, 2} = \frac{1}{\kappa} \sqrt{\frac{T}{2} }\frac{1}{\sqrt{\hat{p}^2 - \hat{q}^2 -1}} ~, \qquad E_{\hat{q}\, 3} = \sqrt{\frac{T}{2} } \frac{1}{\sqrt{\hat{p}^2 - \hat{q}^2 -1}} ~,
\]
satisfying $G_\ind{MN} = E_{\ind{M}a} E_{\ind{N}b} \eta^{ab}$.
The spin connection, which we will need to obtain the RR fields, is given in terms of the vielbein
\[
\omega_{\ind{M} ab} = & \frac{1}{2} E^\ind{N}{}_b (\partial_\ind{M} E_{\ind{N} a} - \partial_\ind{N} E_{\ind{M} a})
\\ & -\frac{1}{2} E^\ind{N}{}_{a} (\partial_\ind{M} E_{\ind{N} b} - \partial_\ind{N} E_{\ind{M} b} )+ \frac{1}{2} E^\ind{R}{}_a E^\ind{S}{}_b (\partial_\ind{R} E_{\ind{S} c} - \partial_\ind{S} E_{\ind{R} c}) E_\ind{M}{}^c ~,
\]
and has the following non-vanishing components
\[
\omega_{\check{p} 01} = +\frac{1}{\kappa} \frac{\check{q}}{\check{p}^2+\check{q}^2-1}~, \qquad \omega_{\check{q} 01} = +\kappa \frac{\check{p}}{\check{p}^2+\check{q}^2-1}~, \\
\omega_{\hat{p} 23} = -\frac{1}{\kappa} \frac{\hat{q}}{\hat{p}^2-\hat{q}^2-1}~, \qquad \omega_{\hat{q} 23} = -\kappa \frac{\hat{p}}{\hat{p}^2-\hat{q}^2-1}~.
\]

\paragraph{Relation to the metric of the $\lambda$-deformed model on $\AdS_2 \times \Sp^2$.}
The metric \eqref{eq:cfmet} is related to the metric of the $\lambda$-deformed model on $\AdS_2 \times \Sp^2$ by an analytic continuation of both the parameters and the coordinates \cite{Hoare:2015gda,Klimcik:1996np,Sfetsos:1999zm}.
Applying the transformation rules of \cite{Hoare:2015gda}
\[
\label{eq:transfo_rules}
T = \frac{k \kappa}{i \pi}~,\qquad \kappa = i\frac{1-\lambda^2}{1+\lambda^2} ~, \qquad \check{q} \rightarrow i \check{q}~, \qquad \hat{q} \rightarrow i \hat{q}~,
\]
we find the following metric
\[
2 \pi k^{-1} \extder s^2 = & \frac{1}{1-\check{p}^2+\check{q}^2} \Big( - \frac{1+\lambda^2}{1-\lambda^2} \extder \check{p}^2 + \frac{1-\lambda^2}{1+\lambda^2} \extder \check{q}^2 \Big)
\\ & + \frac{1}{1-\hat{p}^2-\hat{q}^2} \Big( \frac{1+\lambda^2}{1-\lambda^2} \extder \hat{p}^2 + \frac{1-\lambda^2}{1+\lambda^2} \extder\hat{q}^2 \Big) ~.
\]
This is precisely the metric of the $\lambda$-deformed model on $\AdS_2 \times \Sp^2$ given in \cite{Sfetsos:2014cea},
\[
\label{eq:metricS}
\extder s^2 &= \kst \Big( \frac{1-\lst}{1+\lst} (-\coth^2 \rho \, \extder t^2 + \extder \rho^2) + \frac{4 \lst}{1-\lst^2} (\cosh t \, \extder \rho + \sinh t \coth \rho \, \extder t)^2 \Big) \\
&+ \kst \Big( \frac{1-\lst}{1+\lst} (\cot^2 \omega \, \extder \phi^2 + \extder \omega^2) + \frac{4 \lst}{1-\lst^2} (\cos \phi \, \extder \omega + \sin \phi \cot \omega \, \extder \phi)^2 \Big)~,
\]
where
\[\label{eq:kl}
\kst = \frac{k}{2\pi} ~, \qquad \lst = \lambda^2 ~.
\]
and the coordinates are related as
\[
\check{p} = \cosh \rho \cosh t ~, \qquad \check{q} = \cosh \rho \sinh t ~, \qquad
\hat{p} = \cos \omega \cos \phi ~, \qquad \hat{q} = \cos \omega \sin \phi ~.
\]

%%%%%%%%%%%%%%%%%%%%%%%%%%%%%%%%%%%%%%%%%%%%%%%%%%%%%%%%%%%%%%%%%%%%%%%%%%%%%%%%
\subsection{RR background}\label{ssec:RR}

To obtain the RR fluxes we rewrite the deformed $\AdS_2 \times \Sp^2$ supercoset sigma model in Green-Schwarz form.
The supercoset sigma model has 4 bosonic and 8 fermionic fields compared to the 10 bosonic and 32 fermionic of the type II Green-Schwarz superstring sigma model.
In this subsection we describe a consistent truncation of the latter that can be matched with the former to quadratic order in fermions.
This in turn allows us to derive the RR fluxes of the Poisson-Lie dual of the $\eta$-deformed $\AdS_2 \times \Sp^2 \times \To^6$ superstring with respect to the full bosonic subalgebra $\alg{su}(1,1) \dsum \alg{su}(2)$.

%%%%%%%%%%%%%%%%%%%%%%%%%%%%%%%%%%%%%%%%%%%%%%%%%%%%%%%%%%%%%%%%%%%%%%%%%%%%%%%%
\subsubsection{Truncation of the Green-Schwarz action}

To embed the 4 dimensions of the $\lambdastar$-deformed model on $\AdS_2 \times \Sp^2$ \eqref{eq:cfmet} into 10 dimensions we take the remaining six dimensions to simply be a flat torus, $\To^6$, so that the full 10-dimensional metric is ($i = 4,5,6,7,8,9$)
\[\label{eq:fullmet}
\extder s^2 = \frac{T}{2 \kappa^2} \frac{1}{\check{p}^2+\check{q}^2-1} \Big( - \extder \check{p}^2 + \kappa^2 \extder \check{q}^2 \Big) + \frac{T}{2 \kappa^2} \frac{1}{\hat{p}^2-\hat{q}^2-1} \Big( \extder \hat{p}^2 + \kappa^2 \extder\hat{q}^2 \Big) + \extder X_i \extder X^i ~.
\]
For the truncation of the fermionic fields \cite{Sorokin:2011rr,Borsato:2016zcf} let us start with the action of the type IIB Green-Schwarz superstring at quadratic order in fermions
\unskip\footnote{Since we are primarily interested in extracting the RR background fields we only consider the action of the Green-Schwarz superstring up to quadratic order in the fermions.
Note that the overall factor is proportional to $\sqrt{T}$ as the metric \eqref{eq:cfmet} and vielbein \eqref{eq:vielbein} contain a factor of $T$ and $\sqrt{T}$ respectively.}
\[
\label{eq:GSaction}
\Act = \frac{i}{2}\sqrt{\frac{T}{2} }\int \dx \, \bar{\Theta}^\ind{I} \big(\sigma_1^{\alpha \beta} \delta^\ind{IJ} + \epsilon^{\alpha \beta} \sigma_3^\ind{IJ} \big) E_{\alpha \ind{A}} \Gamma^\ind{A} D_\beta^\ind{JK} \Theta^\ind{K} ~,
\]
where
\[
D_\beta^\ind{JK} = \delta^\ind{JK} \Big( \partial_\beta - \frac{1}{4} \omega_\beta{}_\ind{AB} \Gamma^\ind{AB} \Big) + \frac{1}{8} \sigma_3^\ind{JK} E_{\beta \ind{A}} H^\ind{ABC} \Gamma_\ind{BC} +\frac{1}{8} S^\ind{JK} E_{\beta \ind{A}} \Gamma^\ind{A}~,
\]
\[
S^\ind{JK} = -e^\Phi\big(\epsilon^\ind{JK} \Gamma^\ind{A} F_\ind{A} + \frac{1}{3!} \sigma_1^\ind{JK} \Gamma^\ind{ABC} F_\ind{ABC} + \frac{1}{2 \cdot 5!} \epsilon^\ind{JK} \Gamma^\ind{ABCDE} F_\ind{ABCDE} \big)~.
\]
The action is written in conformal gauge with $\alpha,\beta = +,-$ and the worldsheet light-cone coordinates defined in eq.~\eqref{eq:wslcg}.
The (rescaled) conformal gauge metric and two-index antisymmetric tensor are given by $\sigma_1^{+-}=\sigma_1^{-+}=\epsilon^{+-}=-\epsilon^{-+}=1$.
$E_{\alpha \ind{A}}$ and $\omega_{\alpha \ind{AB}}$ are the pullbacks of the vielbein and corresponding spin connection to the worldsheet, $H$ is the field strength of the B-field, $\Phi$ is the dilaton and $F_\ind{A}$, $F_\ind{ABC}$ and $F_\ind{ABCDE}$ are the RR fluxes.

It is useful to split the tangent space index $A=0,1,2, \dots, 9$ into an index $a=0,1,2,3$ covering the $\AdS_2 \times \Sp^2$ directions and an index $i=4,5,6,7,8,9$ covering the torus directions.
As we are considering the type IIB superstring, the two 32-components spinors $\Theta^\ind{I}$, $I=1,2$ are both Weyl spinors of the same chirality ($\Gamma^{11} \Theta^1 = \Gamma^{11} \Theta^2$) and satisfy the Majorana condition
\unskip\footnote{See \appref{app:gamma_matrices} for the definitions and properties of the gamma matrices and the charge conjugation matrix $\mathcal C$.}
\[
\label{eq:MWcondition}
\bar{\Theta}^\ind{I} = (\Theta^\ind{I})^\dagger \Gamma^0 = (\Theta^\ind{I})^t \mathcal{C}~.
\]

The deformed supercoset sigma model has 8 fermionic fields compared to the 32 of the Green-Schwarz action.
In order to rewrite the action in Green-Schwarz form we embed two 4-component fermions $\hat{\theta}^\ind{I}$ into the two 32-component spinors
\[
\label{eq:relation10d4dfermions}
\Theta^\ind{I} &= \begin{pmatrix} 1 \\ 0 \end{pmatrix} \otimes \hat{\theta}^\ind{I} \otimes \begin{pmatrix} a_\ind{(I)} \\ b_\ind{(I)} \\ c_\ind{(I)} \\ d_\ind{(I)} \end{pmatrix}~, \qquad
\bar{\Theta}^\ind{I} = \begin{pmatrix} 0 & 1 \end{pmatrix} \otimes \bar{\hat{\theta}}^\ind{I} \otimes \begin{pmatrix} a_\ind{(I)}^\star & b_\ind{(I)}^\star & c_\ind{(I)}^\star & d_\ind{(I)}^\star \end{pmatrix}~,
\]
where the complex numbers $a_\ind{(I)}$, $b_\ind{(I)}$, $c_\ind{(I)}$ and $d_\ind{(I)}$ may take different values depending on the grading.
Note that for our choice of gamma matrices the 32-component spinors are of negative chirality, $\Gamma^{11} \Theta^\ind{I} = - \Theta^\ind{I}$.
A priori, there is a non-trivial coupling between the spinors and the torus directions giving rise to terms such as $\bar \Theta \partial X^i \Theta$.
Such terms should not survive the truncation since they do not appear in the deformed supercoset sigma model.
Requiring that this is indeed the case constrains the complex numbers $a_\ind{(I)}$, $b_\ind{(I)}$, $c_\ind{(I)}$ and $d_\ind{(I)}$.

To find the complex numbers we assume that the deformed model is supported only by a self-dual RR five-form flux of the form
\[
\label{eq:five_form_flux}
F_5 = \frac{1}{2} (1+\star) F_2 \wedge \Re \Omega_3~,
\]
where $\Omega_3 = \epsilon_{\tilde\imath\tilde\jmath\tilde{k}} dZ^{\tilde{\imath}} \wedge dZ^{\tilde{\jmath}} \wedge dZ^{\tilde{k}}$ is the holomorphic three-form on the torus,
\unskip\footnote{The complex coordinates are chosen to be $Z^1 = X^4 + i X^5$, $Z^2 = X^6 + i X^7$ and $Z^3 = X^8 + i X^9$.}
$F_2 = \half F_{ab} E^a \wedge E^b$ and $\star$ is the 10-dimensional Hodge dual
\unskip\footnote{The Hodge dual is defined such that
\begin{equation*}
\star ( \extder X^\ind{A_0} \wedge \extder X^\ind{A_1} \wedge \extder X^\ind{A_2} \wedge \extder X^\ind{A_3} \wedge \extder X^\ind{A_4}) = \epsilon^\ind{A_0A_1A_2A_3A_4A_5A_6A_7A_8A_9} \extder X_\ind{A_5} \wedge \extder X_\ind{A_6} \wedge \extder X_\ind{A_7} \wedge \extder X_\ind{A_8} \wedge \extder X_\ind{A_9}~,
\end{equation*}
with $\epsilon^{0123456789}=+1$.}
squaring to one, $\star^2 =+1$.
With this ansatz
\[
S^\ind{IJ} = - \frac{1}{2 \cdot 5!} \epsilon^\ind{IJ} e^\Phi F_\ind{ABCDE} \Gamma^\ind{ABCDE} = - \frac{1}{2} \epsilon^\ind{IJ} e^\Phi F_{ab} \Gamma^{ab} \Gamma^{468} \mathcal P_4 (\identity + \Gamma^{11}) ~,
\]
where
\[
\mathcal P_4 = \frac{1}{4} (\identity - \Gamma^{4567} - \Gamma^{4589} - \Gamma^{6789}) ~,
\]
is a projector, $(\mathcal P_4)^2 = \mathcal P_4$, that can be used to decompose the 32-component spinors
\[
\Theta_\perp^\ind{I} = (\identity - \mathcal P_4) \Theta^\ind{I}~, \qquad \Theta_\parallel^\ind{I} = \mathcal P_4 \Theta^\ind{I} ~.
\]
The additional properties
\[
\mathcal P_4^t = \mathcal P_4~, \quad \com{\mathcal P_4}{\mathcal C}=0~, \quad \com{\mathcal P_4}{\Gamma^a}=0~, \quad \com{\mathcal P_4}{\Gamma^{468}}=0~, \quad \com{\mathcal P_4}{\Gamma^{11}}=0~,\quad \mathcal P_4 \Gamma^{i} \mathcal P_4 =0~,
\]
then imply that there are no linear terms in $\set{\Theta_\perp,X^i}$ in the Green-Schwarz action.
It thus follows that setting $X^i = \Theta_\perp^\ind{I}=0$ is a consistent truncation.
As can be seen from the explicit form of $\mathcal P_4$
\[
\mathcal P_4 = \frac{1}{4} \identity \otimes \identity \otimes \identity \otimes (\identity \otimes \identity - \sum_{\iota=1,2,3} \sigma_\iota \otimes \sigma_\iota)~,
\]
setting $\Theta_\perp^\ind{I} =0$ imposes conditions on the complex numbers $a_\ind{(I)}$, $b_\ind{(I)}$, $c_\ind{(I)}$ and $d_\ind{(I)}$.
In particular, these are satisfied when $a_\ind{(I)}=d_\ind{(I)}=0$, $c_\ind{(I)}=-b_\ind{(I)}$.
Henceforth, we will consider this truncation and take the 32-component spinors and their Dirac conjugates to be
\[
\Theta^\ind{I} &= b_\ind{(I)} \begin{pmatrix} 1 \\ 0 \end{pmatrix} \otimes \hat{\theta}^\ind{I} \otimes \begin{pmatrix} 0 \\ +1 \\ -1 \\ 0 \end{pmatrix}~, \qquad
\bar{\Theta}^\ind{I} = b_\ind{(I)}^\star \begin{pmatrix} 0 & 1 \end{pmatrix} \otimes \bar{\hat{\theta}}^\ind{I} \otimes \begin{pmatrix} 0 & +1 & -1 & 0 \end{pmatrix}~.
\]

Using the 32-dimensional gamma matrices given in \appref{app:gamma_matrices} we have
\[
\label{eq:TGTeqtgt}
\bar{\Theta}^\ind{I} \Gamma^a \Theta^\ind{J} &= 2 b_\ind{(I)}^\star b^{\vphantom{\star}}_\ind{(J)} \bar{\hat{\theta}}^\ind{I} \gamma^a \hat{\theta}^\ind{J} ~, \qquad & \bar{\Theta}^\ind{I} \Gamma^a \Gamma^{bc} \Theta^\ind{J} &= 2 b_\ind{(I)}^\star b^{\vphantom{\star}}_\ind{(J)} \bar{\hat{\theta}}^\ind{I}\gamma^a \gamma^{bc} \hat{\theta}^\ind{J}~, \\ \bar{\Theta}^\ind{I} \Gamma^i \Theta^\ind{J}&=0~, & \bar{\Theta}^\ind{I} \Gamma^i \Gamma^{bc} \Theta^\ind{J}&=0~,
\]
and the Lagrangian of the Green-Schwarz action can be rewritten
\[
\label{eq:4dGS}
\hat{\Lag}_{GS} & = i \sqrt{\frac{T}{2} } b^\star_\ind{(I)} b^{\vphantom{\star}}_\ind{(J)} \bar{\hat{\theta}}^\ind{I} (\sigma_1^{\alpha \beta} \delta^\ind{IJ} + \epsilon^{\alpha \beta} \sigma_3^\ind{IJ} ) E_{\alpha a} \gamma^a \Big(\partial_\beta - \frac{1}{4} \omega_{\beta bc} \gamma^{bc}\Big) \hat{\theta}^\ind{J} \\
& \quad - \frac{i}{16} \sqrt{\frac{T}{2} }\bar{\Theta}_\parallel^\ind{I} (\sigma_1^{\alpha \beta} \delta^\ind{IJ} + \epsilon^{\alpha \beta} \sigma_3^\ind{IJ} ) \epsilon^\ind{JK} e^\Phi E_{\alpha a} \Gamma^a F_{bc} \Gamma^{bc} \Gamma^{468} E_{\beta d}\Gamma^d \Theta_\parallel^\ind{K}~.
\]

Finally, let us comment on the reality condition satisfied by the 4-dimensional fermions $\hat{\theta}^\ind{I}$.
The Majorana condition \eqref{eq:MWcondition} implies
\[
\label{eq:MWcondition_theta}
(\hat{\theta}^\ind{I})^\dagger = \frac{b_\ind{(I)}}{b_\ind{(I)}^\star} (\hat{\theta}^\ind{I})^t~.
\]
The two fermions can thus have different reality conditions depending on the choice of the complex numbers $b_\ind{(I)}$.
We will choose to work with
\[
\label{eq:choice_b}
b_\ind{(1)}= \frac{i}{\sqrt{2}}~, \qquad b_\ind{(2)}=\frac{1}{\sqrt{2}},
\]
such that the Majorana condition becomes
\[
\label{eq:MWcondition_theta_choice_b}
(\hat{\theta}^{1})^\dagger = - (\hat{\theta}^{1})^t~, \qquad (\hat{\theta}^{2})^\dagger = (\hat{\theta}^{2})^t~.
\]

%%%%%%%%%%%%%%%%%%%%%%%%%%%%%%%%%%%%%%%%%%%%%%%%%%%%%%%%%%%%%%%%%%%%%%%%%%%%%%%%
\subsubsection{Field redefinitions}

To match the form of the Green-Schwarz Lagrangian \eqref{eq:4dGS} we start by parametrising the field $k$ of the deformed supercoset sigma model as in eq.~\eqref{eq:param_gen} and expand the action \eqref{eq:actionS1} up to quadratic order in the fermions.
We also use the Polyakov-Wiegmann identity to reduce the Wess-Zumino term to a two-dimensional integral.
The resulting Lagrangian
\[
\Lag = \Lag^0 + \Lag^\partial + \Lag^m + \Lag^{\partial \partial}~,
\]
consists of four distinct parts: $\Lag^0 \sim \partial X \partial X$ is the bosonic Lagrangian giving rise to the metric of the $\lambdastar$-deformed model and is discussed in \namedref{subsec.}{ssec:nsns}, $\Lag^\partial \sim \partial X \theta \partial \theta$ contains the terms with one derivative acting on the fermions, $\Lag^m \sim \partial X \partial X \theta \theta$ are the fermion ``mass'' terms and finally $\Lag^{\partial \partial} \sim \partial \theta \partial \theta$.

Field redefinitions are needed to rewrite $\Lag^\partial$, $\Lag^m$ and $\Lag^{\partial \partial}$ in Green-Schwarz form, in particular matching the consistent truncation \eqref{eq:4dGS}.
In \appref{app:field_redef} we show that $\Lag^{\partial \partial}$ is a total derivative and thus can be ignored.
We then focus on two types of transformations, namely shifts of the bosons $X \rightarrow X + \theta s(X) \theta$ and rotations of the fermions $\theta \rightarrow r(X) \hat{\theta}$, which lead to the following modifications
\[
\Lag^0 \rightarrow \Lag^0 + (\delta \Lag^0)^\partial + (\delta \Lag^0)^m~, \qquad
\Lag^\partial \rightarrow \hat{\Lag}^\partial + (\delta \Lag^\partial)^m~, \qquad \Lag^m \rightarrow \hat{\Lag}^m~.
\]
Therefore, we would like to find functions $s(X)$ and $r(X)$ such that
\begin{equation}\begin{gathered}
\label{eq:conditions_rs}
\hat{\Lag}^\partial + (\delta \Lag^0)^\partial = \hat{\Lag}_{GS}^\partial~, \qquad \hat{\Lag}^m + (\delta \Lag^0)^m + (\delta \Lag^\partial)^m = \hat{\Lag}_{GS}^m~,
\end{gathered}\end{equation}
where the equalities hold up to total derivatives.

In order to find the exact functions $s(X)$ and $r(X)$ satisfying these conditions we follow the procedure outlined in \cite{Arutyunov:2015qva}.
All terms quadratic in the fermions contributing to the Lagrangian can be classified according to their symmetry properties under the exchange of the two fermions.
At quadratic order in fermions the Lagrangian can be written as $\Lag = \Lag_+ + \Lag_-$, where $\Lag_+$ and $\Lag_-$ contain the terms with the symmetry property
\[
\label{eq:sym_prop}
\Lag_\pm \supset \theta^\ind{I} f_\pm^\ind{IJ} \lambda^\ind{J} = \pm \lambda^\ind{I} f_\pm^\ind{IJ} \theta^\ind{J}~.
\]
In the above expression the sum over the spinor indices is understood: $f_\pm^\ind{IJ}$ are $4 \times 4$ matrices depending on the bosons.
In particular, this decomposition can be applied to the terms containing one derivative acting on the fermions, $\Lag^\partial = \Lag_+^\partial + \Lag_-^\partial$.
In the Green-Schwarz Lagrangian we have that $\hat{\Lag}_{GS\,+}^\partial=0$, that is the Lagrangian only contains terms with the symmetry property
\[
\theta^\ind{I} f_-^\ind{IJ} \partial \theta^\ind{J} = - \partial \theta^\ind{I} f_-^\ind{IJ} \theta^\ind{J}~.
\]
Having identified the contributions to $\Lag_\pm ^\partial$ we use a field redefinition to set $\Lag_+^\partial$ equal to zero.
Observing that rotating the fermions does not affect the symmetry property,
\[
\Lag_+^\partial \rightarrow \hat{\Lag}_+^\partial~, \qquad \Lag_-^\partial \rightarrow \hat{\Lag}_-^\partial~,
\]
this can only be achieved with a shift of the bosons and fixes for us the function $s(X)$.
Finally, the rotation of the fermions is implemented to rewrite the remaining terms in $\Lag^\partial$ in Green-Schwarz form.
More details, including the exact field redefinitions $r(X)$ and $s(X)$, are presented in \appref{app:field_redef}.
After using the field redefinitions we find that the conditions \eqref{eq:conditions_rs} are satisfied, the Lagrangian takes precisely the form of the Green-Schwarz Lagrangian \eqref{eq:4dGS} and we can easily read off the RR fluxes, or more precisely the combination $e^\Phi F$.

%%%%%%%%%%%%%%%%%%%%%%%%%%%%%%%%%%%%%%%%%%%%%%%%%%%%%%%%%%%%%%%%%%%%%%%%%%%%%%%%
\subsubsection{RR Fluxes}

Comparing with the form of the Green-Schwarz Lagrangian \eqref{eq:4dGS} we find that the only necessary components of the two-form $F_2$ in eq.~\eqref{eq:five_form_flux} are
\[
\sqrt{\frac{T}{2} } F_{12} = -\sqrt{\frac{T}{2} } F_{21} = -2i \sqrt{1+\kappa^2} e^{-\Phi}~,
\]
which leads to the following RR five-form flux
\[
\label{eq:RRflux}
\sqrt{\frac{T}{2} }  F_5 = -i \sqrt{1+\kappa^2} e^{-\Phi} (1+\star) \, E^1 \wedge E^2 \wedge \Re \Omega_3~.
\]
This flux is imaginary, a consequence of the fact that we have dualised in a timelike direction and are now strictly speaking in type IIB$^\star$ supergravity \cite{Hull:1998vg}.
Since the two spinors $\Theta^\ind{I}$ satisfy the Majorana condition, this implies that the corresponding action of the Green-Schwarz superstring is not real.
However, the Poisson-Lie dual of the $\eta$-deformed $\AdS_2 \times \Sp^2$ supercoset has a real action.
This discrepancy comes from the rotation of the fermions $\theta^\ind{I} = U^\ind{IJ} \hat{\theta}^\ind{J}$, where $\theta^\ind{I}$ are the real 4-dimensional fermions appearing in the deformed supercoset model via the parametrisation \eqref{eq:param_gen}, $U^\ind{IJ}$ is the rotation matrix used to rewrite the action in Green-Schwarz form and the 4-dimensional fermions $\hat{\theta}^\ind{I}$ are related to the 32-components spinors $\Theta^\ind{I}$ as in eq.~\eqref{eq:relation10d4dfermions}.
When choosing the particular coefficients \eqref{eq:choice_b} the Majorana condition implies the reality conditions \eqref{eq:MWcondition_theta_choice_b} for the 4-dimensional fermions $\hat{\theta}^\ind{I}$: $\hat{\theta}^1$ is imaginary and $\hat{\theta}^2$ is real.
However, the rotation matrix $U^\ind{IJ}$ defined in eq.~\eqref{eq:rotation} has only real entries and thus we identify a real and an imaginary fermion, thereby breaking the reality condition.

Implementing the transformation rules \eqref{eq:transfo_rules} and \eqref{eq:kl} we find
\[
\sqrt{\kst} F_5 = - \sqrt{\frac{4 \lst}{1-\lst}} \frac{e^{-\Phi}}{\sqrt{\check{p}^2 - \check{q}^2 -1} \sqrt{\hat{p}^2 + \hat{q}^2 -1}} (1+\star ) \, \extder \check{q} \wedge \extder \hat{p} \wedge \Re \Omega_3 ~,
\]
which is precisely the RR five-form flux supporting the metric of the $\lambda$-deformed model on $\AdS_2 \times \Sp^2 \times \To^6$ found in \cite{Sfetsos:2014cea}.

In conclusion, the Poisson-Lie dual of the $\eta$-deformed $\AdS_2 \times \Sp^2 \times \To^6$ superstring with respect to the full bosonic subalgebra $\alg{su}(1,1) \dsum \alg{su}(2)$, with metric \eqref{eq:fullmet} and RR five-form flux \eqref{eq:RRflux}, solves the standard supergravity equations with the dilaton given by \cite{Sfetsos:2014cea}
\[
e^{\Phi} = \frac{1}{\sqrt{\check{p}^2 + \check{q}^2 - 1}\sqrt{\hat{p}^2 - \hat{q}^2 -1}} ~.
\]

%%%%%%%%%%%%%%%%%%%%%%%%%%%%%%%%%%%%%%%%%%%%%%%%%%%%%%%%%%%%%%%%%%%%%%%%%%%%%%%%
\section{Concluding comments}

In this paper we have investigated Poisson-Lie duals of the $\eta$-deformed $\AdS_2 \times \Sp^2 \times \To^6$ superstring.
While the background of the $\eta$-deformed $\AdS_2 \times \Sp^2 \times \To^6$ superstring satisfies a generalisation of the standard supergravity equations, here we have discussed three Poisson-Lie duals, with respect to \textit{(i)} the full $\alg{psu}(1,1|2)$ superalgebra, \textit{(ii)} the full bosonic subalgebra and $(iii)$ the Cartan subalgebra, for which the corresponding backgrounds are expected to satisfy the standard supergravity equations.
Focusing on the second case we explicitly derived the background, showing agreement with the embedding of the metric of the $\lambda$-deformed model on $\AdS_2 \times \Sp^2$ in supergravity given in \cite{Sfetsos:2014cea} up to an analytic continuation.

There are various interesting open questions extending and developing the results presented here.
Firstly, it would also be useful to confirm some of the remaining conjectures summarised in this paper and extend the results to other backgrounds.
This includes explicitly checking that Poisson-Lie dualising with respect to the full $\mathfrak{psu}(1,1|2)$ superalgebra gives the background of \cite{Borsato:2016zcf} up to analytic continuation and dualising with respect to the Cartan subalgebra gives the two-fold T-dual of \cite{Hoare:2015gda,Hoare:2015wia,Arutyunov:2015mqj}.
Furthermore, candidates for the background of the Poisson-Lie dual of the $\eta$-deformed $\AdS_3 \times \Sp^3 \times \To^4$ and $\AdS_5 \times \Sp^5$ superstrings with respect to the full bosonic subalgebras are given in \cite{Sfetsos:2014cea} and \cite{Demulder:2015lva} respectively, up to analytic continuation.

It is highly probable that the backgrounds discussed in this paper define integrable 2-dimensional sigma models.
The results of \cite{Severa:2017kcs} together with the integrability of the $\eta$-deformed principal chiral, symmetric space sigma and semi-symmetric space sigma models \cite{Klimcik:2008eq,Delduc:2013qra,Delduc:2013fga} suggest that their Poisson-Lie duals are also integrable.
It is important to point out that the semi-symmetric space sigma model on the supercoset \eqref{eq:supercoset} is a truncation of the type II Green-Schwarz superstring on certain $\AdS_2 \times \Sp^2 \times \To^6$ backgrounds.
The integrability of the latter was demonstrated to quadratic order in fermions in \cite{Wulff:2014kja}.
In principle, for a complete analysis of integrability, this analysis should be extended to the deformed backgrounds.

As well as exploring deformations of other $\AdS_2$ integrable string backgrounds \cite{Zarembo:2010sg,Wulff:2015mwa}, it would be interesting to consider some of the ways integrability has been used to study the $\eta$- and $\lambda$-deformed $\AdS_5 \times \Sp^5$ superstrings in the context of the $\eta$-deformed $\AdS_2 \times \Sp^2 \times \To^6$ superstring and its Poisson-Lie duals.
This includes the light-cone gauge S-matrix \cite{Beisert:2008tw,Hoare:2011wr,Arutyunov:2013ega,Arutyunov:2015qva}, which should be a deformation of the S-matrix of \cite{Hoare:2014kma}, and the associated finite-size spectrum \cite{Arutyunov:2012zt,Arutyunov:2012ai,Arutynov:2014ota,Klabbers:2017vtw}.
Another direction would be to investigate how the analysis of the solitons in the $\lambda$-deformed $\AdS_5 \times \Sp^5$ superstring \cite{Appadu:2017xku} is modified when considering the Poisson-Lie dual of the $\eta$-deformation with respect to either the full superalgebra (that is the $\lambdastar$-deformed model, an analytic continuation of the $\lambda$-deformed model) or the full bosonic subalgebra.
This is of relevance to both the $\AdS_5 \times \Sp^5$ and $\AdS_2 \times \Sp^2 \times \To^6$ cases.
One could also ask how the $q$-deformed symmetry of the $\eta$-deformed model \cite{Delduc:2013fga,Delduc:2014kha,Delduc:2016ihq,Delduc:2017brb,Arutyunov:2013ega} and the contraction limits (that is the maximal deformation, $\eta \to 1$, limit) of \cite{Arutynov:2014ota,Arutyunov:2014cra,Pachol:2015mfa} behave under Poisson-Lie duality.

Finally, while there has been much study of quantum aspects of Poisson-Lie duality, as well as its interplay with supergravity and generalised geometry, including, for example, \cite{Tyurin:1995bu,Bossard:2001au,VonUnge:2002xjf,Hlavaty:2004jp,Hlavaty:2004mr,Jurco:2017gii,Alekseev:1995ym,Valent:2009nv,Sfetsos:2009dj,Avramis:2009xi,Sfetsos:2009vt,Hlavaty:2012sg,Hassler:2017yza,Lust:2018jsx}, a systematic understanding of the model on the Drinfel'd double in the path integral and the Weyl anomaly associated to integrating out the degrees of freedom of a non-unimodular algebra, as given for non-abelian duality in \cite{Alvarez:1994np,Elitzur:1994ri}, remains to be found.

%%%%%%%%%%%%%%%%%%%%%%%%%%%%%%%%%%%%%%%%%%%%%%%%%%%%%%%%%%%%%%%%%%%%%%%%%%%%%%%%
\pdfbookmark[1]{Acknowledgements}{ack}
\section*{Acknowledgements}

We thank A.~Tseytlin for comments on the draft.
This work is supported by grant no.~615203 from the European Research Council under the FP7.

%%%%%%%%%%%%%%%%%%%%%%%%%%%%%%%%%%%%%%%%%%%%%%%%%%%%%%%%%%%%%%%%%%%%%%%%%%%%%%%%
\appendix

%%%%%%%%%%%%%%%%%%%%%%%%%%%%%%%%%%%%%%%%%%%%%%%%%%%%%%%%%%%%%%%%%%%%%%%%%%%%%%%%
\section{Dynkin diagrams of \texorpdfstring{$\alg{sl}(2|2; \Complex)$}{sl(2|2;C)} }
\label{app:Different_Dynkin}

The Lie superalgebra $\alg{sl}(2|2; \Complex)$ admits three Dynkin diagrams
\[\label{eq:dynkins}
\Circle - \otimes -\Circle~, \qquad \otimes - \Circle - \otimes~, \qquad \otimes - \otimes - \otimes ~,
\]
where $\Circle$ denotes a bosonic root and $\otimes$ a fermionic root.
In this appendix we present the $\alg{sl}(2|2;\Complex)$ superalgebra and for each Dynkin diagram give a Cartan-Weyl basis, the corresponding Cartan matrix and discuss the unimodularity properties of the Borel subalgebra spanned by the Cartan generators and positive roots.

\paragraph{The $\alg{sl}(2|2;\Complex)$ superalgebra.}
The bosonic subalgebra of $\alg{sl}(2|2;\Complex)$ is $\alg{sl}(2;\Complex) \dsum \alg{sl}(2;\Complex) \dsum \alg{gl}(1;\Complex)$ for which we introduce the corresponding generators $\gen{K}_0, \gen{K}_\pm$, $\gen{L}_0, \gen{L}_\pm$ and $\gen{C}_0$.
We also introduce the eight supercharges $\gen{Q}^{\pm\check{\alpha} \hat{\alpha}}$ where $\check{\alpha} = \pm$ is the spinor index associated to the first copy of $\alg{sl}(2;\Complex)$ and $\hat{\alpha} = \pm$ to the second.
The first index corresponds to the splitting of the supercharges under the $\alg{gl}(1;\Complex)$ outer automorphism generated by $\gen{R}$
\[
\com{\gen{R}}{\gen{Q}^{\pm\check{\alpha} \hat{\alpha}}} = \pm \half \gen{Q}^{\pm\check{\alpha} \hat{\alpha}}~.
\]
The non-vanishing commutation relations are
\[
\com{\gen{K}_0}{\gen{K}_\pm} &= \pm \gen{K}_\pm~, & \com{\gen{K}_+}{\gen{K}_-} &= 2\gen{K}_0~, & \com{\gen{K}_0}{\gen{Q}^{\beta\pm \hat{\alpha}}} & = \pm \half \gen{Q}^{\beta\pm \hat{\alpha}}~,& \com{\gen{K}_\pm}{\gen{Q}^{\beta\mp \hat{\alpha}}} & = \gen{Q}^{\beta\pm \hat{\alpha}}~, \\
\com{\gen{L}_0}{\gen{L}_\pm} &= \pm \gen{L}_\pm~, & \com{\gen{L}_+}{\gen{L}_-} &= 2 \gen{L}_0~, & \com{\gen{L}_0}{\gen{Q}^{\beta \check{\alpha} \pm }} & = \pm \half \gen{Q}^{\beta \check{\alpha} \pm }~,& \com{\gen{L}_\pm}{\gen{Q}^{ \beta\check{\alpha} \mp }} & = \gen{Q}^{\beta\check{\alpha} \pm }~,
\]
while the non-vanishing anticommutation relations for the supercharges read
\[
\anticom{\gen{Q}^{+ \pm +}}{\gen{Q}^{-\pm -}} &= \pm \gen{K}_\pm~, & \anticom{\gen{Q}^{- \pm +}}{\gen{Q}^{+\pm -}} &= \mp \gen{K}_\pm~, & \anticom{\gen{Q}^{\pm+ \pm}}{\gen{Q}^{\mp- \mp}} &= - \gen{K}_0 \pm \gen{L}_0 \mp \gen{C}_0~, \\
\anticom{\gen{Q}^{+ + \pm}}{\gen{Q}^{- - \pm}} &= \mp \gen{L}_\pm~, & \anticom{\gen{Q}^{- + \pm}}{\gen{Q}^{+ - \pm}} &= \pm \gen{L}_\pm~, & \anticom{\gen{Q}^{\mp + \pm}}{\gen{Q}^{\pm - \mp}} &= + \gen{K}_0 \mp \gen{L}_0 \mp \gen{C}_0~,
\]
and the central element $\gen{C}_0$ commutes with all generators.

\paragraph{Cartan-Weyl basis.}
The three Dynkin diagrams of $\alg{sl}(2;\Complex)$ \eqref{eq:dynkins} correspond to inequivalent sets of simple roots.
To identify the roots, let us introduce a generic Cartan-Weyl basis for $\alg{sl}(2|2 ; \Complex)$ composed of the three Cartan generators $\set{h_i}$ and the positive $\set{e_i}$ and negative $\set{f_i}$ simple roots satisfying the defining relations
\[
\com{h_i}{e_j} = a_{ij} e_j ~, \qquad \com{h_i}{f_j} = - a_{ij} f_j ~, \qquad [e_i,f_j\} = \delta_{ij} h_j ~,
\]
where $a_{ij}$ is the symmetrised Cartan matrix.
The non-simple roots $\set{e_\ind{M}}$ are given by
\[
e_{12} &= [e_1,e_2\}~, &\qquad e_{23} &= [e_2,e_3\}~, &\qquad e_{123} &= [e_1,[e_2,e_3\}\}~, \\
f_{21} &= [f_2,f_1\}~, &\qquad f_{32} &= [f_3,f_2\}~, &\qquad f_{321} &= [f_3,[f_2,f_1\}\}~.
\]
The Borel subalgebra is generated by
\[\label{eq:borelsa}
\set{h_i, e_\ind{M} , i e_\ind{M}} ~.
\]

\paragraph{Dynkin diagrams.}
\begin{enumerate}
\item $\Circle-\otimes-\Circle$
\\
In this case there are two bosonic simple roots and one fermionic.
A choice of Cartan generators and positive and negative simple roots is
\[
&h_1 = +2 \gen{K}_0~, &\qquad & e_1 = - \gen{K}_-~, &\qquad & f_1 = + \gen{K}_+~, \\
&h_2 = - \gen{K}_0 - \gen{L}_0 - \gen{C}_0 ~, &\qquad & e_2 = + \gen{Q}^{++-}~, &\qquad & f_2 = - \gen{Q}^{--+} ~, \\
&h_3 = + 2 \gen{L}_0~, &\qquad & e_3 = + \gen{L}_+~, & \qquad & f_3 = + \gen{L}_- ~,
\]
with the corresponding symmetrised Cartan matrix given by
\[
\begin{pmatrix}
-2 & +1 & 0\\
+1 & 0 & -1 \\
0 & -1 & +2
\end{pmatrix}~.
\]
The non-simple roots are
\[
e_{12} &= -\gen{Q}^{+--}~, &\qquad e_{23} &= - \gen{Q}^{+++}~, &\qquad e_{123} &= + \gen{Q}^{+-+}~, \\
f_{21} &= +\gen{Q}^{-++}~, &\qquad f_{32} &= - \gen{Q}^{---}~, &\qquad f_{321} &= + \gen{Q}^{-+-}~.
\]
The Borel subalgebra \eqref{eq:borelsa} is non-unimodular
\[
\tilde{f}^{\gen{K}_0 b}{}_b = -2~, \qquad \tilde{f}^{\gen{L}_0 b}{}_b = +2 ~.
\]
\item $\otimes-\Circle-\otimes$
\\
In this case there are two fermionic simple roots and one bosonic.
The bosonic root can belong either to the first or second copy of $\alg{sl}(2;\Complex)$.
Although the two choices are symmetric, we shall present both for convenience.
\begin{itemize}
\item When the bosonic simple root comes from the first copy of $\alg{sl}(2;\Complex)$ a choice for Cartan generators and positive and negative simple roots is
\[
&h_1 = \gen{K}_0+\gen{L}_0 - \gen{C}_0 ~, &\qquad & e_1 = - \gen{Q}^{+-+ }~, &\qquad & f_1 = +\gen{Q}^{-+-}~, \\
&h_2 = -2 \gen{K}_0 ~, &\qquad & e_2 = +\gen{K}_+ ~, &\qquad & f_2 = -\gen{K}_-~,\\
&h_3 = \gen{K}_0+ \gen{L}_0 + \gen{C}_0 ~, &\qquad & e_3 = -\gen{Q}^{--+ } ~, & \qquad & f_3 = -\gen{Q}^{++- }~,
\]
with the corresponding symmetrised Cartan matrix given by
\[
\begin{pmatrix}
0 & +1 & 0\\
+1 & -2 & +1 \\
0 & +1 & 0
\end{pmatrix}~.
\]
The non-simple roots are
\[
e_{12} &= + \gen{Q}^{+++}~, &\qquad e_{23} &= - \gen{Q}^{-++}~, &\qquad e_{123} &= + \gen{L}_{+}~, \\
f_{21} &= - \gen{Q}^{---}~, &\qquad f_{32} &= - \gen{Q}^{+--} ~, &\qquad f_{321} &= + \gen{L}_- ~.
\]
The Borel subalgebra \eqref{eq:borelsa} is non-unimodular
\[
\tilde{f}^{\gen{K}_0 b}{}_b = +2~, \qquad \tilde{f}^{\gen{L}_0 b}{}_b = +6 ~.
\]
\item When the bosonic simple root comes from the second copy of $\alg{sl}(2;\Complex)$ a choice for Cartan generators and positive and negative simple roots is
\[
&h_1 = -\gen{K}_0-\gen{L}_0 + \gen{C}_0 ~, &\qquad & e_1 = +\gen{Q}^{-+- }~, &\qquad & f_1 = +\gen{Q}^{+-+} ~, \\
&h_2 = +2 \gen{L}_0 ~, &\qquad & e_2 = +\gen{L}_+ ~, &\qquad & f_2 = +\gen{L}_- ~, \\
&h_3 = -\gen{K}_0- \gen{L}_0 - \gen{C}_0 ~, &\qquad & e_3 = +\gen{Q}^{++- } ~, & \qquad & f_3 = -\gen{Q}^{--+} ~,
\]
with the corresponding symmetrised Cartan matrix given by
\[
\begin{pmatrix}
0 & -1 & 0\\
-1 & +2 & -1 \\
0 & -1 & 0
\end{pmatrix} ~.
\]
The non-simple roots are
\[
e_{12} &= -\gen{Q}^{-++}~, &\qquad e_{23} &= + \gen{Q}^{+++}~, &\qquad e_{123} &= + \gen{K}_{+}~, \\
f_{21} &= + \gen{Q}^{+--}~, &\qquad f_{32} &= + \gen{Q}^{---} ~, &\qquad f_{321} &= - \gen{K}_-~.
\]
The Borel subalgebra \eqref{eq:borelsa} is non-unimodular
\[
\tilde{f}^{\gen{K}_0 b}{}_b = +6~, \qquad \tilde{f}^{\gen{L}_0 b}{}_b = +2 ~.
\]
\end{itemize}
\item $\otimes-\otimes-\otimes$
\\
In this case all three simple roots are fermionic.
A choice of Cartan generators and positive and negative simple roots is
\[
&h_1 = +\gen{K}_0+\gen{L}_0+\gen{C}_0 ~, &\qquad & e_1 =+ \gen{Q}^{++- }~, &\qquad & f_1 = +\gen{Q}^{--+} ~, \\
&h_2 = + \gen{K}_0 -\gen{L}_0 - \gen{C}_0~, &\qquad & e_2 = +\gen{Q}^{-++} ~, &\qquad & f_2 = +\gen{Q}^{+--} ~, \\
&h_3 = - \gen{K}_0 - \gen{L}_0+\gen{C}_0~, &\qquad & e_3 = +\gen{Q}^{+-+ } ~, & \qquad & f_3 = +\gen{Q}^{-+- } ~,
\]
with the corresponding symmetrised Cartan matrix given by
\[
\begin{pmatrix}
0 & +1 & 0\\
+1 & 0 & -1 \\
0 & -1 & 0
\end{pmatrix} ~.
\]
The non-simple roots are
\[
e_{12} &= - \gen{K}_+~, &\qquad e_{23} &= + \gen{L}_+~, &\qquad e_{123} &= - \gen{Q}^{+++}~, \\
f_{21} &= + \gen{K}_-~, &\qquad f_{32} &= - \gen{L}_-~, &\qquad f_{321} &= - \gen{Q}^{---}~.
\]
The Borel subalgebra \eqref{eq:borelsa} is non-unimodular
\[
\tilde{f}^{\gen{K}_0 b}{}_b = +4~, \qquad \tilde{f}^{\gen{L}_0 b}{}_b = +4 ~.
\]
\end{enumerate}

\paragraph{The $\eta$-deformed models and their Poisson-Lie duals.}
We conclude this appendix with a few comments on the $\eta$-deformed models that correspond to the different Drinfel'd-Jimbo R-matrices associated to the various Cartan-Weyl bases discussed above, together with their Poisson-Lie duals.
The question of whether these $\eta$-deformations are inequivalent or not has not been previously studied.
However, the Borel subalgebra \eqref{eq:borelsa} is non-unimodular in all three cases and thus we expect the corresponding $\eta$-deformed models to each have a Weyl anomaly.

The backgrounds of the Poisson-Lie duals with respect to the full $\alg{psu}(1,1|2)$ superalgebra and the Cartan subalgebra, which can be considered in all three cases, should solve the supergravity equations as the degrees of freedom that are integrated out are associated to unimodular algebras.
As discussed in \cite{Hoare:2017ukq} one can consider Poisson-Lie duals with respect to subalgebras that correspond to sub-Dynkin diagrams.
The bosonic subalgebra $\alg{su}(1,1) \dsum \alg{su}(2)$ corresponds to a sub-Dynkin diagram for choice 1, but not for choices 2 and 3.
%However, it is worth noting that it should still be possible to Poisson-Lie dualise the $\eta$-deformed models corresponding to choices 2 and 3 with respect to the full bosonic subalgebra by rewriting in terms of the Cartan-Weyl basis of choice 1.
%The same logic holds for any sub-Dynkin diagram of any of the three Dynkin diagrams.

%%%%%%%%%%%%%%%%%%%%%%%%%%%%%%%%%%%%%%%%%%%%%%%%%%%%%%%%%%%%%%%%%%%%%%%%%%%%%%%%
\section{Generators of \texorpdfstring{$\alg{psu}(1,1|2)$}{psu(1,1|2)} and \texorpdfstring{$\alg{pb}(1,1|2)$}{pb(1,1|2)}}
\label{app:sl_generators}

In this appendix we present the matrix realisation of the superalgebras $\alg{psu}(1,1|2)$ and $\alg{pb}(1,1|2)$ that we use in secs.~\ref{sec:etadeformation} and \ref{sec:PLdualityBosonic}.
Our conventions for the former largely follow those of \cite{Borsato:2016zcf}.

%%%%%%%%%%%%%%%%%%%%%%%%%%%%%%%%%%%%%%%%%%%%%%%%%%%%%%%%%%%%%%%%%%%%%%%%%%%%%%%%
\subsection{Generators of \texorpdfstring{$\alg{psu}(1,1|2)$}{psu(1,1|2)}}

The isometry algebra of the $\AdS_2 \times \Sp^2$ supercoset is $\alg{psu}(1,1|2)$.
As a matrix superalgebra, the complexification $\alg{sl}(2|2;\Complex)$ is spanned by $4 \times 4$ matrices of block form
\[
M= \begin{pmatrix}
m & \theta \\
\eta & n
\end{pmatrix} ~,
\]
with vanishing supertrace, $\STr M = \Tr m - \Tr n=0$.
The superalgebra $\alg{su}(1, 1|2)$ is a real form of $\alg{sl}(2|2;\Complex)$ identified by the reality condition
\[ \label{eq:realitycond} M^\dagger H + H M =0~, \qquad H=\begin{pmatrix} \sigma_3 & 0 \\ 0 & \identity_2 \end{pmatrix}~.
\]
This implies
\[ m^\dagger = - \sigma_3 m \sigma_3~, \qquad n^\dagger=-n~,\qquad \eta^\dagger=-\sigma_3 \theta~,
\]
such that $m$ and $n$ span the unitary subalgebras $\alg{u}(1,1)$ and $\alg{u}(2)$ respectively.
The superalgebra $\alg{su}(1,1|2)$ contains the one-dimensional ideal $\alg{u}(1)$ generated by $i \identity_4$.
The superalgebra $\alg{psu}(1,1|2)$ is defined as the quotient algebra of $\alg{su}(1,1|2)$ over this $\alg{u}(1)$ factor.

The automorphism
\[
\Omega(M) = - \begin{pmatrix} \sigma_3 & 0 \\ 0 & \sigma_3 \end{pmatrix}
\begin{pmatrix} m^t & -\eta^t \\ \theta^t & n^t \end{pmatrix}
\begin{pmatrix} \sigma_3 & 0 \\ 0 & \sigma_3 \end{pmatrix} ~,
\]
endows the $\alg{psu}(1,1|2)$ algebra with a $\Integer_4$ grading and the elements of grade $k$ satisfy $\Omega(M)=i^k M$.
The generators below are chosen so that they belong to a specific grading.

\paragraph{Bosonic generators.}
Our choice for the three $\alg{su}(1,1)$ generators is
\begin{align}
P_0 &= -\begin{pmatrix} i \sigma_3 & 0 \\ 0 & 0 \end{pmatrix} ~,
&\quad
P_1 &= -\begin{pmatrix} \sigma_2 & 0 \\ 0 & 0 \end{pmatrix} ~,
&\quad
J_{01} &= + \frac{1}{2} \com{P_0}{P_1}= + \begin{pmatrix} \sigma_1 & 0 \\ 0 & 0 \end{pmatrix}~, \\
\intertext{and for the three $\alg{su}(2)$ generators}
P_2 &= -\begin{pmatrix} 0 & 0 \\ 0 & i \sigma_3 \end{pmatrix} ~,
&\quad
P_3 &= -\begin{pmatrix} 0 & 0 \\ 0 & i \sigma_2 \end{pmatrix} ~,
&\quad J_{23} &= -\frac{1}{2} \com{P_2}{P_3}= - \begin{pmatrix} 0 & 0 \\ 0 & i \sigma_1 \end{pmatrix}~.
\end{align}
Here $J_{01}$ and $J_{23}$ generate the $\alg{so}(1,1) \dsum \alg{so}(2)$ grade 0 subalgebra.
The other bosonic generators $P_a$, $a=0,1,2,3$ are of grade 2.

\paragraph{Fermionic generators.}
The $\alg{psu}(1,1|2)$ superalgebra also contains eight fermionic generators $Q_{\ind{I} \check{\alpha} \hat{\alpha}}$, where $I=1,2$ is the grading, $\check{\alpha}=1,2$ is the $\alg{su}(1,1)$ index and $\hat{\alpha}=1,2$ is the $\alg{su}(2)$ index.
To define their $4\times4$ matrix representation, we use the following basis of $\Mat(2; \Complex)$
\[
(N_\ind{\check{\alpha} \hat{\alpha}})_\ind{\check{\beta} \hat{\beta}} = \delta_\ind{\check{\alpha} \check{\beta}} \delta_\ind{\hat{\alpha} \hat{\beta}}~, \qquad \check{\alpha},\check{\beta},\hat{\alpha},\hat{\beta}=1,2 ~,
\]
such that
\[
Q_{1 \check{\alpha} \hat{\alpha}} = e^{- (-1)^{\hat{\alpha}} i \pi/4} \begin{pmatrix} 0 & N_{\check{\alpha} \hat{\alpha}} \\ i \sigma_3 (N_{\check{\alpha} \hat{\alpha}})^t \sigma_3 & 0 \end{pmatrix}~, \qquad Q_{2 \check{\alpha} \hat{\alpha}} = e^{-(-1)^{\hat{\alpha}} i \pi/4} \begin{pmatrix} 0 & i N_{\check{\alpha} \hat{\alpha}} \\ \sigma_3 (N_{\check{\alpha} \hat{\alpha}})^t \sigma_3 & 0 \end{pmatrix}~.
\]
The four generators $Q_{1\check{\alpha} \hat{\alpha}}$ belong to the grade 1 subspace and $Q_{2 \check{\alpha} \hat{\alpha}}$ to the grade 3 subspace.
They satisfy the reality condition \eqref{eq:realitycond} and therefore we use real fermions $\theta^\ind{I}$ to construct the Grassmann envelope, $\theta^\ind{I \check{\alpha} \hat{\alpha}} Q_{\ind{I} \check{\alpha} \hat{\alpha}}$.
For our conjugation conventions \eqref{eq:conjugation}, we have $( \textsf{c} \, \theta_1 \theta_2)^\star = - \textsf{c}^\star \, \theta_1 \theta_2$ for real fermions.
Imposing this quantity to be real fixes the phase $\textsf{c}=i$ as in eq.~\eqref{eq:cei}.

\paragraph{Commutation relations.}
The commutation relations of the $\alg{su}(1,1|2)$ generators are ($J_{bc}=J_{01}, J_{23}$)
\unskip\footnote{For the definitions of the gamma matrices refer to \appref{app:gamma_matrices}.}
\begin{equation}\begin{gathered}
\com{P_0}{P_1} = 2 J_{01} ~, \qquad \com{P_2}{P_3} = -2 J_{23} ~, \qquad \com{P_a}{J_{bc}} =2 (\eta_{ab} P_c - \eta_{ac} P_b)~,
\\
\com{\theta^\ind{I} Q_\ind{I}}{P_a} = -i \epsilon^\ind{IJ} Q_\ind{J} \gamma_a \theta^\ind{I}~,\qquad
\com{\theta^\ind{I} Q_\ind{I}}{J_{ab}} = - \delta^\ind{IJ} Q_\ind{J} \gamma_{ab} \theta^\ind{I}~, \\
\com{\theta^\ind{I} Q_\ind{I}}{\lambda^\ind{J} Q_\ind{J}} = i \delta^\ind{IJ} \theta^\ind{I} \gamma^0 \gamma^a P_a \lambda^\ind{J} + \epsilon^\ind{IJ} \theta^\ind{I} \gamma^0 (- \gamma^{01} J_{01} + \gamma^{23} J_{23}) \lambda^\ind{J} -i \delta^\ind{IJ} \theta^\ind{I} \gamma^0 \identity_4 \lambda^\ind{J}~.
\end{gathered}\end{equation}
For the superalgebra $\alg{psu}(1,1|2)$, the term proportional to the identity in the final commutator is projected out.

%%%%%%%%%%%%%%%%%%%%%%%%%%%%%%%%%%%%%%%%%%%%%%%%%%%%%%%%%%%%%%%%%%%%%%%%%%%%%%%%
\subsection{Generators of \texorpdfstring{$\alg{pb}(1,1|2)$}{pb(1,1|2)}}

The projected Borel subalgebra $\alg{pb}(1,1|2)$ is spanned by the Cartan generators and the positive roots.
For the Dynkin diagram $\Circle - \otimes - \Circle$ with the matrix realisation of $\alg{psu}(1,1|2)$ given above, these can be chosen to be upper triangular matrices.
The duals of the $\alg{psu}(1,1|2)$ generators can then be identified using the inner product \eqref{eq:innerproduct}.

\paragraph{Bosonic generators.}
The six bosonic generators of $\alg{pb}(1,1|2)$ are given by
\[
\tilde{P}^0 &= \frac{1}{2} \begin{pmatrix} \sigma_3 & 0 \\ 0 & 0 \end{pmatrix}~, &\quad \tilde{P}^1 &= \begin{pmatrix} \sigma_+ & 0 \\ 0 & 0 \end{pmatrix} ~, &\quad \tilde{J}^{01} &= -i\com{\tilde{P}^0}{\tilde{P}^1} = \begin{pmatrix} -i\sigma_+ & 0 \\ 0 & 0 \end{pmatrix} ~, \\
\tilde{P}^2 &= \frac{1}{2} \begin{pmatrix} 0 & 0 \\ 0 & - \sigma_3 \end{pmatrix}~, &\quad \tilde{P}^3 &= \begin{pmatrix} 0 & 0 \\ 0 & i \sigma_+ \end{pmatrix} ~, &\quad \tilde{J}^{23} &= -i\com{\tilde{P}^2}{\tilde{P}^3} = \begin{pmatrix} 0 & 0 \\ 0 & -\sigma_+ \end{pmatrix} ~.
\]

\paragraph{Fermionic generators.}
The eight fermionic generators are given by
\[
\tilde{Q}^{1 \check{\alpha} \hat{\alpha}} = e^{+ (-1)^{\hat{\alpha}} i \pi/4} \begin{pmatrix} 0 & i \sigma_3 N_{\check{\alpha} \hat{\alpha}} \sigma_3 \\ 0 & 0 \end{pmatrix}~, \qquad \tilde{Q}^{2 \check{\alpha} \hat{\alpha}} = e^{+ (-1)^{\hat{\alpha}} i \pi/4} \begin{pmatrix} 0 & -\sigma_3 N_{\check{\alpha} \hat{\alpha}} \sigma_3 \\ 0 & 0 \end{pmatrix}~.
\]

\paragraph{Commutation relations.}
The bosonic generators satisfy
\[
\com{\tilde{P}^0}{\tilde{P}^1} & = i \tilde{J}^{01} ~, & \qquad
\com{\tilde{P}^0}{\tilde{J}^{01}} &=\tilde{J}^{01}~, &\qquad \com{\tilde{P}^1}{\tilde{J}^{01}}&=0~, \\
\com{\tilde{P}^2}{\tilde{P}^3} & = i \tilde{J}^{23} ~, & \qquad
\com{\tilde{P}^2}{\tilde{J}^{23}} &=\tilde{J}^{23}~, &\qquad \com{\tilde{P}^3}{\tilde{J}^{23}}&=0~.
\]
The remaining commutation relations are
\[
\com{\tilde{\theta}_\ind{I} \tilde{Q}^\ind{I}}{\tilde{P}^a} = - \frac{1}{2} \delta^\ind{IJ} \tilde{\theta}_\ind{I} \tilde{\gamma}^a \tilde{Q}^\ind{J}~, \qquad
\com{\tilde{\theta}_\ind{I} \tilde{Q}^\ind{I}}{\tilde{J}^{ab}} = - \frac{1}{2} \epsilon^\ind{IJ} \tilde{\theta}_\ind{I} \tilde{\gamma}^{ab} \tilde{Q}^\ind{J}~, \qquad
\com{\tilde{\theta}_\ind{I} \tilde{Q}^\ind{I}}{\tilde{\theta}_\ind{J} \tilde{Q}^\ind{J}} = 0~.
\]
Note that the fermionic supercharges anticommute among themselves and the central element does not appear in the commutation relations.

%%%%%%%%%%%%%%%%%%%%%%%%%%%%%%%%%%%%%%%%%%%%%%%%%%%%%%%%%%%%%%%%%%%%%%%%%%%%%%%%
\subsection{Mixed commutation relations}

The Poisson-Lie dual of the $\eta$-deformed $AdS_2 \times \Sp^2$ supercoset with respect to the full bosonic subalgebra $\alg{su}(1,1) \dsum \alg{su}(2)$ discussed in \secref{sec:PLdualityBosonic} follows from integrating out the degrees of freedom associated to the algebra
\[\tilde{\alg{k}} = \set{J_{01},J_{23},P_a,\tilde{Q}^{\ind{I}\check{\alpha}\hat{\alpha}}} ~.\]
To expand the action to quadratic order in fermions the following commutation relations are useful
\[
\com{\theta^\ind{I} Q_\ind{I}}{\tilde{P}^a} =\, &\delta^\ind{IJ} \tilde{Q}^\ind{J} \gamma^0 \gamma^a \theta^\ind{I}+ \frac{1}{2} \delta^\ind{IJ} Q_\ind{J} \tilde{\gamma}^a \theta^\ind{I}~, \\
\com{\theta^\ind{I} Q_\ind{I}}{\tilde{J}^{ab}}=\, &i \epsilon^\ind{IJ} \tilde{Q}^\ind{J} \gamma^0 \gamma_{ab} \theta^\ind{I} - \frac{1}{2} \epsilon^\ind{IJ} Q_\ind{J} \tilde{\gamma}^{ab} \theta^\ind{I}~, \\
\com{\theta^\ind{I} Q_\ind{I}}{\tilde{\theta}_\ind{J} \tilde{Q}^\ind{J}} = &-i \delta^\ind{IJ} \tilde{\theta}_\ind{J} \Big( \gamma_{12} \tilde{J}^{12} + \gamma_{34} \tilde{J}^{34} - \frac{1}{2} \tilde{\gamma}^a P_a \Big) \theta^\ind{I} \\
&-\epsilon^\ind{IJ} \tilde{\theta}_\ind{J} \Big(\frac{i}{2} \tilde{\gamma}^{12} J_{12} + \frac{i}{2}\tilde{\gamma}^{34} J_{34} - \gamma_a \tilde{P}^a \Big) \theta^\ind{I}~.
\]

%%%%%%%%%%%%%%%%%%%%%%%%%%%%%%%%%%%%%%%%%%%%%%%%%%%%%%%%%%%%%%%%%%%%%%%%%%%%%%%%
\section{4- and 32-dimensional gamma matrices}
\label{app:gamma_matrices}

%%%%%%%%%%%%%%%%%%%%%%%%%%%%%%%%%%%%%%%%%%%%%%%%%%%%%%%%%%%%%%%%%%%%%%%%%%%%%%%%
\subsection{4-dimensional gamma matrices}

Using the Pauli matrices
\[
\sigma_1 = \begin{pmatrix} 0 & 1 \\ 1 & 0 \end{pmatrix}~, \quad \sigma_2 = \begin{pmatrix} 0 & -i \\ i & 0 \end{pmatrix}~, \quad \sigma_3 = \begin{pmatrix} 1 & 0 \\ 0 & -1 \end{pmatrix}~, \quad \sigma_+ = \begin{pmatrix} 0 & 1 \\ 0 & 0 \end{pmatrix}~, \quad \sigma_- = \begin{pmatrix} 0 & 0 \\ 1 & 0 \end{pmatrix}~,
\]
we define the 4-dimensional gamma matrices
\[
\gamma^0 &= -i \sigma_3 \otimes \identity~, &\qquad \gamma^1 &= \sigma_2 \otimes \identity~, &\qquad \gamma^{01} &= + \frac{1}{2} \com{\gamma^0}{\gamma^1}~, \\
\gamma^2 &= -\identity \otimes i\sigma_3~, &\qquad \gamma^3 &= -\identity \otimes i\sigma_1~, &\qquad \gamma^{23} &= - \frac{1}{2} \com{\gamma^2}{\gamma^3}~.
\]
They do not satisfy the Clifford algebra in $1+3$ dimensions, but $\set{\gamma^0,\gamma^1}$ satisfy the Clifford algebra in $1+1$ dimensions and $\set{\gamma^2,\gamma^3}$ in $2$ dimensions.
Dirac conjugation acts on real 4-dimensional fermions as $\bar{\theta} = \theta^\dagger \gamma^0 = \theta^t \gamma^0$.
It is also be useful to introduce
\[
\tilde{\gamma}^0 &= i \gamma^0 ~, &\qquad \tilde{\gamma}^1 &= i \gamma^1 + \gamma^{01}~, &\qquad \tilde{\gamma}^{01} &= \frac{1}{2} \com{\tilde{\gamma}^0}{\tilde{\gamma}^1}~, \\
\tilde{\gamma}^2 &= i\gamma^2~, &\qquad \tilde{\gamma}^3 &= i \gamma^3 + \gamma^{23}~, &\qquad \tilde{\gamma}^{23} &= \frac{1}{2} \com{\tilde{\gamma}^2}{\tilde{\gamma}^3}~.
\]

%%%%%%%%%%%%%%%%%%%%%%%%%%%%%%%%%%%%%%%%%%%%%%%%%%%%%%%%%%%%%%%%%%%%%%%%%%%%%%%%
\subsection{32-dimensional gamma matrices}

We choose the following representation for the ten 32-dimensional gamma matrices appearing in the Green Schwarz action:
\[
\Gamma^0 &= -i \sigma_1 \otimes \sigma_3 \otimes \identity \otimes \identity \otimes \identity~, & \qquad
\Gamma^1 &= \sigma_1 \otimes \sigma_2 \otimes \identity \otimes \identity \otimes \identity~, \\
\Gamma^2 &= -\sigma_2 \otimes \identity \otimes \sigma_3 \otimes \identity \otimes \identity~, & \qquad
\Gamma^3 &= -\sigma_2 \otimes \identity \otimes \sigma_1 \otimes \identity \otimes \identity~, \\
\Gamma^4 &= \sigma_2 \otimes \identity \otimes \sigma_2 \otimes \identity \otimes \sigma_1~, & \qquad
\Gamma^5 &= \sigma_1 \otimes \sigma_1 \otimes \identity \otimes \sigma_1 \otimes \identity~, \\
\Gamma^6 &= \sigma_2 \otimes \identity \otimes \sigma_2 \otimes \identity \otimes \sigma_2~, & \qquad
\Gamma^7 &= \sigma_1 \otimes \sigma_1 \otimes \identity \otimes \sigma_2 \otimes \identity~, \\
\Gamma^8 &= \sigma_2 \otimes \identity \otimes \sigma_2 \otimes \identity \otimes \sigma_3~, & \qquad
\Gamma^9 &= \sigma_1 \otimes \sigma_1 \otimes \identity \otimes \sigma_3 \otimes \identity ~.
\]
They satisfy the Clifford algebra in $1+9$ dimensions, $\anticom{\Gamma^\ind{A}}{\Gamma^\ind{B}}=2\eta^\ind{AB}$, and are related to the 4-dimensional gamma matrices by
\[
\Gamma^a = \sigma_1 \otimes \gamma^a \otimes \identity_4 ~, \quad a=0,1 ~,
\qquad \Gamma^a = -i \sigma_2 \otimes \gamma^a \otimes \identity_4~,\quad a=2,3~.
\]
Furthermore,
\[
\Gamma^{11} = \Gamma^{0} \Gamma^{1} \Gamma^{2} \Gamma^{3} \Gamma^{4} \Gamma^{5} \Gamma^{6} \Gamma^{7} \Gamma^{8} \Gamma^{9} = \begin{pmatrix} -\identity & 0 \\ 0 & \identity \end{pmatrix}~.
\]
Dirac conjugation acts on 32-component spinors as $\bar{\Theta} = \Theta^\dagger \Gamma^0$ and the Majorana condition is
\[
\bar{\Theta}= \Theta^t \mathcal C ~,\]
where the charge conjugation is defined as
\[
\mathcal{C} = i \sigma_1 \otimes \sigma_3 \otimes \identity \otimes \sigma_2 \otimes \sigma_2~, \qquad
\mathcal C^2 =-\identity_{32}~, \qquad (\mathcal{C} \Gamma^a)^t = \mathcal C \Gamma^a~.
\]

%%%%%%%%%%%%%%%%%%%%%%%%%%%%%%%%%%%%%%%%%%%%%%%%%%%%%%%%%%%%%%%%%%%%%%%%%%%%%%%%
\section{Field redefinitions}
\label{app:field_redef}

In this appendix we present the field redefinitions that bring the Poisson-Lie dual of the $\eta$-deformed $\AdS_2 \times \Sp^2$ supercoset with respect to the full bosonic subalgebra $\alg{su}(1,1) \dsum \alg{su}(2)$ to Green-Schwarz form.
Following the same notation as in \namedref{subsec}{ssec:RR} we split the Lagrangian into four distinct parts, $\Lag = \Lag^0 + \Lag^\partial + \Lag^m + \Lag^{\partial \partial}$.
The terms quadratic in the fermions can also be split according to their symmetry properties under the exchange of the two fermions, $\Lag^\partial = \Lag^\partial_+ + \Lag^\partial_-$.
To match the form of the Green-Schwarz Lagrangian we first show that $\Lag^{\partial \partial}$ is a total derivative and thus can be ignored.
We then identify $\Lag_\pm^\partial$ and find the appropriate shift of the bosons cancelling the terms with the wrong symmetry property, that is so that $\Lag_+^\partial+(\delta \Lag^0)^\partial_+ =0$.
Finally, we rewrite the remaining terms, $\Lag_-^\partial + (\delta \Lag^0)^\partial_-$, in Green-Schwarz form with a rotation of the fermions.

\paragraph{Contribution to $\Lag^{\partial \partial}$.}
Parametrising the field of the deformed supercoset sigma model as in eq.~\eqref{eq:param_gen}
\[
k=\tilde{g}_0 e^\chi ~, \qquad \tilde{g}_0 \in \tilde{\grp{G}}_0 ~, \qquad \chi \equiv \theta^{\ind{I} \check{\alpha} \hat{\alpha}} Q_{\ind{I} \check{\alpha} \hat{\alpha}} ~,
\]
we expand the action \eqref{eq:actionS1} to quadratic order in $\chi$ and find
\[\label{eq:lpartialpartial}
\Lag^{\partial \partial} & = -\frac{T}{\kappa} \< \partial_+ \chi, \proj_{\tilde{\alg{m}}} \En^{-1} \proj_{\alg{m}}\partial_- \chi\>~.
\]
On an element of the Grassmann envelope the operator $\proj_{\tilde{\alg{m}}} \En^{-1} \proj_{\alg{m}}$ acts as
\[
\proj_{\tilde{\alg{m}}} \En^{-1} \proj_{\alg{m}}(\theta^\ind{I} Q_\ind{I}) = \frac{i}{2\eta} \theta^\ind{I} ( \eta \, \delta^\ind{IJ} + \sigma_1^\ind{IJ} )\gamma^0 \tilde{Q}^\ind{J}~,
\]
and is thus antisymmetric with respect to the inner product.
Moreover, since it does not depend on the bosons, $\Lag^{\partial\partial}$ can be rewritten
\[
\Lag^{\partial \partial} = - \frac{T}{2 \kappa} \epsilon^{\alpha \beta} \partial_\alpha \< \chi, \proj_{\tilde{\alg{m}}} \En^{-1} \proj_{\alg{m}}\partial_\beta\chi\>~,
\]
where $\alpha, \beta = +,-$ and $\epsilon^{+-} = - \epsilon^{-+} =1$, thus showing that $\Lag^{\partial \partial}$ is a total derivative.

\paragraph{Contribution to $\Lag^\partial$.}
The terms in the Lagrangian containing one derivative acting on the fermions are
\[\label{eq:lpartial}
\Lag^\partial  = &- \frac{T}{\kappa} \Big(  \< \tilde{g}_0^{-1} \partial_+ \tilde{g}_0, \proj_{\alg{g}_0} ( \En^{-1} + \Pi(\tilde{g}_0) )^{-1} ( \proj_{\tilde{\alg{g}}_0} + \proj_{\tilde{\alg{g}}_0} \En^{-1} \proj_{\alg{g}_0} ) \ad_\chi ( \proj_{\tilde{\alg{m}}} \En^{-1} \proj_{\alg{m}} - \half \proj_{\alg{m}} )\partial_- \chi\>
\\
&+ \< \partial_+ \chi , (\proj_{\tilde{\alg{m}}} \En^{-1} \proj_{\alg{m}} + \half \proj_{\tilde{\alg{m}}} ) \ad_\chi (\proj_{\alg{g}_0} - \proj_{\tilde{\alg{g}}_0} \En^{-1} \proj_{\alg{g}_0} )( \En^{-1} + \Pi(\tilde{g}_0) )^{-1} \proj_{\tilde{\alg{g}}_0} \tilde{g}_0^{-1} \partial_- \tilde{g}_0\> \Big)~.
\]
To bring this expression closer to Green-Schwarz form we rewrite it in terms of the fermions $\theta^\ind{I}$.
We start by expanding the bosonic currents
\unskip\footnote{In this appendix we use $T_\ind{A}$ for the bosonic generators of $\alg{psu}(1,1|2)$.
The index $A$ takes the values $A=a, 01, 23$, with $T_a = P_a$, $T_{01}=J_{01}$ and $T_{23}=J_{23}$.
Similarly, $\tilde{T}^\ind{A}$ are the bosonic generators of $\alg{pb}(1,1|2)$, with $\tilde{T}^a = \tilde{P}^a$, $\tilde{T}^{01}=\tilde{J}^{01}$, $\tilde{T}^{23} = \tilde{J}^{23}$.}
\[
\label{eq:bosonic_current}
\tilde{g}_0^{-1} \partial_\alpha \tilde{g}_0 = e_{\alpha a} \tilde{P}^a + e_{\alpha 01} \tilde{J}^{01} + e_{\alpha 23} \tilde{J}^{23} \equiv e_{\alpha \ind{A}} \tilde{T}^\ind{A} ~,
\]
where we have introduced $e_{\alpha \ind{A}} = e_\ind{MA} \partial_\alpha X^\ind{M}$ and $e_{\alpha \ind{AB}} = e_\ind{MAB} \partial_\alpha X^\ind{M}$.
The bosonic coordinates $X^\ind{M}$ parametrise the group element $\tilde{g}_0$.
We also define the operator
\[
F &= \proj_{\alg{g}_0} ( \En^{-1} + \Pi(\tilde{g}_0) )^{-1} \proj_{\tilde{\alg{g}}_0}~, \qquad F(\tilde{T}^\ind{A}) = F^\ind{BA} T_\ind{B}~,
\]
which can be written as $F=G+B$, where $G$ is the symmetric and $B$ the antisymmetric part of $F$ with respect of the inner product.
Then using the action of the operator $\En^{-1}: \alg{g} \rightarrow \tilde{\alg{g}}$ on the Grassmann envelope
\unskip\footnote{The indices $a=0,1,2,3$ are raised and lowered with $\eta_{ab} = \diag(-1,+1,+1,+1)_{ab}$.}
\[
\En^{-1}(J_{ab}) = 0~, \qquad \En^{-1}(P_{a}) = - \frac{4 \eta}{1-\eta^2} \tilde{P}_a~, \qquad \En^{-1} (\theta^\ind{I} Q_\ind{I}) = \frac{i}{2\eta} \theta^\ind{I} ( \eta \, \delta^\ind{IJ} + \sigma_1^\ind{IJ} )\gamma^0 \tilde{Q}^\ind{J}~,
\]
together with the commutation relations between elements of $\alg{psu}(1,1|2)$ and $\alg{pb}(1,1|2)$ given in \appref{app:sl_generators}, the part of the Lagrangian given in eq.~\eqref{eq:lpartial} can be written in the form
\[
\label{eq:Lag_partial}
\Lag^\partial &= \theta^\ind{I} \big( \sigma_1^{\alpha \beta} f^\ind{IJ}_{\alpha , -} + \epsilon^{\alpha \beta} g^\ind{IJ}_{\alpha ,-} + \sigma_1^{\alpha \beta} f^\ind{IJ}_{\alpha,+} + \epsilon^{\alpha \beta} g^\ind{IJ}_{\alpha,+} \big) \partial_\beta \theta^\ind{J} \\
&\equiv \Lag_-^{\partial, \sigma_1} + \Lag_-^{\partial, \epsilon} + \Lag_+^{\partial,\sigma_1} + \Lag_+^{\partial,\epsilon} ~,
\]
where $f^\ind{IJ}_{\alpha , \pm} = f^\ind{IJ}_{\ind{M} , \pm} \partial_\alpha X^\ind{M}$, $g^\ind{IJ}_{\alpha , \pm} = g^\ind{IJ}_{\ind{M} , \pm} \partial_\alpha X^\ind{M}$  and
\[
f^\ind{IJ}_{\alpha,-} &= -i\frac{T}{2} e_{\alpha \ind{A}} \gamma^0 \Big( - \sigma_3^\ind{IJ} G^{\ind{A} a} + \big(\frac{1+\eta^2}{1-\eta^2} \delta^\ind{IJ} - \frac{2 \eta}{1-\eta^2} \sigma_1^\ind{IJ}\big) B^{\ind{A} a} \Big) \gamma_a~, \\
g^\ind{IJ}_{\alpha,-} &= -i\frac{T}{2}  e_{\alpha \ind{A}} \gamma^0 \Big( - \sigma_3^\ind{IJ} B^{\ind{A} a} + \big(\frac{1+\eta^2}{1-\eta^2} \delta^\ind{IJ} - \frac{2 \eta}{1-\eta^2} \sigma_1^\ind{IJ}\big) G^{\ind{A} a} \Big) \gamma_a~, \\
f^\ind{IJ}_{\alpha,+} &= -i\frac{T}{2}  e_{\alpha \ind{A}} \gamma^0 \bigg( \epsilon^\ind{IJ} \eta G^{\ind{A} a} \gamma_a \\
&\qquad +i \big( \eta \delta^\ind{IJ} - \sigma_1^\ind{IJ} \big) \Big( \big(G^{\ind{A} 01} - \frac{2 \eta}{1-\eta^2} B^{\ind{A} 1} \big) \gamma_{01} + \big( G^{\ind{A} 23} + \frac{2 \eta}{1-\eta^2} B^{\ind{A} 3 } \big) \gamma_{23} \Big)
\bigg)~, \\
g^\ind{IJ}_{\alpha,+} &= -i\frac{T}{2}  e_{\alpha \ind{A}} \gamma^0 \bigg( \epsilon^\ind{IJ} \eta B^{\ind{A} a} \gamma_a \\
&\qquad +i \big( \eta \delta^\ind{IJ} - \sigma_1^\ind{IJ} \big) \Big( \big(B^{\ind{A} 01} - \frac{2 \eta}{1-\eta^2} G^{\ind{A} 1} \big) \gamma_{01} + \big( B^{\ind{A} 23} + \frac{2 \eta}{1-\eta^2} G^{\ind{A} 3 } \big) \gamma_{23} \Big)
\bigg)~.
\]
The field redefinitions should then be chosen such that $\Lag_+^{\partial,\sigma_1}+\Lag_+^{\partial,\epsilon} + (\delta \Lag^0)^\partial_+=0$ and $\Lag_-^{\partial,\sigma_1}+\Lag_-^{\partial,\epsilon} + (\delta \Lag^0)^\partial_-=\hat{\Lag}^\partial_{GS}$ up to total derivatives.
The first condition can be satisfied with a shift of the bosons, while the second requires a rotation of the fermions.

\paragraph{Shift of the bosons.}
Under a shift of the bosons
\[
\label{eq:field_redef_B}
X^\ind{M} \rightarrow X^\ind{M} - \frac{1}{2} G^\ind{MN} \theta^\ind{I} f^\ind{IJ}_{ \ind{N},+} \theta^\ind{J} ~ , \qquad M=0,1,2,3 ~,
\]
where $G^\ind{MN}$ is the inverse of the metric of the $\lambda^\star$-deformed model on $\AdS^2 \times \Sp^2$, new terms with one derivative acting on the fermions and new ``mass'' terms arise from the bosonic Lagrangian
\[
\Lag^0 \rightarrow \Lag^0 + (\delta \Lag^0)^{\partial,\sigma_1}_+ + (\delta \Lag^0)^m_+~.
\]
In the particular example we are interested in, namely the $\lambda^\star$-deformed model on $\AdS_2 \times \Sp^2$, the antisymmetric B-field vanishes and these additional terms are given by
\[
(\delta \Lag^0)^{\partial,\sigma_1}_+ = &- \sigma_1^{\alpha \beta} \theta^\ind{I} f^\ind{IJ}_{\alpha,+}\partial_\beta \theta^\ind{J}~, \\
(\delta \Lag^0)^m_+ = & - \frac{1}{2} \sigma_1^{\alpha \beta} \partial_\alpha X^\ind{M} \theta^\ind{I} \Big(G_\ind{MN} \partial_\beta( G^\ind{NQ} f^\ind{IJ}_{\ind{Q},+}) + \frac{1}{2} \partial_\ind{P} G_\ind{MN} \partial_\beta X^\ind{N} G^\ind{PQ} f^\ind{IJ}_{\ind{Q}, +}\Big) \theta^\ind{J} ~.
\]
The shift \eqref{eq:field_redef_B} thus guarantees
\[
\Lag_+^{\partial,\sigma_1} + (\delta \Lag^0)^{\partial, \sigma_1}_+ =0~.
\]
It is not sufficient to cancel $\Lag^{\partial,\epsilon}_+$, however using integration by parts one can rewrite these terms such that the derivatives act only on the bosons, giving new ``mass'' terms.
Explicitly, one has
\[
\Lag_+^{\partial,\epsilon} = \epsilon^{\alpha \beta} \theta^\ind{I} g^\ind{IJ}_{\alpha, +} \partial_\beta \theta^\ind{J}
& = \frac{1}{2} \partial_\beta \big( \epsilon^{\alpha \beta} \theta^\ind{I} g^\ind{IJ}_{\alpha, +} \theta^\ind{J} \big) - \frac{1}{2} \epsilon^{\alpha \beta} \theta^\ind{I} \partial_\beta g^\ind{IJ}_{\alpha, +} \theta^\ind{J}
\\ & = \text{total derivative} + (\delta \Lag^\partial)_+^m~.
\]
Finally, after performing the field redefinition \eqref{eq:field_redef_B} the only surviving terms are
\[
\label{eq:Lag_partial_m}
\Lag^\partial &= \Lag_-^\partial = \Lag^{\partial, \sigma_1}_- + \Lag^{\partial, \epsilon}_-~,
\]
with
\[
\Lag^{\partial, \sigma_1}_- &= -i\frac{T}{2} \sigma_1^{\alpha \beta} e_{\alpha \ind{A}} \theta^\ind{I} \gamma^0 \Big( - \sigma_3^\ind{IJ} G^{\ind{A} a} + \big(\frac{1+\eta^2}{1-\eta^2} \delta^\ind{IJ} - \frac{2 \eta}{1-\eta^2} \sigma_1^\ind{IJ}\big) B^{\ind{A} a} \Big) \gamma_a \, \partial_\beta \theta^\ind{J} ~, \\
\Lag^{\partial, \epsilon}_- &= -i\frac{T}{2}  \epsilon^{\alpha \beta} e_{\alpha \ind{A}} \theta^\ind{I} \gamma^0 \Big( - \sigma_3^\ind{IJ} B^{\ind{A} a} + \big(\frac{1+\eta^2}{1-\eta^2} \delta^\ind{IJ} - \frac{2 \eta}{1-\eta^2} \sigma_1^\ind{IJ}\big) G^{\ind{A} a}\Big) \gamma_a \, \partial_\beta \theta^\ind{J}~, \\
\]
and one should add $(\delta \Lag^0)_+^m + (\delta \Lag^\partial)_+^m$ to the ``mass'' terms.

\paragraph{Rotation of the fermions.}
The final task is then to bring the remaining terms in $\Lag^\partial$ to Green-Schwarz form
\[
\hat{\Lag}_{GS}^\partial =  i \sqrt{\frac{T}{2}}  b_\ind{(I)} b_\ind{(J)} \hat{\theta}^\ind{I} \gamma^0 \big( \delta^\ind{IJ} \sigma_1^{\alpha \beta} + \sigma_3^\ind{IJ} \epsilon^{\alpha \beta} \big) E_{\alpha a} \gamma^a \partial_\beta\hat{\theta}^\ind{J}
~,\]
where we have used the Majorana condition \eqref{eq:MWcondition_theta}.
To proceed we compare the terms proportional to $\sigma_1^{\alpha \beta}$ and $\epsilon^{\alpha \beta}$.
Adding and subtracting the two resulting equations leads to
\[
\sqrt{\frac{T}{2}} e_{\alpha \ind{A}} (G+B)^{\ind{A} a} \theta^\ind{I} \gamma^0 \Big( \sigma_3^\ind{IJ} - \frac{1+\eta^2}{1-\eta^2} \delta^\ind{IJ} + \frac{2 \eta}{1-\eta^2} \sigma_1^\ind{IJ} \Big) \gamma_a \, \partial_\beta \theta^\ind{J} &= 2 (b_\ind{(1)})^2 E_{\alpha a} \hat{\theta}^1 \gamma^0 \gamma^a \partial_\beta\hat{\theta}^1~,\\
\sqrt{\frac{T}{2}} e_{\alpha \ind{A}} (G-B)^{\ind{A} a} \theta^\ind{I} \gamma^0 \Big( \sigma_3^\ind{IJ} + \frac{1+\eta^2}{1-\eta^2} \delta^\ind{IJ} - \frac{2 \eta}{1-\eta^2} \sigma_1^\ind{IJ} \Big) \gamma_a \, \partial_\beta \theta^\ind{J} &= 2 (b_\ind{(2)})^2 E_{\alpha a} \hat{\theta}^2 \gamma^0 \gamma^a \partial_\beta\hat{\theta}^2~,
\]
where we note that only $\hat{\theta}^1$ or $\hat{\theta}^2$ appear on the right-hand side.
To match this structure on the left-hand side we perform the following rotation of the fermions
\[
\label{eq:rotation}
\theta^\ind{I} = U^\ind{IJ} \hat{\theta}^\ind{J}~, \qquad
U^\ind{IJ} =  \frac{1}{\sqrt{1-\eta^2}} (\eta \delta^\ind{IJ} + \sigma_1^\ind{IJ})U_\ind{(J)} ~,
\]
where $U_\ind{(J)}, \, J=1,2,$ are two $4 \times 4$ matrices that are to be determined.
This redefinition may lead to new ``mass'' terms $\sim \hat{\theta}^\ind{K} U^\ind{IK} \partial_\beta U^\ind{JL} \hat{\theta}^\ind{L}$ as the rotation matrices may in principle depend on the bosons.
We will see that this is not the case here: the two rotation matrices $U_\ind{(J)}$ do not depend on the bosons.
Defining $\bar{U}_\ind{(I)} = \gamma^0 U_\ind{(I)}^t \gamma^0$ and implementing the transformation \eqref{eq:rotation} the equations we are left to solve are
\[\label{eq:ets}
\sqrt{\frac{T}{2}} \frac{1}{(b_\ind{(1)})^2} e_{\alpha \ind{A}} (G+B)^{\ind{A} a} \bar{U}_\ind{(1)} \gamma_a U_\ind{(1)} &= E_{\alpha a} \gamma^a ~, \\
\sqrt{\frac{T}{2}} \frac{1}{(b_\ind{(2)})^2} e_{\alpha \ind{A}} (G-B)^{\ind{A} a} \bar{U}_\ind{(2)} \gamma_a U_\ind{(2)} &= - E_{\alpha a} \gamma^a ~.
\]
Let us now choose the specific parametrisation of $\tilde{g}_0$ in eq.~\eqref{eq:paramg0} together with the coordinate redefinition \eqref{eq:coord_redef}.
Enumerating the bosonic coordinates as $X^0 = \check{p}$, $X^1 = \check{q}$, $X^2 = \hat{p}$ and $X^3=\hat{q}$, we find
\[
e_{\check{p} 0} &= - \frac{1}{\sqrt{\check{p}^2+\check{q}^2 -1}} ~, &\qquad e_{\hat{p} 2} &= \frac{1}{\sqrt{\hat{p}^2-\hat{q}^2 -1}} ~,\\
e_{\check{q} 0} &= \frac{1}{\sqrt{\check{p}^2 + \check{q}^2 -1}} \frac{\check{q}}{ \sqrt{\check{p}^2 + \check{q}^2 -1}-\check{p}}~, &\qquad e_{\hat{q} 2} &= \frac{1}{\sqrt{\hat{p}^2-\hat{q}^2 -1}} \frac{\hat{q}}{\sqrt{\hat{p}^2-\hat{q}^2 -1} -\hat{p}} ~, \\
e_{\check{q} 01} &= \frac{1}{ \sqrt{\check{p}^2 + \check{q}^2 -1}-\check{p}}~, &\qquad e_{\hat{q} 23} &= \frac{1}{\sqrt{\hat{p}^2-\hat{q}^2 -1} -\hat{p}} ~,
\]
while the actions of the operator $F=G+B$ and its transpose $F^t=G-B$ on $\tilde{P}^a$ are
\[
(G \pm B)(\tilde{P}^0) &= \frac{1}{2 \kappa} P_0 - \frac{1}{2 \kappa} \frac{\check{q}}{\sqrt{\check{p}^2 + \check{q}^2-1}} J_{01}~,\\
(G \pm B )(\tilde{P}^1) &= \pm \frac{1}{2} \big( \frac{\check{p}}{\sqrt{\check{p}^2 + \check{q}^2-1}} -1 \big)J_{01}~, \\
(G \pm B)(\tilde{P}^2) &= - \frac{1}{2 \kappa} P_2 + \frac{1}{2 \kappa} \frac{\hat{q}}{\sqrt{\hat{p}^2 - \hat{q}^2-1}} J_{23}~, \\
(G \pm B )(\tilde{P}^3) &= \pm \frac{1}{2} \big( 1- \frac{\hat{p}}{\sqrt{\hat{p}^2 - \hat{q}^2-1}} \big)J_{23}~.
\]
From this it follows that
\[
e_{\alpha \ind{A}} (G+B)^{\ind{A} a} = \frac{1}{\sqrt{2T}}  (\sigma_3 \otimes \sigma_3 )^{a}{}_b \eta^{bc} E_{\alpha c} ~, \quad e_{\alpha \ind{A}} (G-B)^{\ind{A} a} =\frac{1}{\sqrt{2T}}   (\sigma_3 \otimes \identity)^{a}{}_b \eta^{bc} E_{\alpha c} ~,
\]
and eqs.~\eqref{eq:ets} then take the form
\[
\frac{1}{2(b_\ind{(1)})^2} \, \bar{U}_\ind{(1)} (\sigma_3 \otimes \sigma_3)^{a}{}_b \gamma_a U_\ind{(1)} = + \gamma_b ~, \qquad
\frac{1}{2 (b_\ind{(2)})^2} \, \bar{U}_\ind{(2)} (\sigma_3 \otimes \identity)^{a}{}_b \gamma_a U_\ind{(2)} = - \gamma_b ~.
\]
These equations do not admit solutions for arbitrary coefficients $b_\ind{(I)}$.
Working with the choice \eqref{eq:choice_b} we find
\[
\bar{U}_\ind{(1)} \gamma_a U_\ind{(1)} &= - \gamma_a~, \quad a=0,3~, \qquad &\bar{U}_\ind{(1)} \gamma_a U_\ind{(1)}&= \gamma_a~, \quad a=1,2~, \\
\bar{U}_\ind{(2)} \gamma_a U_\ind{(2)} &= - \gamma_a~, \quad a=0,1~, \qquad &\bar{U}_\ind{(2)} \gamma_a U_\ind{(2)}&= \gamma_a~, \quad a=2,3~,
\]
for which a solution is given by
\[
U_\ind{(1)} = \gamma^0 \gamma^3 ~, \qquad U_\ind{(2)} = \gamma^0 \gamma^1~.
\]
Therefore, the field redefinition \eqref{eq:rotation} does not depend on the bosons and there are no additional ``mass'' terms coming from the rotation of the fermions.

%%%%%%%%%%%%%%%%%%%%%%%%%%%%%%%%%%%%%%%%%%%%%%%%%%%%%%%%%%%%%%%%%%%%%%%%%%%%%%%%
\begin{bibtex}[\jobname]

@article{Hoare:2017ukq,
author         = "Hoare, Ben and Seibold, Fiona K.",
title          = "{Poisson-Lie duals of the $\eta$ deformed symmetric space sigma model}",
journal        = "JHEP",
volume         = "11",
year           = "2017",
pages          = "014",
doi            = "10.1007/JHEP11(2017)014",
eprint         = "1709.01448",
archivePrefix  = "arXiv",
primaryClass   = "hep-th",
SLACcitation   = "%%CITATION = ARXIV:1709.01448;%%"
}

@article{Klimcik:1995dy,
author         = "Klim\v{c}\'{i}k, C. and \v{S}evera, P.",
title          = "{Poisson-Lie T-duality and loop groups of Drinfel'd doubles}",
journal        = "Phys. Lett.",
volume         = "B372",
year           = "1996",
pages          = "65-71",
doi            = "10.1016/0370-2693(96)00025-1",
eprint         = "hep-th/9512040",
archivePrefix  = "arXiv",
primaryClass   = "hep-th",
reportNumber   = "CERN-TH-95-330",
SLACcitation   = "%%CITATION = HEP-TH/9512040;%%"
}

@article{Klimcik:1996nq,
author         = "Klim\v{c}\'{i}k, C. and \v{S}evera, P.",
title          = "{Non-abelian momentum-winding exchange}",
journal        = "Phys. Lett.",
volume         = "B383",
year           = "1996",
pages          = "281-286",
doi            = "10.1016/0370-2693(96)00755-1",
eprint         = "hep-th/9605212",
archivePrefix  = "arXiv",
primaryClass   = "hep-th",
reportNumber   = "CERN-TH-96-142",
SLACcitation   = "%%CITATION = HEP-TH/9605212;%%"
}

@article{Arutyunov:2015qva,
author         = "Arutyunov, Gleb and Borsato, Riccardo and Frolov, Sergey",
title          = "{Puzzles of $\eta$-deformed $AdS_5 \times S^5$}",
journal        = "JHEP",
volume         = "12",
year           = "2015",
pages          = "049",
doi            = "10.1007/JHEP12(2015)049",
eprint         = "1507.04239",
archivePrefix  = "arXiv",
primaryClass   = "hep-th",
reportNumber   = "ITP-UU-15-10, TCD-MATH-15-05, ZMP-HH-15-19",
SLACcitation   = "%%CITATION = ARXIV:1507.04239;%%"
}

%journal
@article{Araujo:2018rbc,
author         = "Araujo, Thiago and Colg\'{a}in, Eoin \'{O}. and Yavartanoo, Hossein",
title          = "{Embedding the modified CYBE in supergravity}",
year           = "2018",
eprint         = "1806.02602",
archivePrefix  = "arXiv",
primaryClass   = "hep-th",
reportNumber   = "APCTP Pre2018-004, APCTP-PRE2018-004",
SLACcitation   = "%%CITATION = ARXIV:1806.02602;%%"
}

@article{Borsato:2016zcf,
author         = "Borsato, R. and Tseytlin, A. A. and Wulff, L.",
title          = "{Supergravity background of $\lambda$-deformed model for $AdS_2 \times S^2$ supercoset}",
journal        = "Nucl. Phys.",
volume         = "B905",
year           = "2016",
pages          = "264-292",
doi            = "10.1016/j.nuclphysb.2016.02.018",
eprint         = "1601.08192",
archivePrefix  = "arXiv",
primaryClass   = "hep-th",
reportNumber   = "IMPERIAL-TP-RB-2016-01",
SLACcitation   = "%%CITATION = ARXIV:1601.08192;%%"
}

@article{Sfetsos:2014cea,
author         = "Sfetsos, Konstantinos and Thompson, Daniel C.",
title          = "{Spacetimes for $\lambda$-deformations}",
journal        = "JHEP",
volume         = "12",
year           = "2014",
pages          = "164",
doi            = "10.1007/JHEP12(2014)164",
eprint         = "1410.1886",
archivePrefix  = "arXiv",
primaryClass   = "hep-th",
SLACcitation   = "%%CITATION = ARXIV:1410.1886;%%"
}

@article{Hoare:2015gda,
author         = "Hoare, B. and Tseytlin, A. A.",
title          = "{On integrable deformations of superstring sigma models related to $AdS_n \times S^n$ supercosets}",
journal        = "Nucl. Phys.",
volume         = "B897",
year           = "2015",
pages          = "448-478",
doi            = "10.1016/j.nuclphysb.2015.06.001",
eprint         = "1504.07213",
archivePrefix  = "arXiv",
primaryClass   = "hep-th",
reportNumber   = "IMPERIAL-TP-AT-2015-02, HU-EP-15-21",
SLACcitation   = "%%CITATION = ARXIV:1504.07213;%%"
}

@article{Klimcik:2002zj,
author         = "Klim\v{c}\'{i}k, Ctirad",
title          = "{Yang-Baxter $\sigma$-models and dS/AdS T-duality}",
journal        = "JHEP",
volume         = "12",
year           = "2002",
pages          = "051",
doi            = "10.1088/1126-6708/2002/12/051",
eprint         = "hep-th/0210095",
archivePrefix  = "arXiv",
primaryClass   = "hep-th",
reportNumber   = "IML-02-XY",
SLACcitation   = "%%CITATION = HEP-TH/0210095;%%"
}

@article{Klimcik:2008eq,
author         = "Klim\v{c}\'{i}k, Ctirad",
title          = "{On integrability of the Yang-Baxter $\sigma$-model}",
journal        = "J. Math. Phys.",
volume         = "50",
year           = "2009",
pages          = "043508",
doi            = "10.1063/1.3116242",
eprint         = "0802.3518",
archivePrefix  = "arXiv",
primaryClass   = "hep-th",
SLACcitation   = "%%CITATION = ARXIV:0802.3518;%%"
}

@article{Delduc:2013fga,
author         = "Delduc, Francois and Magro, Marc and Vicedo, Benoit",
title          = "{On classical $q$-deformations of integrable $\sigma$-models}",
journal        = "JHEP",
volume         = "11",
year           = "2013",
pages          = "192",
doi            = "10.1007/JHEP11(2013)192",
eprint         = "1308.3581",
archivePrefix  = "arXiv",
primaryClass   = "hep-th",
SLACcitation   = "%%CITATION = ARXIV:1308.3581;%%"
}

@article{Eichenherr:1979ci,
author         = "Eichenherr, H. and Forger, M.",
title          = "{On the dual symmetry of the non-linear sigma models}",
journal        = "Nucl. Phys.",
volume         = "B155",
year           = "1979",
pages          = "381-393",
doi            = "10.1016/0550-3213(79)90276-1",
reportNumber   = "FREIBURG-THEP 79/2a",
SLACcitation   = "%%CITATION = NUPHA,B155,381;%%"
}

@article{Eichenherr:1979hz,
author         = "Eichenherr, H. and Forger, M.",
title          = "{More about non-linear sigma models on symmetric spaces}",
journal        = "Nucl. Phys.",
volume         = "B164",
year           = "1980",
pages          = "528-535",
doi            = "10.1016/0550-3213(87)90706-1,10.1016/0550-3213(80)90525-8",
note           = "[Erratum: Nucl. Phys.B282,745(1987)]",
reportNumber   = "FREIBURG-THEP-79-7",
SLACcitation   = "%%CITATION = NUPHA,B164,528;%%"
}

@article{Metsaev:1998it,
author         = "Metsaev, R. R. and Tseytlin, Arkady A.",
title          = "{Type IIB superstring action in $AdS_5 \times S^5$ background}",
journal        = "Nucl. Phys.",
volume         = "B533",
year           = "1998",
pages          = "109-126",
doi            = "10.1016/S0550-3213(98)00570-7",
eprint         = "hep-th/9805028",
archivePrefix  = "arXiv",
primaryClass   = "hep-th",
reportNumber   = "FIAN-TD-98-21, IMPERIAL-TP-97-98-44, NSF-ITP-98-055",
SLACcitation   = "%%CITATION = HEP-TH/9805028;%%"
}

@article{Zhou:1999sm,
author         = "Zhou, Jian-Ge",
title          = "{Super 0-brane and GS superstring actions on $AdS_2 \times S^2$}",
journal        = "Nucl. Phys.",
volume         = "B559",
year           = "1999",
pages          = "92-102",
doi            = "10.1016/S0550-3213(99)00462-9",
eprint         = "hep-th/9906013",
archivePrefix  = "arXiv",
primaryClass   = "hep-th",
SLACcitation   = "%%CITATION = HEP-TH/9906013;%%"
}

@article{Berkovits:1999zq,
author         = "Berkovits, N. and Bershadsky, M. and Hauer, T. and Zhukov, S. and Zwiebach, B.",
title          = "{Superstring theory on $AdS_2 \times S^2$ as a coset supermanifold}",
journal        = "Nucl. Phys.",
volume         = "B567",
year           = "2000",
pages          = "61-86",
doi            = "10.1016/S0550-3213(99)00683-5",
eprint         = "hep-th/9907200",
archivePrefix  = "arXiv",
primaryClass   = "hep-th",
reportNumber   = "IFT-P-060-99, HUTP-99-A044, MIT-CTP-2878, CTP-MIT-2878",
SLACcitation   = "%%CITATION = HEP-TH/9907200;%%"
}

@article{Zarembo:2010sg,
author         = "Zarembo, K.",
title          = "{Strings on semisymmetric superspaces}",
journal        = "JHEP",
volume         = "05",
year           = "2010",
pages          = "002",
doi            = "10.1007/JHEP05(2010)002",
eprint         = "1003.0465",
archivePrefix  = "arXiv",
primaryClass   = "hep-th",
reportNumber   = "ITEP-TH-12-10, LPTENS-10-12, UUITP-05-10",
SLACcitation   = "%%CITATION = ARXIV:1003.0465;%%"
}

@article{Delduc:2013qra,
author         = "Delduc, Francois and Magro, Marc and Vicedo, Benoit",
title          = "{An integrable deformation of the $AdS_5 \times S^5$ superstring action}",
journal        = "Phys. Rev. Lett.",
volume         = "112",
year           = "2014",
number         = "5",
pages          = "051601",
doi            = "10.1103/PhysRevLett.112.051601",
eprint         = "1309.5850",
archivePrefix  = "arXiv",
primaryClass   = "hep-th",
SLACcitation   = "%%CITATION = ARXIV:1309.5850;%%"
}

@article{Delduc:2014kha,
author         = "Delduc, Francois and Magro, Marc and Vicedo, Benoit",
title          = "{Derivation of the action and symmetries of the $q$-deformed $AdS_5 \times S^5$ superstring}",
journal        = "JHEP",
volume         = "10",
year           = "2014",
pages          = "132",
doi            = "10.1007/JHEP10(2014)132",
eprint         = "1406.6286",
archivePrefix  = "arXiv",
primaryClass   = "hep-th",
SLACcitation   = "%%CITATION = ARXIV:1406.6286;%%"
}

@article{Green:1983wt,
author         = "Green, Michael B. and Schwarz, John H.",
title          = "{Covariant description of superstrings}",
journal        = "Phys. Lett.",
volume         = "136B",
year           = "1984",
pages          = "367-370",
doi            = "10.1016/0370-2693(84)92021-5",
reportNumber   = "QMC-83-7",
SLACcitation   = "%%CITATION = PHLTA,136B,367;%%"
}

@article{Green:1983sg,
author         = "Green, Michael B. and Schwarz, John H.",
title          = "{Properties of the covariant formulation of superstring theories}",
journal        = "Nucl. Phys.",
volume         = "B243",
year           = "1984",
pages          = "285-306",
doi            = "10.1016/0550-3213(84)90030-0",
reportNumber   = "Print-84-0264 (QUEEN MARY COLL.)",
SLACcitation   = "%%CITATION = NUPHA,B243,285;%%"
}

@article{Witten:1985nt,
author         = "Witten, Edward",
title          = "{Twistor-like transform in ten dimensions}",
journal        = "Nucl. Phys.",
volume         = "B266",
year           = "1986",
pages          = "245-264",
doi            = "10.1016/0550-3213(86)90090-8",
reportNumber   = "PRINT-85-0458 (PRINCETON)",
SLACcitation   = "%%CITATION = NUPHA,B266,245;%%"
}

@article{Grisaru:1985fv,
author         = "Grisaru, Marcus T. and Howe, Paul S. and Mezincescu, L. and Nilsson, B. and Townsend, P. K.",
title          = "{$N=2$ superstrings in a supergravity background}",
journal        = "Phys. Lett.",
volume         = "162B",
year           = "1985",
pages          = "116-120",
doi            = "10.1016/0370-2693(85)91071-8",
reportNumber   = "Print-85-0603 (CAMBRIDGE)",
SLACcitation   = "%%CITATION = PHLTA,162B,116;%%"
}

@article{Sorokin:2011rr,
author         = "Sorokin, Dmitri and Tseytlin, Arkady and Wulff, Linus and Zarembo, Konstantin",
title          = "{Superstrings in $AdS_2 \times S^2 \times T^6$}",
journal        = "J. Phys.",
volume         = "A44",
year           = "2011",
pages          = "275401",
doi            = "10.1088/1751-8113/44/27/275401",
eprint         = "1104.1793",
archivePrefix  = "arXiv",
primaryClass   = "hep-th",
reportNumber   = "MIFPA-11-11, NORDITA-2011-30, IMPERIAL-TP-AT-2011-2",
SLACcitation   = "%%CITATION = ARXIV:1104.1793;%%"
}

@article{Klimcik:1996np,
author         = "Klim\v{c}\'{i}k, C. and \v{S}evera, P.",
title          = "{Dressing cosets}",
journal        = "Phys. Lett.",
volume         = "B381",
year           = "1996",
pages          = "56-61",
doi            = "10.1016/0370-2693(96)00669-7",
eprint         = "hep-th/9602162",
archivePrefix  = "arXiv",
primaryClass   = "hep-th",
reportNumber   = "CERN-TH-96-43",
SLACcitation   = "%%CITATION = HEP-TH/9602162;%%"
}

@article{Squellari:2011dg,
author         = "Squellari, R.",
title          = "{Dressing cosets revisited}",
journal        = "Nucl. Phys.",
volume         = "B853",
year           = "2011",
pages          = "379-403",
doi            = "10.1016/j.nuclphysb.2011.07.025",
eprint         = "1105.0162",
archivePrefix  = "arXiv",
primaryClass   = "hep-th",
SLACcitation   = "%%CITATION = ARXIV:1105.0162;%%"
}

@article{Sfetsos:1999zm,
author         = "Sfetsos, Konstadinos",
title          = "{Duality-invariant class of two-dimensional field theories}",
journal        = "Nucl. Phys.",
volume         = "B561",
year           = "1999",
pages          = "316-340",
doi            = "10.1016/S0550-3213(99)00485-X",
eprint         = "hep-th/9904188",
archivePrefix  = "arXiv",
primaryClass   = "hep-th",
reportNumber   = "CERN-TH-99-112",
SLACcitation   = "%%CITATION = HEP-TH/9904188;%%"
}

@article{Hoare:2016ibq,
author         = "Hoare, Ben and van Tongeren, Stijn J.",
title          = "{Non-split and split deformations of $AdS_5$}",
journal        = "J. Phys.",
volume         = "A49",
year           = "2016",
number         = "48",
pages          = "484003",
doi            = "10.1088/1751-8113/49/48/484003",
eprint         = "1605.03552",
archivePrefix  = "arXiv",
primaryClass   = "hep-th",
SLACcitation   = "%%CITATION = ARXIV:1605.03552;%%"
}

@article{Araujo:2017enj,
author         = "Araujo, T. and Colg\'{a}in, E. \'{O} and Sakamoto, J. and Sheikh-Jabbari, M. M. and Yoshida, K.",
title          = "{$I$ in generalized supergravity}",
journal        = "Eur. Phys. J.",
volume         = "C77",
year           = "2017",
number         = "11",
pages          = "739",
doi            = "10.1140/epjc/s10052-017-5316-5",
eprint         = "1708.03163",
archivePrefix  = "arXiv",
primaryClass   = "hep-th",
reportNumber   = "APCTP-PRE2017---015, KUNS-2696, IPM-P-2017-024,IPM-P-2017-028",
SLACcitation   = "%%CITATION = ARXIV:1708.03163;%%"
}

@article{Sfetsos:2013wia,
author         = "Sfetsos, Konstadinos",
title          = "{Integrable interpolations: from exact CFTs to non-abelian T-duals}",
journal        = "Nucl. Phys.",
volume         = "B880",
year           = "2014",
pages          = "225-246",
doi            = "10.1016/j.nuclphysb.2014.01.004",
eprint         = "1312.4560",
archivePrefix  = "arXiv",
primaryClass   = "hep-th",
reportNumber   = "DMUS-MP-13-23, DMUS--MP--13-23",
SLACcitation   = "%%CITATION = ARXIV:1312.4560;%%"
}

@article{Hollowood:2014rla,
author         = "Hollowood, Timothy J. and Miramontes, J. Luis and Schmidtt, David M.",
title          = "{Integrable deformations of strings on symmetric spaces}",
journal        = "JHEP",
volume         = "11",
year           = "2014",
pages          = "009",
doi            = "10.1007/JHEP11(2014)009",
eprint         = "1407.2840",
archivePrefix  = "arXiv",
primaryClass   = "hep-th",
SLACcitation   = "%%CITATION = ARXIV:1407.2840;%%"
}

@article{Hollowood:2014qma,
author         = "Hollowood, Timothy J. and Miramontes, J. Luis and Schmidtt, David M.",
title          = "{An integrable deformation of the $AdS_5 \times S^5$ superstring}",
journal        = "J. Phys.",
volume         = "A47",
year           = "2014",
number         = "49",
pages          = "495402",
doi            = "10.1088/1751-8113/47/49/495402",
eprint         = "1409.1538",
archivePrefix  = "arXiv",
primaryClass   = "hep-th",
SLACcitation   = "%%CITATION = ARXIV:1409.1538;%%"
}

@article{Vicedo:2015pna,
author         = "Vicedo, Benoit",
title          = "{Deformed integrable $\sigma$-models, classical R-matrices and classical exchange algebra on Drinfel'd doubles}",
journal        = "J. Phys.",
volume         = "A48",
year           = "2015",
number         = "35",
pages          = "355203",
doi            = "10.1088/1751-8113/48/35/355203",
eprint         = "1504.06303",
archivePrefix  = "arXiv",
primaryClass   = "hep-th",
SLACcitation   = "%%CITATION = ARXIV:1504.06303;%%"
}

@article{Sfetsos:2015nya,
author         = "Sfetsos, Konstantinos and Siampos, Konstantinos and Thompson, Daniel C.",
title          = "{Generalised integrable $\lambda$- and $\eta$-deformations and their relation}",
journal        = "Nucl. Phys.",
volume         = "B899",
year           = "2015",
pages          = "489-512",
doi            = "10.1016/j.nuclphysb.2015.08.015",
eprint         = "1506.05784",
archivePrefix  = "arXiv",
primaryClass   = "hep-th",
SLACcitation   = "%%CITATION = ARXIV:1506.05784;%%"
}

@article{Klimcik:2015gba,
author         = "Klim\v{c}\'{i}k, Ctirad",
title          = "{$\eta$ and $\lambda$ deformations as $\mathcal{E}$-models}",
journal        = "Nucl. Phys.",
volume         = "B900",
year           = "2015",
pages          = "259-272",
doi            = "10.1016/j.nuclphysb.2015.09.011",
eprint         = "1508.05832",
archivePrefix  = "arXiv",
primaryClass   = "hep-th",
SLACcitation   = "%%CITATION = ARXIV:1508.05832;%%"
}

@article{Arutyunov:2015mqj,
author         = "Arutyunov, G. and Frolov, S. and Hoare, B. and Roiban, R. and Tseytlin, A. A.",
title          = "{Scale invariance of the $\eta$-deformed $AdS_5 \times S^5$ superstring, T-duality and modified type II equations}",
journal        = "Nucl. Phys.",
volume         = "B903",
year           = "2016",
pages          = "262-303",
doi            = "10.1016/j.nuclphysb.2015.12.012",
eprint         = "1511.05795",
archivePrefix  = "arXiv",
primaryClass   = "hep-th",
reportNumber   = "ZMP-HH-15-27, TCDMATH-15-12, IMPERIAL-TP-AT-2015-08",
SLACcitation   = "%%CITATION = ARXIV:1511.05795;%%"
}

@article{Wulff:2016tju,
author         = "Wulff, L. and Tseytlin, A. A.",
title          = "{Kappa-symmetry of superstring sigma model and generalized 10d supergravity equations}",
journal        = "JHEP",
volume         = "06",
year           = "2016",
pages          = "174",
doi            = "10.1007/JHEP06(2016)174",
eprint         = "1605.04884",
archivePrefix  = "arXiv",
primaryClass   = "hep-th",
reportNumber   = "IMPERIAL-TP-LW-2016-02",
SLACcitation   = "%%CITATION = ARXIV:1605.04884;%%"
}

@article{Bakhmatov:2017joy,
author         = "Bakhmatov, I. and Kelekci, {\"O}. and Colg\'{a}in, E. \'{O}. and Sheikh-Jabbari, M. M.",
title          = "{Classical Yang-Baxter equation from supergravity}",
journal        = "Phys. Rev.",
volume         = "D98",
year           = "2018",
number         = "2",
pages          = "021901",
doi            = "10.1103/PhysRevD.98.021901",
eprint         = "1710.06784",
archivePrefix  = "arXiv",
primaryClass   = "hep-th",
reportNumber   = "APCTP-PRE2017-017",
SLACcitation   = "%%CITATION = ARXIV:1710.06784;%%"
}

@article{Hoare:2015wia,
author         = "Hoare, B. and Tseytlin, A. A.",
title          = "{Type IIB supergravity solution for the T-dual of the $\eta$-deformed $AdS_5 \times S^5$ superstring}",
journal        = "JHEP",
volume         = "10",
year           = "2015",
pages          = "060",
doi            = "10.1007/JHEP10(2015)060",
eprint         = "1508.01150",
archivePrefix  = "arXiv",
primaryClass   = "hep-th",
reportNumber   = "HU-EP-15-34, IMPERIAL-TP-AT-2015-05",
SLACcitation   = "%%CITATION = ARXIV:1508.01150;%%"
}

@article{Alvarez:1994np,
author         = "\'{A}lvarez, Enrique and \'{A}lvarez-Gaum\'{e}, Luis and Lozano, Yolanda",
title          = "{On non-abelian duality}",
journal        = "Nucl. Phys.",
volume         = "B424",
year           = "1994",
pages          = "155-183",
doi            = "10.1016/0550-3213(94)90093-0",
eprint         = "hep-th/9403155",
archivePrefix  = "arXiv",
primaryClass   = "hep-th",
reportNumber   = "CERN-TH-7204-94",
SLACcitation   = "%%CITATION = HEP-TH/9403155;%%"
}

@article{Elitzur:1994ri,
author         = "Elitzur, S. and Giveon, A. and Rabinovici, E. and Schwimmer, A. and Veneziano, G.",
title          = "{Remarks on non-abelian duality}",
journal        = "Nucl. Phys.",
volume         = "B435",
year           = "1995",
pages          = "147-171",
doi            = "10.1016/0550-3213(94)00426-F",
eprint         = "hep-th/9409011",
archivePrefix  = "arXiv",
primaryClass   = "hep-th",
reportNumber   = "CERN-TH-7414-94, RI-9-94, WIS-7-94",
SLACcitation   = "%%CITATION = HEP-TH/9409011;%%"
}

@article{Tyurin:1995bu,
author         = "Tyurin, Eugene and von Unge, Rikard",
title          = "{Poisson-Lie T-duality: the path-integral derivation}",
journal        = "Phys. Lett.",
volume         = "B382",
year           = "1996",
pages          = "233-240",
doi            = "10.1016/0370-2693(96)00680-6",
eprint         = "hep-th/9512025",
archivePrefix  = "arXiv",
primaryClass   = "hep-th",
reportNumber   = "ITP-SB-95-50, USITP-95-11",
SLACcitation   = "%%CITATION = HEP-TH/9512025;%%"
}

@article{Bossard:2001au,
author         = "Bossard, A. and Mohammedi, N.",
title          = "{Poisson-Lie duality in the string effective action}",
journal        = "Nucl. Phys.",
volume         = "B619",
year           = "2001",
pages          = "128-154",
doi            = "10.1016/S0550-3213(01)00541-7",
eprint         = "hep-th/0106211",
archivePrefix  = "arXiv",
primaryClass   = "hep-th",
SLACcitation   = "%%CITATION = HEP-TH/0106211;%%"
}

@article{VonUnge:2002xjf,
author         = "von Unge, Rikard",
title          = "{Poisson-Lie T-plurality}",
journal        = "JHEP",
volume         = "07",
year           = "2002",
pages          = "014",
doi            = "10.1088/1126-6708/2002/07/014",
eprint         = "hep-th/0205245",
archivePrefix  = "arXiv",
primaryClass   = "hep-th",
SLACcitation   = "%%CITATION = HEP-TH/0205245;%%"
}

@article{Hlavaty:2004jp,
author         = "Hlavat\'{y}, L. and \v{S}nobl, L.",
title          = "{Poisson-Lie T plurality of three-dimensional conformally invariant sigma models}",
journal        = "JHEP",
volume         = "05",
year           = "2004",
pages          = "010",
doi            = "10.1088/1126-6708/2004/05/010",
eprint         = "hep-th/0403164",
archivePrefix  = "arXiv",
primaryClass   = "hep-th",
SLACcitation   = "%%CITATION = HEP-TH/0403164;%%"
}

@article{Hlavaty:2004mr,
author         = "Hlavat\'{y}, L. and \v{S}nobl, L.",
title          = "{Poisson-Lie T-plurality of three-dimensional conformally invariant sigma models. II. Nondiagonal metrics and dilaton puzzle}",
journal        = "JHEP",
volume         = "10",
year           = "2004",
pages          = "045",
doi            = "10.1088/1126-6708/2004/10/045",
eprint         = "hep-th/0408126",
archivePrefix  = "arXiv",
primaryClass   = "hep-th",
SLACcitation   = "%%CITATION = HEP-TH/0408126;%%"
}

@article{Jurco:2017gii,
author         = "Jur\v{c}o, Branislav and Vysok\'{y}, Jan",
title          = "{Poisson-Lie T-duality of string effective actions: A new approach to the dilaton puzzle}",
journal        = "J. Geom. Phys.",
volume         = "130",
year           = "2018",
pages          = "1-26",
doi            = "10.1016/j.geomphys.2018.03.019",
eprint         = "1708.04079",
archivePrefix  = "arXiv",
primaryClass   = "hep-th",
SLACcitation   = "%%CITATION = ARXIV:1708.04079;%%"
}

@article{Borsato:2016ose,
author         = "Borsato, Riccardo and Wulff, Linus",
title          = "{Target space supergeometry of $\eta$ and $\lambda$-deformed strings}",
journal        = "JHEP",
volume         = "10",
year           = "2016",
pages          = "045",
doi            = "10.1007/JHEP10(2016)045",
eprint         = "1608.03570",
archivePrefix  = "arXiv",
primaryClass   = "hep-th",
reportNumber   = "IMPERIAL-TP-LW-2016-03",
SLACcitation   = "%%CITATION = ARXIV:1608.03570;%%"
}

@article{Sakatani:2016fvh,
author         = "Sakatani, Yuho and Uehara, Shozo and Yoshida, Kentaroh",
title          = "{Generalized gravity from modified DFT}",
journal        = "JHEP",
volume         = "04",
year           = "2017",
pages          = "123",
doi            = "10.1007/JHEP04(2017)123",
eprint         = "1611.05856",
archivePrefix  = "arXiv",
primaryClass   = "hep-th",
reportNumber   = "KUNS-2653",
SLACcitation   = "%%CITATION = ARXIV:1611.05856;%%"
}

@article{Baguet:2016prz,
author         = "Baguet, Arnaud and Magro, Marc and Samtleben, Henning",
title          = "{Generalized IIB supergravity from exceptional field theory}",
journal        = "JHEP",
volume         = "03",
year           = "2017",
pages          = "100",
doi            = "10.1007/JHEP03(2017)100",
eprint         = "1612.07210",
archivePrefix  = "arXiv",
primaryClass   = "hep-th",
SLACcitation   = "%%CITATION = ARXIV:1612.07210;%%"
}

@article{Sakamoto:2017wor,
author         = "Sakamoto, Jun-ichi and Sakatani, Yuho and Yoshida, Kentaroh",
title          = "{Weyl invariance for generalized supergravity backgrounds from the doubled formalism}",
journal        = "PTEP",
volume         = "2017",
year           = "2017",
number         = "5",
pages          = "053B07",
doi            = "10.1093/ptep/ptx067",
eprint         = "1703.09213",
archivePrefix  = "arXiv",
primaryClass   = "hep-th",
reportNumber   = "KUNS-2668",
SLACcitation   = "%%CITATION = ARXIV:1703.09213;%%"
}

@article{Fernandez-Melgarejo:2017oyu,
author         = "Fernandez-Melgarejo, Jose J. and Sakamoto, Jun-ichi and Sakatani, Yuho and Yoshida, Kentaroh",
title          = "{$T$-folds from Yang-Baxter deformations}",
journal        = "JHEP",
volume         = "12",
year           = "2017",
pages          = "108",
doi            = "10.1007/JHEP12(2017)108",
eprint         = "1710.06849",
archivePrefix  = "arXiv",
primaryClass   = "hep-th",
reportNumber   = "KUNS-2682, YITP-17-110",
SLACcitation   = "%%CITATION = ARXIV:1710.06849;%%"
}

@article{Lust:2018jsx,
author         = "L{\"u}st, Dieter and Osten, David",
title          = "{Generalised fluxes, Yang-Baxter deformations and the O(d,d) structure of non-abelian T-duality}",
journal        = "JHEP",
volume         = "05",
year           = "2018",
pages          = "165",
doi            = "10.1007/JHEP05(2018)165",
eprint         = "1803.03971",
archivePrefix  = "arXiv",
primaryClass   = "hep-th",
reportNumber   = "LMU-ASC 11/18, MPP-2018-35, LMU-ASC-11-18, MPP-2018-35",
SLACcitation   = "%%CITATION = ARXIV:1803.03971;%%"
}

@article{Sakamoto:2018krs,
author         = "Sakamoto, Jun-ichi and Sakatani, Yuho",
title          = "{Local $\beta$-deformations and Yang-Baxter sigma model}",
journal        = "JHEP",
volume         = "06",
year           = "2018",
pages          = "147",
doi            = "10.1007/JHEP06(2018)147",
eprint         = "1803.05903",
archivePrefix  = "arXiv",
primaryClass   = "hep-th",
reportNumber   = "KUNS-2717, KUNS-2717",
SLACcitation   = "%%CITATION = ARXIV:1803.05903;%%"
}

@article{Severa:2017kcs,
author         = "\v{S}evera, Pavol",
title          = "{On integrability of 2-dimensional $\sigma$-models of Poisson-Lie type}",
journal        = "JHEP",
volume         = "11",
year           = "2017",
pages          = "015",
doi            = "10.1007/JHEP11(2017)015",
eprint         = "1709.02213",
archivePrefix  = "arXiv",
primaryClass   = "hep-th",
SLACcitation   = "%%CITATION = ARXIV:1709.02213;%%"
}

@article{Demulder:2015lva,
author         = "Demulder, Saskia and Sfetsos, Konstantinos and Thompson, Daniel C.",
title          = "{Integrable $\lambda$-deformations: squashing coset CFTs and $AdS_5 \times S^5$}",
journal        = "JHEP",
volume         = "07",
year           = "2015",
pages          = "019",
doi            = "10.1007/JHEP07(2015)019",
eprint         = "1504.02781",
archivePrefix  = "arXiv",
primaryClass   = "hep-th",
SLACcitation   = "%%CITATION = ARXIV:1504.02781;%%"
}

@article{Appadu:2017xku,
author         = "Appadu, Calan and Hollowood, Timothy J. and Miramontes, J. Luis and Price, Dafydd and Schmidtt, David M.",
title          = "{Giant magnons of string theory in the lambda background}",
journal        = "JHEP",
volume         = "07",
year           = "2017",
pages          = "098",
doi            = "10.1007/JHEP07(2017)098",
eprint         = "1704.05437",
archivePrefix  = "arXiv",
primaryClass   = "hep-th",
}

@article{Alekseev:1995ym,
author         = "Alekseev, A. {\relax Yu}. and Klim\v{c}\'{i}k, C. and Tseytlin, Arkady A.",
title          = "{Quantum Poisson-Lie T-duality and WZNW model}",
journal        = "Nucl. Phys.",
volume         = "B458",
year           = "1996",
pages          = "430-444",
doi            = "10.1016/0550-3213(95)00575-7",
eprint         = "hep-th/9509123",
archivePrefix  = "arXiv",
primaryClass   = "hep-th",
reportNumber   = "CERN-TH-95-251, ETH-TH-95-26, IMPERIAL-TP-94-95-61, UUITP-16-95",
SLACcitation   = "%%CITATION = HEP-TH/9509123;%%"
}

@article{Valent:2009nv,
author         = "Valent, Galliano and Klim\v{c}\'{i}k, Ctirad and Squellari, Romain",
title          = "{One loop renormalizability of the Poisson-Lie sigma models}",
journal        = "Phys. Lett.",
volume         = "B678",
year           = "2009",
pages          = "143-148",
doi            = "10.1016/j.physletb.2009.06.001",
eprint         = "0902.1459",
archivePrefix  = "arXiv",
primaryClass   = "hep-th",
SLACcitation   = "%%CITATION = ARXIV:0902.1459;%%"
}

@article{Sfetsos:2009dj,
author         = "Sfetsos, Konstadinos and Siampos, Konstadinos",
title          = "{Quantum equivalence in Poisson-Lie T-duality}",
journal        = "JHEP",
volume         = "06",
year           = "2009",
pages          = "082",
doi            = "10.1088/1126-6708/2009/06/082",
eprint         = "0904.4248",
archivePrefix  = "arXiv",
primaryClass   = "hep-th",
SLACcitation   = "%%CITATION = ARXIV:0904.4248;%%"
}

@article{Avramis:2009xi,
author         = "Avramis, Spyros D. and Derendinger, Jean-Pierre and Prezas, Nikolaos",
title          = "{Conformal chiral boson models on twisted doubled tori and non-geometric string vacua}",
journal        = "Nucl. Phys.",
volume         = "B827",
year           = "2010",
pages          = "281-310",
doi            = "10.1016/j.nuclphysb.2009.11.003",
eprint         = "0910.0431",
archivePrefix  = "arXiv",
primaryClass   = "hep-th",
SLACcitation   = "%%CITATION = ARXIV:0910.0431;%%"
}

@article{Sfetsos:2009vt,
author         = "Sfetsos, K. and Siampos, K. and Thompson, Daniel C.",
title          = "{Renormalization of Lorentz non-invariant actions and manifest T-duality}",
journal        = "Nucl. Phys.",
volume         = "B827",
year           = "2010",
pages          = "545-564",
doi            = "10.1016/j.nuclphysb.2009.11.001",
eprint         = "0910.1345",
archivePrefix  = "arXiv",
primaryClass   = "hep-th",
reportNumber   = "QMUL-PH-09-22",
SLACcitation   = "%%CITATION = ARXIV:0910.1345;%%"
}

@article{Hlavaty:2012sg,
author         = "Hlavat\'{y}, Ladislav and Navr\'{a}til, Josef and \v{S}nobl, Libor",
title          = "{On renormalization of Poisson-Lie T-plural sigma models}",
journal        = "Acta Polytech.",
volume         = "53",
year           = "2013",
number         = "5",
pages          = "433-437",
doi            = "10.14311/AP.2013.53.0433",
eprint         = "1212.5936",
archivePrefix  = "arXiv",
primaryClass   = "hep-th",
SLACcitation   = "%%CITATION = ARXIV:1212.5936;%%"
}

%journal
@article{Hassler:2017yza,
author         = "Hassler, Falk",
title          = "{Poisson-Lie T-duality in double field theory}",
year           = "2017",
eprint         = "1707.08624",
archivePrefix  = "arXiv",
primaryClass   = "hep-th",
SLACcitation   = "%%CITATION = ARXIV:1707.08624;%%"
}

@article{Delduc:2016ihq,
author         = "Delduc, Francois and Lacroix, Sylvain and Magro, Marc and Vicedo, Benoit",
title          = "{On $q$-deformed symmetries as Poisson-Lie symmetries and application to Yang-Baxter type models}",
journal        = "J. Phys.",
volume         = "A49",
year           = "2016",
number         = "41",
pages          = "415402",
doi            = "10.1088/1751-8113/49/41/415402",
eprint         = "1606.01712",
archivePrefix  = "arXiv",
primaryClass   = "hep-th",
SLACcitation   = "%%CITATION = ARXIV:1606.01712;%%"
}

@article{Arutyunov:2013ega,
author         = "Arutyunov, Gleb and Borsato, Riccardo and Frolov, Sergey",
title          = "{S-matrix for strings on $\eta$-deformed $AdS_5 \times S^5$}",
journal        = "JHEP",
volume         = "04",
year           = "2014",
pages          = "002",
doi            = "10.1007/JHEP04(2014)002",
eprint         = "1312.3542",
archivePrefix  = "arXiv",
primaryClass   = "hep-th",
reportNumber   = "ITP-UU-13-31, SPIN-13-23, HU-MATHEMATIK-2013-24, TCD-MATH-13-16",
SLACcitation   = "%%CITATION = ARXIV:1312.3542;%%"
}

@article{Klimcik:1995ux,
author         = "Klim\v{c}\'{i}k, C. and \v{S}evera, P.",
title          = "{Dual non-abelian duality and the Drinfel'd double}",
journal        = "Phys. Lett.",
volume         = "B351",
year           = "1995",
pages          = "455-462",
doi            = "10.1016/0370-2693(95)00451-P",
eprint         = "hep-th/9502122",
archivePrefix  = "arXiv",
primaryClass   = "hep-th",
reportNumber   = "CERN-TH-95-39, CERN-TH-95-039",
SLACcitation   = "%%CITATION = HEP-TH/9502122;%%"
}

@article{Klimcik:1995jn,
author         = "Klim\v{c}\'{i}k, C.",
title          = "{Poisson-Lie T-duality}",
journal        = "Nucl. Phys. Proc. Suppl.",
volume         = "46",
year           = "1996",
pages          = "116-121",
doi            = "10.1016/0920-5632(96)00013-8",
eprint         = "hep-th/9509095",
archivePrefix  = "arXiv",
primaryClass   = "hep-th",
reportNumber   = "CERN-TH-95-248",
SLACcitation   = "%%CITATION = HEP-TH/9509095;%%"
}

@article{Hoare:2014pna,
author         = "Hoare, B. and Roiban, R. and Tseytlin, A. A.",
title          = "{On deformations of $AdS_n \times S^n$ supercosets}",
journal        = "JHEP",
volume         = "06",
year           = "2014",
pages          = "002",
doi            = "10.1007/JHEP06(2014)002",
eprint         = "1403.5517",
archivePrefix  = "arXiv",
primaryClass   = "hep-th",
reportNumber   = "IMPERIAL-TP-AT-2014-02, HU-EP-14-10",
SLACcitation   = "%%CITATION = ARXIV:1403.5517;%%"
}

@article{Delduc:2017brb,
author         = "Delduc, Francois and Kameyama, Takashi and Magro, Marc and Vicedo, Benoit",
title          = "{Affine $q$-deformed symmetry and the classical Yang-Baxter $\sigma$-model}",
journal        = "JHEP",
volume         = "03",
year           = "2017",
pages          = "126",
doi            = "10.1007/JHEP03(2017)126",
eprint         = "1701.03691",
archivePrefix  = "arXiv",
primaryClass   = "hep-th",
SLACcitation   = "%%CITATION = ARXIV:1701.03691;%%"
}

@article{Fateev:1992tk,
author         = "Fateev, V. A. and Onofri, E. and Zamolodchikov, Alexei B.",
title          = "{The sausage model (integrable deformations of O(3) sigma model)}",
journal        = "Nucl. Phys.",
volume         = "B406",
year           = "1993",
pages          = "521-565",
doi            = "10.1016/0550-3213(93)90001-6",
reportNumber   = "PAR-LPTHE-92-46, LPTHE-92-46",
SLACcitation   = "%%CITATION = NUPHA,B406,521;%%"
}

@article{Arutyunov:2012zt,
author         = "Arutyunov, Gleb and de Leeuw, Marius and van Tongeren, Stijn J.",
title          = "{The Quantum Deformed Mirror TBA I}",
journal        = "JHEP",
volume         = "10",
year           = "2012",
pages          = "090",
doi            = "10.1007/JHEP10(2012)090",
eprint         = "1208.3478",
archivePrefix  = "arXiv",
primaryClass   = "hep-th",
SLACcitation   = "%%CITATION = ARXIV:1208.3478;%%"
}

@article{Arutyunov:2012ai,
author         = "Arutyunov, Gleb and de Leeuw, Marius and van Tongeren, Stijn J.",
title          = "{The Quantum Deformed Mirror TBA II}",
journal        = "JHEP",
volume         = "02",
year           = "2013",
pages          = "012",
doi            = "10.1007/JHEP02(2013)012",
eprint         = "1210.8185",
archivePrefix  = "arXiv",
primaryClass   = "hep-th",
reportNumber   = "ITP-UU-12-34, SPIN-12-32",
SLACcitation   = "%%CITATION = ARXIV:1210.8185;%%"
}

@article{Arutynov:2014ota,
author         = "Arutyunov, Gleb and de Leeuw, Marius and van Tongeren, Stijn J.",
title          = "{The exact spectrum and mirror duality of the $(AdS_5 \times S^5)_\eta$ superstring}",
journal        = "Theor. Math. Phys.",
volume         = "182",
year           = "2015",
number         = "1",
pages          = "23-51",
doi            = "10.1007/s11232-015-0243-9",
note           = "[Teor. Mat. Fiz. 182, 28 (2014)]",
eprint         = "1403.6104",
archivePrefix  = "arXiv",
primaryClass   = "hep-th",
SLACcitation   = "%%CITATION = ARXIV:1403.6104;%%"
}

@article{Klabbers:2017vtw,
author         = "Klabbers, Rob and van Tongeren, Stijn J.",
title          = "{Quantum Spectral Curve for the eta-deformed $AdS_5 \times S^5$ superstring}",
journal        = "Nucl. Phys.",
volume         = "B925",
year           = "2017",
pages          = "252-318",
doi            = "10.1016/j.nuclphysb.2017.10.005",
eprint         = "1708.02894",
archivePrefix  = "arXiv",
primaryClass   = "hep-th",
reportNumber   = "ZMP-HH-17-25, HU-EP-17-21",
SLACcitation   = "%%CITATION = ARXIV:1708.02894;%%"
}

@article{Arutyunov:2014cra,
author         = "Arutyunov, Gleb and van Tongeren, Stijn J.",
title          = "{$AdS_5 \times S^5$ mirror model as a string sigma model}",
journal        = "Phys. Rev. Lett.",
volume         = "113",
year           = "2014",
pages          = "261605",
doi            = "10.1103/PhysRevLett.113.261605",
eprint         = "1406.2304",
archivePrefix  = "arXiv",
primaryClass   = "hep-th",
reportNumber   = "HU-EP-14-21, HU-MATH-14-12, ITP-UU-14-18, SPIN-14-16",
SLACcitation   = "%%CITATION = ARXIV:1406.2304;%%"
}

@article{Pachol:2015mfa,
author         = "Pacho\l{}, Anna and van Tongeren, Stijn J.",
title          = "{Quantum deformations of the flat space superstring}",
journal        = "Phys. Rev.",
volume         = "D93",
year           = "2016",
pages          = "026008",
doi            = "10.1103/PhysRevD.93.026008",
eprint         = "1510.02389",
archivePrefix  = "arXiv",
primaryClass   = "hep-th",
reportNumber   = "HU-EP-15-48, HU-MATH-15-13",
SLACcitation   = "%%CITATION = ARXIV:1510.02389;%%"
}

@article{Beisert:2008tw,
author         = "Beisert, Niklas and Koroteev, Peter",
title          = "{Quantum deformations of the one-dimensional Hubbard model}",
journal        = "J. Phys.",
volume         = "A41",
year           = "2008",
pages          = "255204",
doi            = "10.1088/1751-8113/41/25/255204",
eprint         = "0802.0777",
archivePrefix  = "arXiv",
primaryClass   = "hep-th",
reportNumber   = "AEI-2008-003, ITEP-TH-06-08",
SLACcitation   = "%%CITATION = ARXIV:0802.0777;%%"
}

@article{Hoare:2011wr,
author         = "Hoare, Ben and Hollowood, Timothy J. and Miramontes, J. Luis",
title          = "{q-deformation of the $AdS_5 \times S^5$ superstring S-matrix and its relativistic limit}",
journal        = "JHEP",
volume         = "03",
year           = "2012",
pages          = "015",
doi            = "10.1007/JHEP03(2012)015",
eprint         = "1112.4485",
archivePrefix  = "arXiv",
primaryClass   = "hep-th",
reportNumber   = "IMPERIAL-TP-11-BH-03",
SLACcitation   = "%%CITATION = ARXIV:1112.4485;%%"
}

@article{Hoare:2014kma,
author         = "Hoare, Ben and Pittelli, Antonio and Torrielli, Alessandro",
title          = "{Integrable S-matrices, massive and massless modes and the $AdS_2 \times S^2$ superstring}",
journal        = "JHEP",
volume         = "11",
year           = "2014",
pages          = "051",
doi            = "10.1007/JHEP11(2014)051",
eprint         = "1407.0303",
archivePrefix  = "arXiv",
primaryClass   = "hep-th",
reportNumber   = "HU-EP-14-28, DMUS--MP--14-05",
SLACcitation   = "%%CITATION = ARXIV:1407.0303;%%"
}

@article{Wulff:2014kja,
author         = "Wulff, Linus",
title          = "{Superisometries and integrability of superstrings}",
journal        = "JHEP",
volume         = "05",
year           = "2014",
pages          = "115",
doi            = "10.1007/JHEP05(2014)115",
eprint         = "1402.3122",
archivePrefix  = "arXiv",
primaryClass   = "hep-th",
reportNumber   = "IMPERIAL-TP-LW-2014-01",
SLACcitation   = "%%CITATION = ARXIV:1402.3122;%%"
}

@article{Wulff:2015mwa,
author         = "Wulff, Linus",
title          = "{On integrability of strings on symmetric spaces}",
journal        = "JHEP",
volume         = "09",
year           = "2015",
pages          = "115",
doi            = "10.1007/JHEP09(2015)115",
eprint         = "1505.03525",
archivePrefix  = "arXiv",
primaryClass   = "hep-th",
reportNumber   = "IMPERIAL-TP-LW-2015-01",
SLACcitation   = "%%CITATION = ARXIV:1505.03525;%%"
}

%journal
@article{Driezen:2018glg,
author         = "Driezen, Sibylle and Sevrin, Alexander and Thompson, Daniel C.",
title          = "{D-branes in $\lambda$-deformations}",
year           = "2018",
eprint         = "1806.10712",
archivePrefix  = "arXiv",
primaryClass   = "hep-th",
SLACcitation   = "%%CITATION = ARXIV:1806.10712;%%"
}

@article{Hull:1998vg,
author         = "Hull, C. M.",
title          = "{Timelike T-duality, de Sitter space, large N gauge theories and topological field theory}",
journal        = "JHEP",
volume         = "07",
year           = "1998",
pages          = "021",
doi            = "10.1088/1126-6708/1998/07/021",
eprint         = "hep-th/9806146",
archivePrefix  = "arXiv",
primaryClass   = "hep-th",
reportNumber   = "QMW-PH-98-28",
SLACcitation   = "%%CITATION = HEP-TH/9806146;%%"
}

@article{Lunin:2014tsa,
author         = "Lunin, O. and Roiban, R. and Tseytlin, A. A.",
title          = "{Supergravity backgrounds for deformations of $AdS_n \times S^n$ supercoset string models}",
journal        = "Nucl. Phys.",
volume         = "B891",
year           = "2015",
pages          = "106-127",
doi            = "10.1016/j.nuclphysb.2014.12.006",
eprint         = "1411.1066",
archivePrefix  = "arXiv",
primaryClass   = "hep-th",
reportNumber   = "IMPERIAL-TP-AT-2014-07",
SLACcitation   = "%%CITATION = ARXIV:1411.1066;%%"
}

@article{Hoare:2016wsk,
author         = "Hoare, B. and Tseytlin, A. A.",
title          = "{Homogeneous Yang-Baxter deformations as non-abelian duals of the $AdS_5$ sigma-model}",
journal        = "J. Phys.",
volume         = "A49",
year           = "2016",
number         = "49",
pages          = "494001",
doi            = "10.1088/1751-8113/49/49/494001",
eprint         = "1609.02550",
archivePrefix  = "arXiv",
primaryClass   = "hep-th",
reportNumber   = "IMPERIAL-TP-AT-2016-03",
SLACcitation   = "%%CITATION = ARXIV:1609.02550;%%"
}

@article{Hong:2018tlp,
author         = "Hong, Moonju and Kim, Yoonsoo and Colg\'{a}in, Eoin \'{O}",
title          = "{On non-Abelian T-duality for non-semisimple groups}",
year           = "2018",
eprint         = "1801.09567",
archivePrefix  = "arXiv",
primaryClass   = "hep-th",
SLACcitation   = "%%CITATION = ARXIV:1801.09567;%%"
}

@article{Wulff:2018aku,
author         = "Wulff, Linus",
title          = "{Trivial solutions of generalized supergravity vs non-abelian T-duality anomaly}",
journal        = "Phys. Lett.",
volume         = "B781",
year           = "2018",
pages          = "417-422",
doi            = "10.1016/j.physletb.2018.04.025",
eprint         = "1803.07391",
archivePrefix  = "arXiv",
primaryClass   = "hep-th",
SLACcitation   = "%%CITATION = ARXIV:1803.07391;%%"
}

\end{bibtex}

\bibliographystyle{nb}
\bibliography{\jobname}

\end{document}